\documentclass[%
reprint,
 amsmath,amssymb,
 aps,
]{revtex4-2}

\usepackage{graphicx}
\usepackage{dcolumn}
\usepackage{bm}

\pdfoutput=1
\usepackage[colorlinks=true, pdfstartview=FitV, linkcolor=blue, citecolor=blue, urlcolor=blue]{hyperref}
\usepackage[export]{adjustbox}    
\usepackage{lineno}

\usepackage{soul,xcolor}

\begin{document}

\preprint{ET12011}

\setstcolor{red}

\title{Stochastic Evolutionary Dynamics of Trust Games with Asymmetric Parameters}

\author{Ik Soo Lim}%
 \email{i.s.lim@bangor.ac.uk}
\affiliation{%
School of Computer Science and Electrical Engineering, Bangor University, Dean Street, Bangor, Gwynedd, LL57 1UT, UK
}%

\date{\today}

\begin{abstract}

Trusting in others and reciprocating that trust with trustworthy actions  are crucial to successful and prosperous societies.
The Trust Game has been widely used to quantitatively study trust and trustworthiness, involving a sequential exchange between an investor and a trustee.
The deterministic evolutionary game theory predicts no trust and no trustworthiness whereas the behavioural experiments with the one-shot anonymous Trust Game show that people substantially trust and respond trustworthily.
To explain these discrepancies, previous works often turn to additional mechanisms, which are borrowed from other games such as Prisoner's Dilemma.
Although these mechanisms lead to the evolution of trust and trustworthiness to an extent, the optimal or the most common strategy often involves no trustworthiness.
In this paper, 
we study the impact of asymmetric demographic parameters (e.g.\,different population sizes) on game dynamics of the Trust Game.
We show that, in weak-mutation limit, stochastic evolutionary dynamics with the asymmetric parameters  can lead to the evolution of high trust and high trustworthiness without any additional mechanisms in well-mixed finite populations.
Even full trust and near full trustworthiness can be the most common strategy. 
These results are qualitatively different from those of the previous works.
Our results thereby demonstrate rich evolutionary dynamics of the asymmetric Trust Game.

\end{abstract}

\maketitle


\section{Introduction}

	Prosocial behaviour is an important aspect of human interactions. 
Emergence  and maintenance of prosocial behaviours among self-interested individuals is 
a considerable focus of research across
 various disciplines including physics  \cite{nowak2006five}\cite{Van-Segbroeck:2009uq}\cite{Helbing:2010bx}\cite{Perc:2017aa}\cite{Lim:2018aa}\cite{Mittal:2020aa}. 
For instance, the evolution of cooperation in social dilemma situations such as Prisoner's Dilemma (PD) has attracted lots of attention.
	Evolutionary game theory is widely used to provide a theoretical framework to study the evolution of prosocial behaviours or strategies, 
where successful strategies are spread by reproduction in genetic evolution and imitation in cultural evolution \cite{SMITH:1973aa}\cite{Taylor:1978aa}\cite{hofbauer1998evolutionary}.
To explain the evolution of cooperation 
that is seemingly irrational and altruistic,
for instance,
various mechanisms  have been proposed; 
network reciprocity \cite{Szabo:2007tg}\cite{Lieberman:2005aa}, reputation \cite{Fu:2008fk}\cite{Nowak:1998aa}, and uncertainty-led stochastic dynamics \cite{traulsen2006stochastic}\cite{Taylor:2004aa}, etc.

Trusting in others and reciprocating that trust with trustworthy actions are central components of 
successful social and economic interactions among humans.
Higher levels of trusting and trustworthy behaviours have been associated with more efficient judicial systems, higher quality government bureaucracies, lower corruption, greater financial development, and better economic outcomes among other benefits for the society \cite{Johnson:2011aa}.

		To study trust and trustworthiness in  quantitative manners,
the Trust Game (TG) has been widely used in various disciplines
 \cite{Johnson:2011aa}\cite{Manapat:2013aa}\cite{Tarnita:2015aa}\cite{Berg:1995aa}\cite{McNamara:2009aa}\cite{Fehr:2009aa}\cite{Tzieropoulos:2013aa}\cite{Abbass:2016ac}.	
The TG involves a sequential exchange between an investor and a trustee without any contract to enforce agreements \cite{Berg:1995aa}. 
The investor starts with a stake of one monetary unit and invests or transfers some fraction of it to the trustee,
which measures a degree of trust. 
To represent the value created by interactions based on trust, the invested amount is multiplied by a factor.
The trustee then returns a certain fraction  of the enhanced investment to the investor,
which measures a degree of trustworthiness.

		According to the logic of the classical economic theory,
rational self-interest leads to no trust and no trustworthiness in a one-shot anonymous TG;
a self-interested trustee would not return anything 
and, therefore, a self-interested investor would not invest
\cite{Johnson:2011aa}.
Thus, the potential gains of trust and exchange are lost.
Deterministic models of evolutionary game theory yield the same outcome as the dooming prediction of classical economic theory \cite{Tarnita:2015aa}.
According to behavioural experiments with the TG, 
however,
people are willing to trust and reciprocate trust;
investors make transfers and trustees return substantial amounts to investors \cite{Johnson:2011aa}.

To explain this discrepancy between the theoretical predictions and  the behavioural experiment results,
additional mechanisms have been proposed in the framework of  evolutionary game theory.
Reputation about trustees can boost the evolution of  trust and trustworthiness \cite{Manapat:2013aa}\cite{King-Casas:2005aa}\cite{Masuda:2012aa}. 
In the limit of weak selection, stochastic evolutionary dynamics
due to randomness in finite populations can evolve some degrees of trust and trustworthiness
\cite{Manapat:2013aa}\cite{Tarnita:2015aa}. 
Allowing interactions and imitation with only neighbouring players,  
the networked structure of populations boosts the evolution of trust and trustworthiness, selecting for more trusting and trustworthy strategies than well-mixed populations \cite{Tarnita:2015aa}.
Note that these mechanisms have been originally proposed to evolve other prosocial behaviours such as
cooperation in 
the PD game.

			Compared to the symmetric games (e.g.\,PD) which have been extensively studied			with single-population models,
the TG has an additional complexity due to the two different roles of the game players,
each of which has its own set of strategies;
one for an investor and the other for a trustee.
Despite this asymmetric nature,
however, 
the TG is often symmetrised
\cite{Tarnita:2015aa}\cite{McNamara:2009aa}. 
In the symmetric TG,
each player takes turns playing investor and trustee roles
and, thus, a strategy of a player has  two components;
one for an investor role and the other, a trustee.
One of the key motivations behind the symmetrisation is that the TG can be studied with a single-population model
and thus the mechanisms to evolve prosocial behaviours in other games can be used for the TG 
as well
\cite{Tarnita:2015aa}.

	Due to the asymmetric nature of the TG,
however,
a two-population model is more natural for it;
one population for investors and the other for trustees,
each player having a single role.
The TG is then played between a player  from the investor population and a player from the trustee population, 
whereas imitation takes place between players  in the same populations.
Although there were previous attempts for this, they are limited 
or not asymmetric enough
in the sense that symmetric parameters were used
(e.g.\,the same selection strengths and population sizes between the two populations) \cite{Manapat:2013aa}\cite{Tarnita:2015aa}.
Studying both the single- and two-population models,
it was even asserted that the two-population model led to the same prediction as that of the single-population model \cite{Tarnita:2015aa}.
However, this conclusion is premature 
in that the two-population model was still based on the symmetric parameters.
It missed potentially richer evolutionary dynamics stemming from the asymmetric parameters between the populations.

	In this paper, we introduce a two-population model of the TG with asymmetric demographic parameters.
We show that stochastic evolutionary dynamics with the asymmetric parameters 
yields evolutionary outcomes richer than those of the symmetric parameters or the single-population models.
In particular, a combination of  
stronger selection in the investor population 
and weak selection in the trustee population can lead to the evolution of high trust and high trustworthiness without any additional mechanisms in well-mixed populations.
Even the most common strategy can involve
high trustworthiness.
These outcomes are significantly different from the previous works on the TG,
which predicted that 
null trustworthiness is the most common.

\section{Model and Methods}

\subsection{Trust Games}

	In the TG,
a pair of players have an investor--trustee transaction.
The investor starts with an initial stake of one unit
and transfers some fraction $0 \le p \le 1$ of it to the trustee.
The trustee receives the transferred amount multiplied by a factor $b > 1$, 
the latter of which represents the value generated by trust-based interactions.
	The trustee then returns some fraction $0\le r\le 1$ of the enhanced transfer amount $pb$ to the investor.
The payoffs of the investor and the trustee from a transaction are respectively given by
\[
\pi_{\scriptscriptstyle I}(p,r) =
1-p +p b r,	\quad
\pi_{\scriptscriptstyle T}(p,r)
=pb(1-r).
\]

The fitness of a player playing the TG is given by
\[
f=1 +\beta \pi,
\]
where $\beta$ denotes the selection strength
and $\pi$, the mean payoff of a player.

\subsection{Moran Process}

	We consider the stochastic evolutionary game dynamics
in finite populations.
For the evolutionary process,
we use the Moran process \cite{moran1962statistical}.
In each time step,
an individual is picked at random to switch strategy.
The focal individual imitates a strategy of
another individual that is picked with probability proportional to fitness.
With probability $u$, a mutation occurs and the focal individual instead switches to a random strategy.
The Moran process has been used as a model of biological evolution as well as imitation learning \cite{nowak2006evolutionary}.

\subsection{Discretisation of Strategy Space}
 
 	We discretise the continuous strategies $0 \le p\le  1$ and $0\le r\le 1$ in increments of $1/L_p$ and $1/L_r$, respectively,
where $L_p$ and $L_r$ are positive  integers;
$p \in \mathcal{S}_{\scriptscriptstyle I}=\{p_m | m=1,2,\ldots,L_p+1\}=\{0,1/L_p,2/L_p,\ldots,(L_p -1)/L_p,1\}$ for investors
and $r\in \mathcal{S}_{\scriptscriptstyle T}= \{r_n | n=1,2,\ldots,L_r+1\} =\{0,1/L_r,2/L_r,\ldots,(L_r -1)/L_r,1\}$ for trustees.
With the discretisation,
we can use methods assuming discrete strategies such as weak-mutation limit.

\subsection{Weak-Mutation Limit}

	In a finite population, with no mutation, imitation-led stochastic dynamics yields fixation 
and, thus, the population state becomes `pure' or homogeneous;
i.e.\,all individuals in the population use the same strategy.

	With mutation,
we use the weak-mutation limit $u \rightarrow 0$,
which is a common assumption in evolutionary game theory \cite{Fudenberg:2006aa}\cite{Veller:2016aa}\cite{Hauert:2007fk}\cite{Van-Segbroeck:2009uq}\cite{Veller:2017aa}.
A population consists of one or two types of strategies at any time;
a single mutant in the otherwise pure population will either  perish or completely take over the resident population
before another mutant occurs.
	The evolutionary process, therefore, simplifies to an  embedded dynamics over just the pure population states $\{s_1 ,\ldots , s_i, \ldots,s_K\}$,
where $s_i =(p_m,r_n)\in \mathcal{S}_{\scriptscriptstyle I} \otimes \mathcal{S}_{\scriptscriptstyle T}$ and  
$K=|\mathcal{S}_{\scriptscriptstyle I}| \times |\mathcal{S}_{\scriptscriptstyle T}|$.	
More specifically,
we use the scheme
$i = (m-1)|\mathcal{S}_{\scriptscriptstyle T}| +n$,
$1\le m\le |\mathcal{S}_{\scriptscriptstyle I}|$ and $1\le n\le |\mathcal{S}_{\scriptscriptstyle T}|$.
In the embedded dynamics,
a population transitions between the pure states with probabilities determined by the relative frequency of mutant appearance and the fixation probabilities of these mutants \cite{Fudenberg:2006aa}.
Given the stochastic dynamics,
what we are interested in is the stationary distribution $\lambda=(\lambda_1,\ldots,\lambda_K)$, i.e.\,the proportion of time spent in each of pure states in the long run
or, equivalently, the stationary abundance of these discrete strategies.
The stationary distribution $\lambda$ is uniquely determined and can be obtained by solving a left eigenequation
\begin{equation}
\lambda \Lambda =\lambda,
\label{eq_eigen}
\end{equation}
where $\Lambda$ is a transition matrix for an ergodic Markov chain over the pure state space $\{s_1 ,\ldots , s_K\}$.
The entries of the $K\times K$ matrix $\Lambda$ are given by
\[
	\Lambda_{ij} =\mu_{ij}  \rho_{ij} \text{ for } i\ne j,
	\quad \Lambda_{ii} =1 -\sum_{j\ne i} \mu_{ij}  \rho_{ij},
\]
where	 $\mu_{ij}$ is the probability that, in pure state $i$, a single mutant of type $j$
arises
and $\rho_{ij}$ is the fixation probability that this mutant takes over the resident population,
leading to pure state $j$.	
	Note that,
with the weak-mutation limit $\mu_{ij} \rightarrow 0$,
every diagonal entry $\Lambda_{ii}$ is non-negative.

\subsection{Single-Population Formulation}

	We start with a single-population model of the symmetric TG in the weak-mutation limit,
whereas a previous work studied the TG in a weak-selection limit but not a weak-mutation limit \cite{Tarnita:2015aa}.
In a population of size $N$, each  player can be both an investor and a trustee with equal probability.
In other words,
given a pair of players,
they play the TG, each taking turns playing investor and trustee roles.
We specify a player's strategy as a tuple $s=(p,r) \in\mathcal{S}_{\scriptscriptstyle I} \otimes \mathcal{S}_{\scriptscriptstyle T}$.
The (mean) payoff $\pi_s(s^\prime)$ that a player with strategy $s=(p,r)$ gets from an interaction with another player with strategy $s^\prime = (p^\prime, r^\prime)$ is given by	
\begin{equation}
\pi_{s}(s^\prime) =\frac{1}{2} \pi_{\scriptscriptstyle I}(p,r^\prime) 
+\frac{1}{2} \pi_{\scriptscriptstyle T}(p^\prime,r).
\label{eq_payoff_single}
\end{equation}

\subsubsection{Weak--Mutation Limit}

In the weak-mutation limit,
there are at most two types of strategies present in a population.
Let $s_i$ and $s_j$ denote resident and mutant strategies, 
respectively, where $1\le i,j\le K$.
The mean payoff $\pi_i(k)$ of an $i$-player
and $\pi_j(k)$ of a $j$-player in a population consisting of $N-k$ $i$-players and $k$ $j$-players are given by
\begin{align*}
        \pi_i(k) &=\frac{k}{N-1}\pi_{s_i}(s_j) +\frac{N-k-1}{N-1}\pi_{s_i}(s_i),\\
        \pi_j(k) &=\frac{k-1}{N-1}\pi_{s_j}(s_j) +\frac{N-k}{N-1}\pi_{s_j}(s_i).
\end{align*}

For $i\ne j$,	the probability that, in
  a population of $i$-players,  a single mutant of $j$-player reaches fixation is given by 
\[
	\rho_{ij} =\rho_{i\rightarrow j} =\frac{1}{1 +\sum_{q=1}^{N-1}\Pi _{k=1}^q \frac{f_i(k)}{f_j(k)}},
\]
where 
$f_i(k)$ and $f_j(k)$ are the fitness of an $i$-player
and a $j$-player, respectively,
when there are $k$  $j$-players in the population \cite{Taylor:2004aa}.
In the embedded dynamics, 
 thus, $\rho_{ij}$ is the probability
that pure state $i$ switches to pure state $j$
given a single mutant of $j$-player arising in a population of $i$-players.
With $\mu_{ij}=Nu/K$,
the transition matrix $\Lambda$ is given by
\[
\Lambda_{ij} 
=\frac{N u}{K}\left(1 +\sum_{q=1}^{N-1}\Pi _{k=1}^q \frac{f_i(k)}{f_j(k)}\right)^{-1} \text{ for } i\ne j.
\]

\subsection{Two-Population Formulation}

	In the asymmetric TG, 
each individual plays a single role of either an investor or a trustee, exclusively.
Thus, we have two populations;
one consisting of investors and the other of trustees.
Interactions of playing the TG  are inter-population events,
whereas imitations of strategies are intra-population events.
An individual from the investor population plays the TG with an individual from the trustee population.
An investor imitates the strategy of another investor,
whereas a trustee imitates that of another trustee.

\subsubsection{Weak-mutation Limit}

Under the weak-mutation limit,
there are at most three types of strategies in the two-population system.
Both populations are in pure states
unless in a transition period due to a rare mutation.
If a mutant arises in one of the populations,
the extinction or fixation of it  is settled 
before another mutant appears either in the same or the other population.

Unlike the single-population model,
the payoffs of resident and mutant  strategies in one population are constant 
during the extinction-fixation period \cite{Veller:2016aa}.
This is so because the payoffs of them
depend on 
the state of the other population
that is 
in the same pure state during the extinction-fixation period.	
Hence, mean payoff $\pi_l$ of a resident player and $\pi_l^{\prime}$ of a mutant player in population $l \in\{I,T\}$ are given by
\begin{align*}
\pi_{\scriptscriptstyle I} &=\pi_{\scriptscriptstyle I}(p,r), & \pi_{\scriptscriptstyle I}^{\prime} &=\pi_{\scriptscriptstyle I}(p^\prime,r)
& \text{for an investor mutation},
\\
\pi_{\scriptscriptstyle T} &=\pi_{\scriptscriptstyle T}(p,r), 
&\pi_{\scriptscriptstyle T}^{\prime} &=\pi_{\scriptscriptstyle T}(p,r^\prime)
& \text{for a trustee mutation}.
\end{align*}

The fitness of a player in population $l$ is given by
\[
f_l(p,r) =1+\beta_l \pi_l(p,r),
\]
where $\beta_l$ is selection strength in population $l$.

\subsubsection{Fixation Probabilities}

Note that
pure state $s_i =(p_m, r_n)\in\mathcal{S}_{\scriptscriptstyle I}\otimes\mathcal{S}_{\scriptscriptstyle T}$ of the two-population system 
is a tuple of pure state $p_m\in\mathcal{S}_{\scriptscriptstyle I}$ of the investor population and pure state $r_n\in\mathcal{S}_{\scriptscriptstyle T}$ of the trustee population.
The embedded dynamics over pure states of the two-population system is formally equivalent to that of the single population
in the sense that each of them
can be viewed as the dynamics over the same finite strategy space $\mathcal{S}_{\scriptscriptstyle I} \otimes \mathcal{S}_{\scriptscriptstyle T}$.
They differ only in values of the transition matrix entries.

The frequency-independent selection in the two-population model allows us to use the well-known formula for fixation probability under
the Moran process \cite{nowak2006evolutionary}.
The fixation probability that, in population $l$ otherwise pure for a resident strategy of fitness $f_l$, 
a single mutant of fitness $f_l^{\prime}$ takes over the population
is given by
\[
\rho_l(f_l,f_l^{\prime})=
\left\{
\begin{aligned}
&\frac{1-f_l/f_l^{\prime}}{1-\left(f_l/f_l^{\prime}\right)^{N_l}}
&\text{\quad for } f_l^{\prime}\ne f_l \\
&\frac{1}{N_l} 
&\text{\quad for } f_l^{\prime} =f_l,
\end{aligned}
\right.
\]
where $N_l$ is the size of population $l$.

\subsubsection{Transition Matrix}

For $i\ne j$,
the (one-step) transition probabilities $\Lambda_{ij}$ in the embedded dynamics are given by
\begin{widetext}
\begin{equation} 
\Lambda_{ij} =\left\{
\begin{aligned}
\Lambda_{i(p_m,r_n)j(p_{m^\prime},r_n)} 
&=\frac{ N_{\scriptscriptstyle I} u_{\scriptscriptstyle I}}{g_{\scriptscriptstyle I}|\mathcal{S}_{\scriptscriptstyle I}|} \rho_{\scriptscriptstyle I}\left(f_{\scriptscriptstyle I}(p_m,r_n), f_{\scriptscriptstyle I}(p_{m^\prime},r_n)\right) 
&\text{ for } p_m\ne p_{m^\prime}
\\
\Lambda_{i(p_m,r_n)j(p_m,r_{n^\prime})} 
&=\frac{ N_{\scriptscriptstyle T} u_{\scriptscriptstyle T}}{g_{\scriptscriptstyle T}|\mathcal{S}_{\scriptscriptstyle T}|} \rho_{\scriptscriptstyle T}\left(f_{\scriptscriptstyle T}(p_m,r_n),f_{\scriptscriptstyle T}(p_m,r_{n^\prime})\right)  
& \text{ for } r_n\ne r_{n^\prime}
\\
\Lambda_{i(p_m,r_n)j(p_{m^\prime},r_{n^\prime})} &=0 \quad
& \text{ for } p_m\ne p_{m^\prime} \text{ and } r_n\ne r_{n^\prime},
\end{aligned}
\right.
\label{eq_transition_prob}
\end{equation}
\end{widetext}
where $u_l$ and $g_l$
are mutation rate per indivdiual and generation time in population $l\in\{I,T\}$, respectively.
	
	Note that the weak-mutation limit in the two-population model constrains a one-step transition in specific ways.
There are only $|\mathcal{S}_{\scriptscriptstyle I}|+|\mathcal{S}_{\scriptscriptstyle T}| -1$ states available  for a one-step transition from a pure state $s_i
=s_{i(p_m,r_n)}$
since mutation exclusively occurs in either the investor or trustee populations but not both under the weak-mutation limit.
This contrasts to the single-population model,
where there are $|\mathcal{S}_{\scriptscriptstyle I}| \times |\mathcal{S}_{\scriptscriptstyle T}| -1$ states available for the transition without the constraint.
More importantly, in the two-population model,
asymmetric parameters are naturally set
since each population $l$ has its own parameters 
($N_l, u_l, g_l,  \beta_l$ and $|\mathcal{S}_l|$)
and their values can be different from 
corresponding values 
of the other population, in general.
The asymmetry in these parameters can  
lead to stationary distributions qualitatively different from those stemming from symmetric parameters or the single-population model.

\subsubsection{Weak-selection Limit}

We can expand the fixation probability $\rho(f,f^{\prime})$ by Taylor series
\[
\rho(f,f^{\prime})
=\frac{1-f/f^{\prime}}{1-\left(f/f^{\prime}\right)^{N}}
= \frac{1}{N}+\frac{(N -1) \triangle \pi}{2 N}\beta   +O\left(\beta^2\right),
\]
where $\triangle \pi =\pi^\prime -\pi$ denotes the difference of mutant (mean) payoff $\pi^\prime$ and resident payoff $\pi$.
Note that  $\triangle \pi$ is constant during the extinction-fixation period.
For weak selection $\beta_l \ll 1$ and $N_l \gg 1$,
we can approximate $N_l 
\rho_l(f_l,f_l^{\prime})$ by
\begin{equation}
N_l 
\rho_l(f_l,f_l^{\prime})
\approx 
1 +\frac{1}{2} \beta_l  N_l \triangle \pi. 
\label{eq_selection_times_population}
\end{equation}
The transition probability  $\Lambda_{ij} =\frac{u_l}{g_l|\mathcal{S}_l|} N_l 
\rho_l(f_l,f_l^{\prime})$  is then a function of $\beta_l N_l$.
In other words,
the product $\beta_l N_l$ effectively acts as a single parameter
as far as the transition probability is concerned.

\section{Results}

\subsection{Symmetric Games in One Population}

\begin{figure}[t]  
\begin{center}  
\begin{tabular}{ll}
\includegraphics[width=0.19\textwidth]{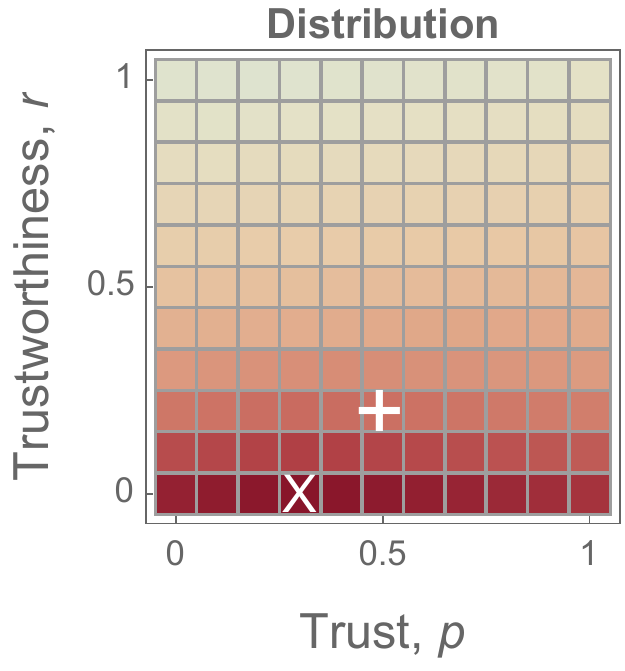}
&
\includegraphics[width=0.19\textwidth]{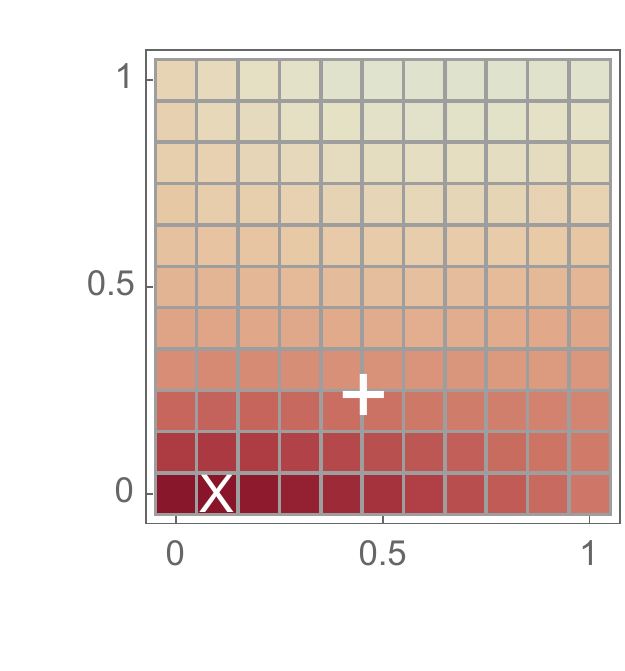} 
 \hspace{-0.1cm}\adjustbox{raise=7ex}
{\includegraphics[height=0.11\textwidth]{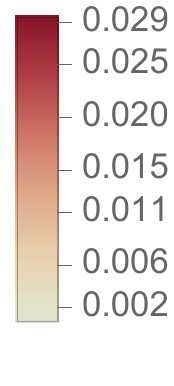}}
\\
\includegraphics[width=0.19\textwidth]{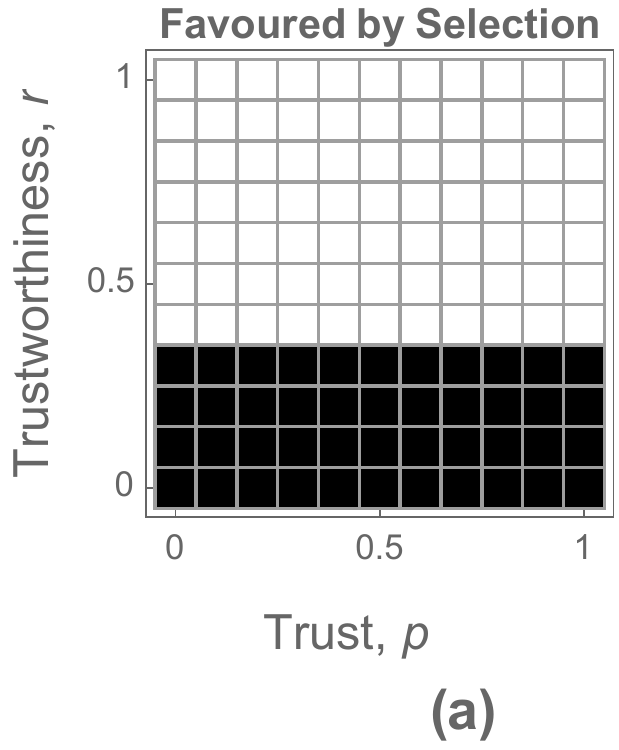} &
\includegraphics[width=0.19\textwidth]{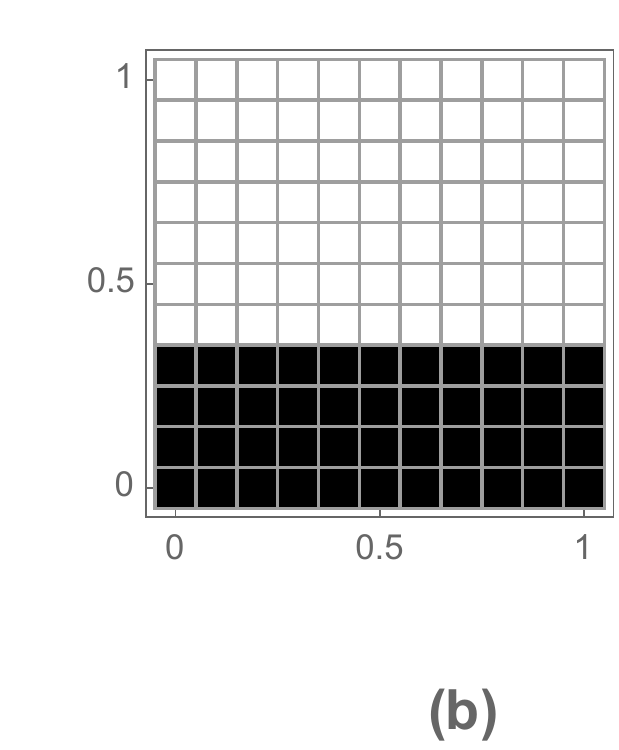} 
\end{tabular}   
\end{center}  
\caption{
(\textit{Top row}) The stationary distribution of the pure population states. The times `$\times$' sign denotes the mode of the stationary distribution 
and the plus `$+$' sign, the mean.
(\textit{Bottom}) The strategies favoured by selection, i.e.\,those more frequent than  $1/K$ are in black.
(a) The single-population model of the symmetric TG.
 $N=100, \beta=0.05, u=10^{-3}$,
$b = 3$, and $K=|\mathcal{S}_{\scriptscriptstyle I}| \times |\mathcal{S}_{\scriptscriptstyle T}| =11^2 =121$.
(b) 
The two-population model of the asymmetric TG with symmetric parameters.
$N_{\scriptscriptstyle I}=N_{\scriptscriptstyle T}=50, \beta_{\scriptscriptstyle I}=\beta_{\scriptscriptstyle T}=0.05, u_{\scriptscriptstyle I}=u_{\scriptscriptstyle T}=10^{-3}$, $g_{\scriptscriptstyle I} =g_{\scriptscriptstyle T} =1$, and $|\mathcal{S}_{\scriptscriptstyle I}| =|\mathcal{S}_{\scriptscriptstyle T}|=11$.
With the symmetric parameters,
even the two-population model yields evolutionary outcomes similar to those of the single-population model.
}    
\label{fig_base_line}
\end{figure}

We first present the stationary distribution of a single-population model of the symmetric TG (Fig.\,\ref{fig_base_line}\,(a)).
The modal (or most common) strategy in the distribution
is (a tuple of) low trust and null trustworthiness.
The mean strategy is mid trust and low trustworthiness.
A strategy is said to be selected for or favoured by selection
if its frequency (or abundance) exceeds   $1/K$
that would be the frequency of each strategy
if there were no differences in fitness between the strategies
\cite{Rand:2013aa}\cite{Tarnita:2011aa}\cite{Manapat:2012ab}\cite{Hilbe:2013aa}\cite{du2015aspiration}\cite{Hilbe:2018ab}.
The strategies favoured by selection
(i.e.\,those more frequent than $1/K$) can include a wide range of trust and low trustworthiness.

\subsection{Asymmetric Games in Two Populations}

	For the two-population model of the asymmetric TG, we start with symmetric parameters as a baseline.
All the parameters of generation times, mutation rates, selection strengths, population sizes, and discretisation resolutions  are the same between the two populations
(Fig.\,\ref{fig_base_line}\,(b)). 

\begin{figure}[t]  
\begin{center}  
\begin{tabular}{l}
\includegraphics[width=0.235\textwidth]{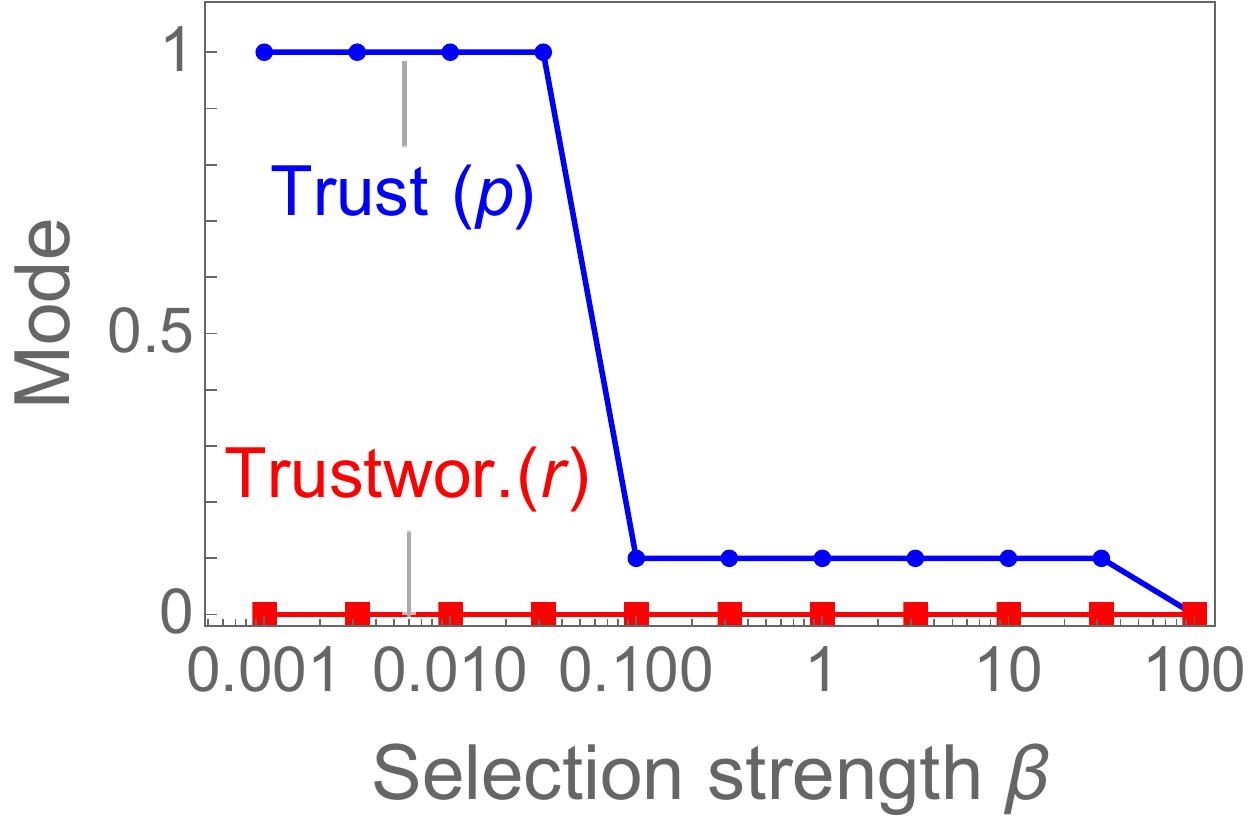} 
\includegraphics[width=0.235\textwidth]{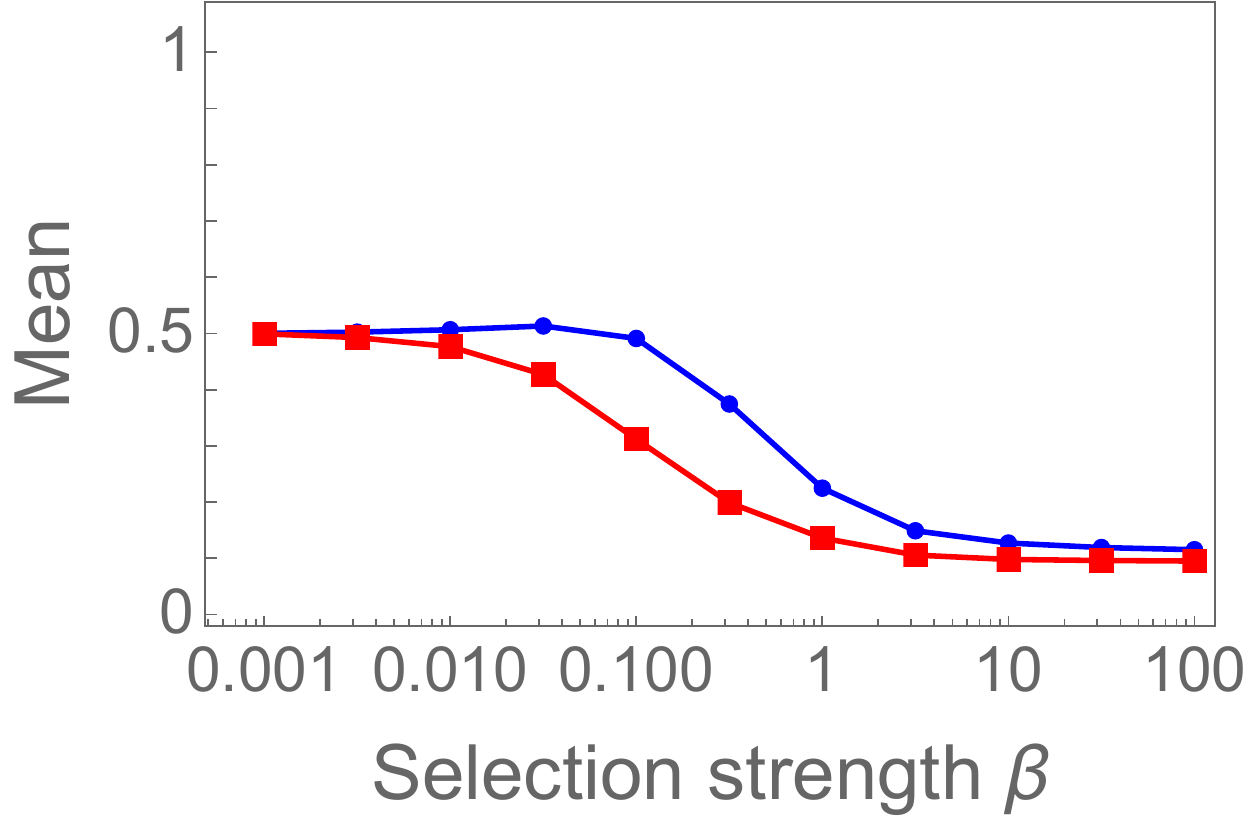} 
\end{tabular}   
\end{center}  
\caption{
Although evolution results in null trustworthy $r=0$ for the modal strategy,
it yields low-to-mid trustworthiness for the mean strategy.
}   
\label{fig_mean_mode_vs_selection}
\end{figure}

With the symmetric parameters,
even the two-population model of the asymmetric TG
yields evolutionary outcomes similar to those of  the single-population.
\begin{figure*}  
\begin{center}  
\begin{tabular}{llll}
\includegraphics[width=0.19\textwidth]{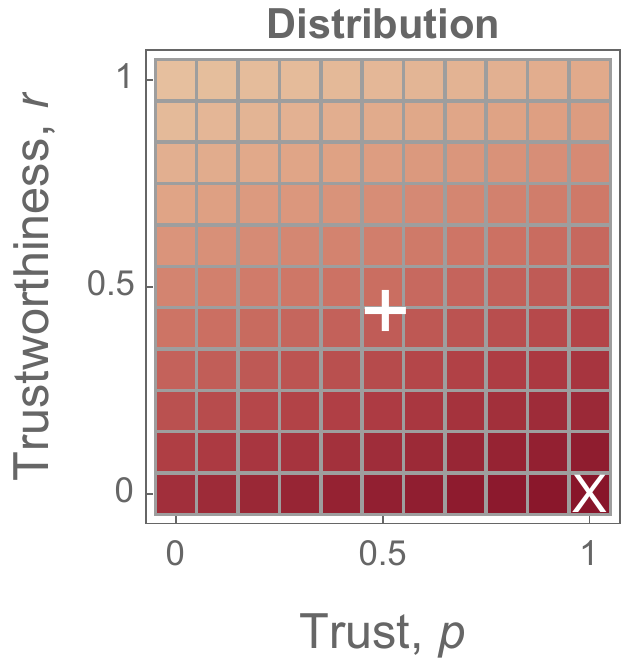}
\hspace{-0.1cm}\adjustbox{raise=7ex}{\includegraphics[height=0.11\textwidth]{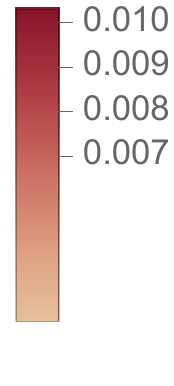}}    &
\includegraphics[width=0.19\textwidth]{AsymmetricTrustGames-429} 
 \hspace{-0.1cm}\adjustbox{raise=7ex}{\includegraphics[height=0.11\textwidth]{AsymmetricTrustGames-327}}  &
 \includegraphics[width=0.19\textwidth]{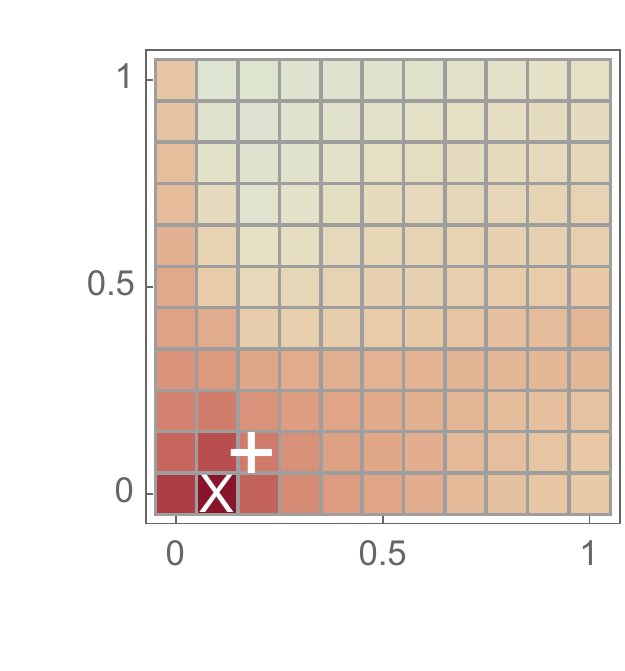}
\hspace{-0.1cm}\adjustbox{raise=7ex}{\includegraphics[height=0.11\textwidth]{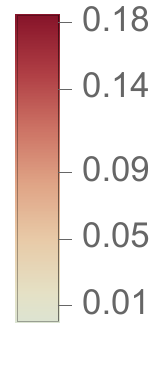}}    &
 \includegraphics[width=0.19\textwidth]{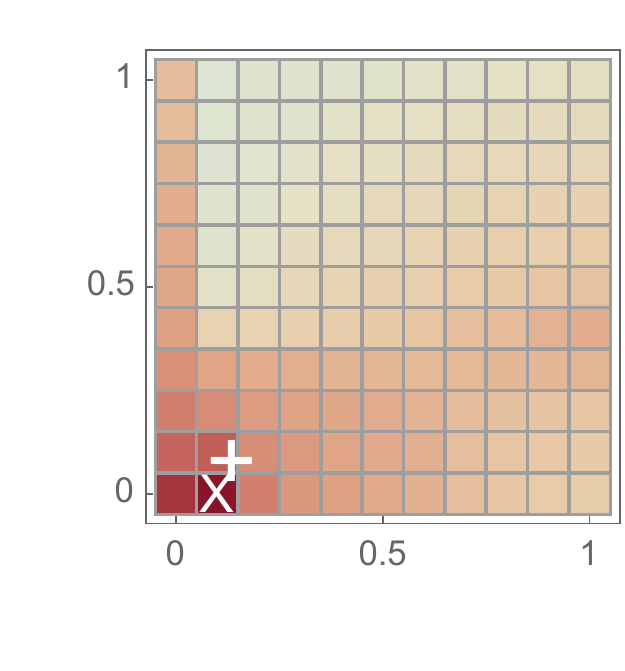}
\hspace{-0.1cm}\adjustbox{raise=7ex}{\includegraphics[height=0.11\textwidth]{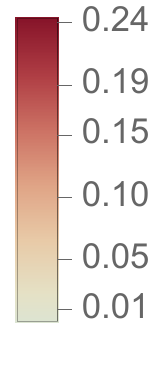}}    
\\
\includegraphics[width=0.19\textwidth]{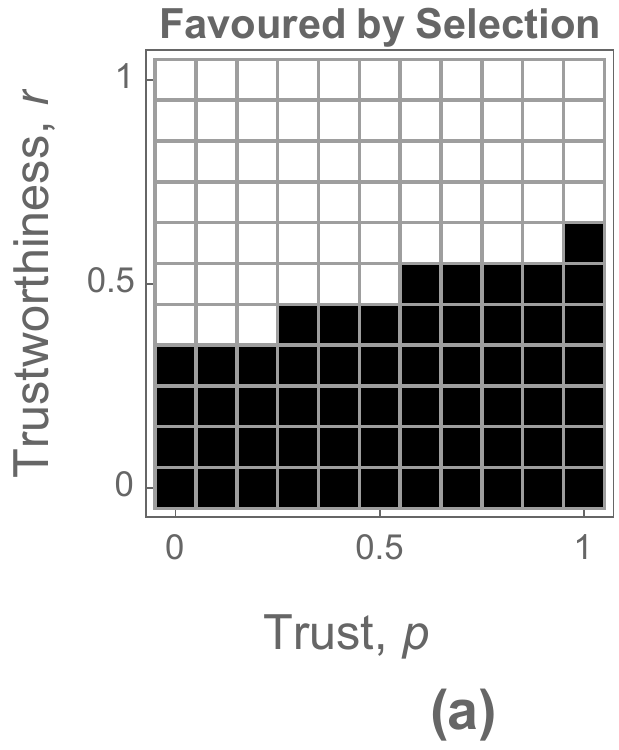} &
\includegraphics[width=0.19\textwidth]{AsymmetricTrustGames-430} &
\includegraphics[width=0.19\textwidth]{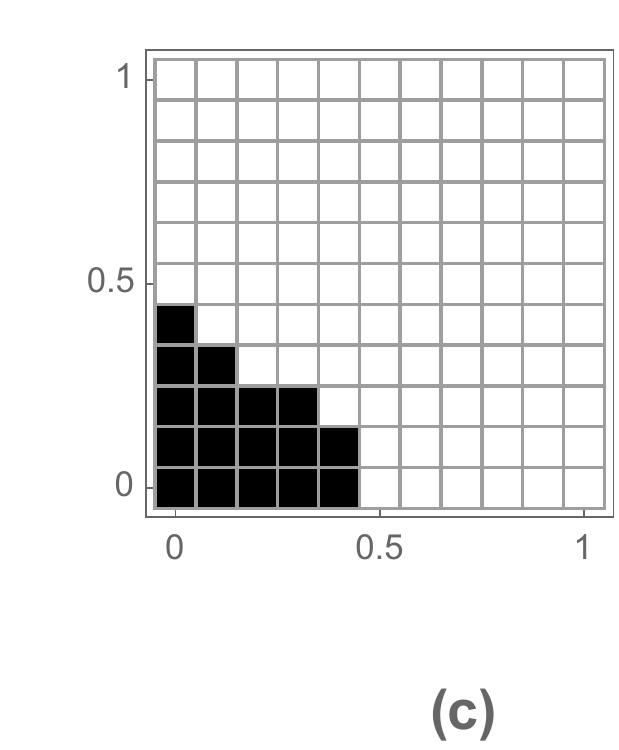} &
\includegraphics[width=0.19\textwidth]{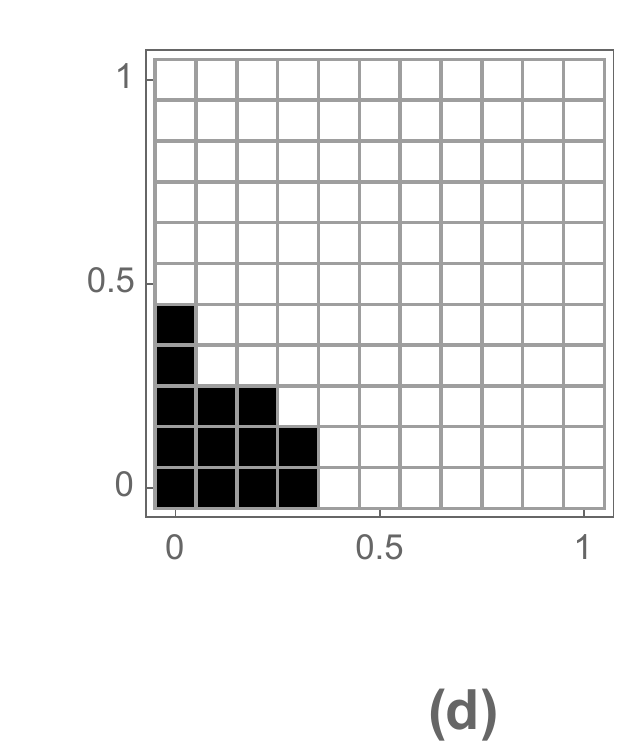} 
\end{tabular}   
\end{center}  
\caption{Varying selection strengths in the two-population model with the symmetric parameters.
(a)  Selection strength $\beta_{\scriptscriptstyle I}=\beta_{\scriptscriptstyle T} =0.005$.
(b)   $\beta_{\scriptscriptstyle I}=\beta_{\scriptscriptstyle T} =0.05$. 
(c)   $\beta_{\scriptscriptstyle I}=\beta_{\scriptscriptstyle T} =0.5$.
(d)   $\beta_{\scriptscriptstyle I}=\beta_{\scriptscriptstyle T} =1.5$.
As the selection strength decreases
($\beta_{\scriptscriptstyle I}=\beta_{\scriptscriptstyle T} \rightarrow 0$),
the mean strategy converges to mid trust and mid trustworthiness 
while more of high trust and mid trustworthiness are  favoured by selection.
This contrasts to the modal strategy that involves null trustworthiness, regardless of the selection strength.
The remaining parameters are the same as those in Fig.\,\ref{fig_base_line}\,(b).
}   
\label{fig_selection_two}.
\end{figure*} 
Note that the modal and mean strategies reveal a different aspect of evolution.
Specifically,
while null trustworthiness $r=0$ is the modal strategy regardless of selection strengths $\beta_{\scriptscriptstyle I}=\beta_{\scriptscriptstyle T}$,
low-to-mid trustworthiness $r>0$ evolves on average for low selection strength
(Fig.\,\ref{fig_selection_two} and  \ref{fig_mean_mode_vs_selection}).

For asymmetric demographics parameters, 
we take the all-else-equal approach in varying the parameters;
values of a parameter differ between the two populations, 
whereas the remaining parameter values are the same between the populations \cite{Veller:2017aa}.

\subsubsection{Asymmetric Ratios of Mutation Rate to Generation Time}

\begin{figure*}  
\begin{center}  
\begin{tabular}{lll}
 \includegraphics[width=0.19\textwidth]{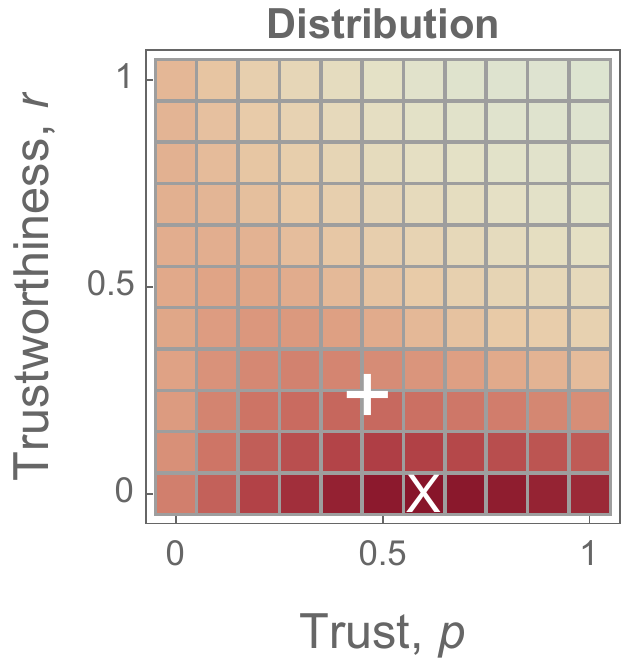}   
\hspace{-0.13cm}\adjustbox{raise=7ex}{\includegraphics[height=0.11\textwidth]{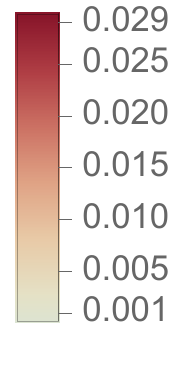}} &
\includegraphics[width=0.19\textwidth]{AsymmetricTrustGames-429}   
\hspace{-0.13cm}\adjustbox{raise=7ex}{\includegraphics[height=0.11\textwidth]{AsymmetricTrustGames-327}} &
\includegraphics[width=0.19\textwidth]{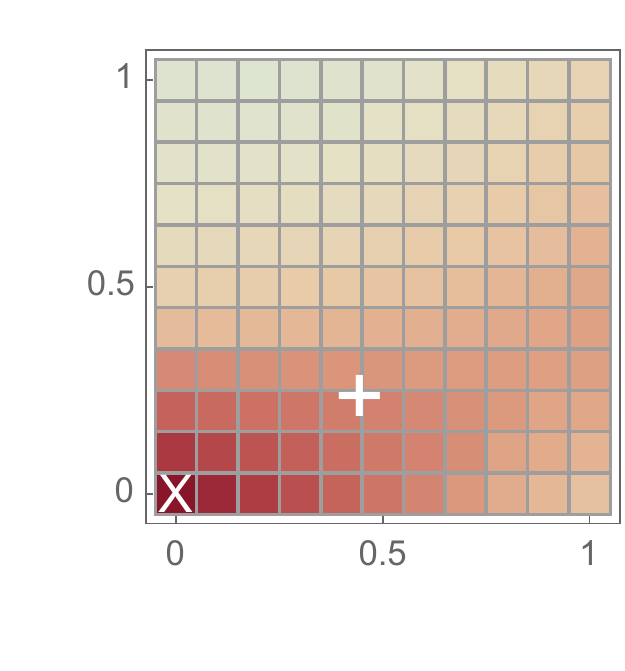}  
 \hspace{-0.13cm}\adjustbox{raise=7ex}{\includegraphics[height=0.11\textwidth]{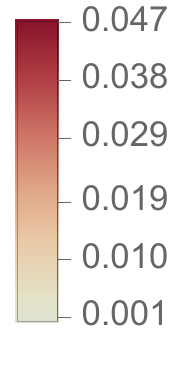}} 
 \\
\includegraphics[width=0.19\textwidth]{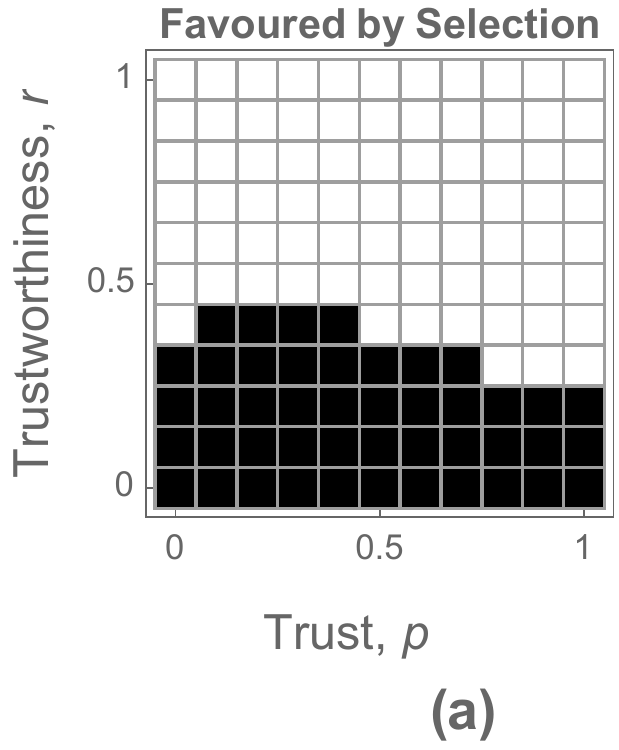}  & \includegraphics[width=0.19\textwidth]{AsymmetricTrustGames-430}  & \includegraphics[width=0.19\textwidth]{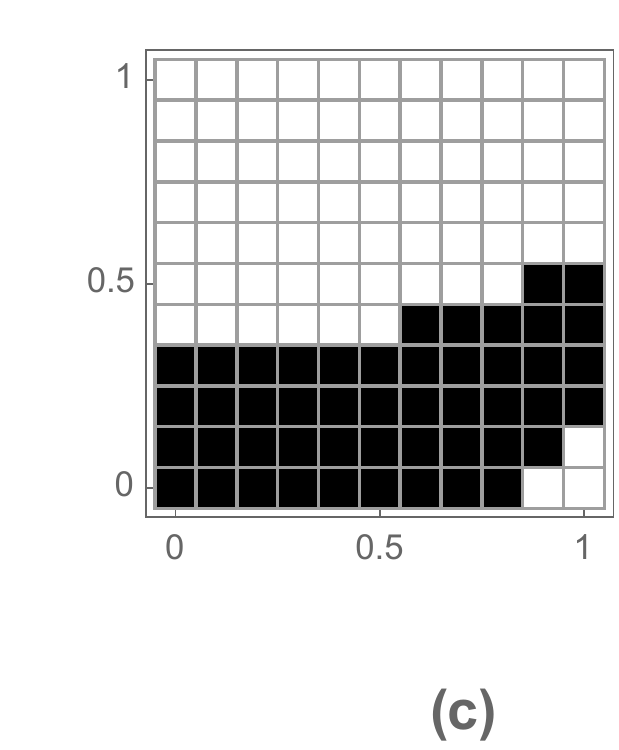} 
\end{tabular}   
\end{center}  
\caption{Asymmetry in the ratio of mutation rate $u$ to generation time $g$. 
(a) $u_{\scriptscriptstyle I} / g_{\scriptscriptstyle I}= 10^{-4}, u_{\scriptscriptstyle T} /g_{\scriptscriptstyle T} =10^{-3}$.  
(b)  $u_{\scriptscriptstyle I} / g_{\scriptscriptstyle I} =u_{\scriptscriptstyle T} / g_{\scriptscriptstyle T}  =10^{-3}$. 
(c) $u_{\scriptscriptstyle I} / g_{\scriptscriptstyle I}=10^{-3}, u_{\scriptscriptstyle T} / g_{\scriptscriptstyle T}   = 10^{-4}$.
Compared to that of the symmetric ratio in (b),
the asymmetric ratios in (a) and (c) do not yield a substantial difference in the mean strategy nor the strategies favoured by selection.
The remaining parameters are the same as those in Fig.\,\ref{fig_base_line}\,(b).   
}   
\label{fig_mutation_generation_difference}
\end{figure*}

	A previous paper treated and varied mutation rate $u_l$ and generation time $g_l$, independently for the all-else-equal comparison \cite{Veller:2017aa}.
From the definition of the transition probability in Eq.\,\eqref{eq_transition_prob},
however,
what matters is the ratio $u_l/g_l$ but not individuals of them.
For the all-else-equal comparison, thus, 
we treat and vary the ratio $u_l/g_l$ as if a single parameter.
Compared to the symmetric case,
the asymmetric ratios (an order-of-magnitude difference) between the two populations do not yield substantial differences in the stationary distributions,
especially, the mean strategy and the strategies favoured by selection
(Fig.\,\ref{fig_mutation_generation_difference}).
Thus, the asymmetry in the ratio of mutation rate to generation time does not substantially promote trust nor trustworthiness.

\subsubsection{Asymmetric Selection Strengths}

\begin{figure*}  
\begin{center}  
\begin{tabular}{lll}
\includegraphics[width=0.19\textwidth]{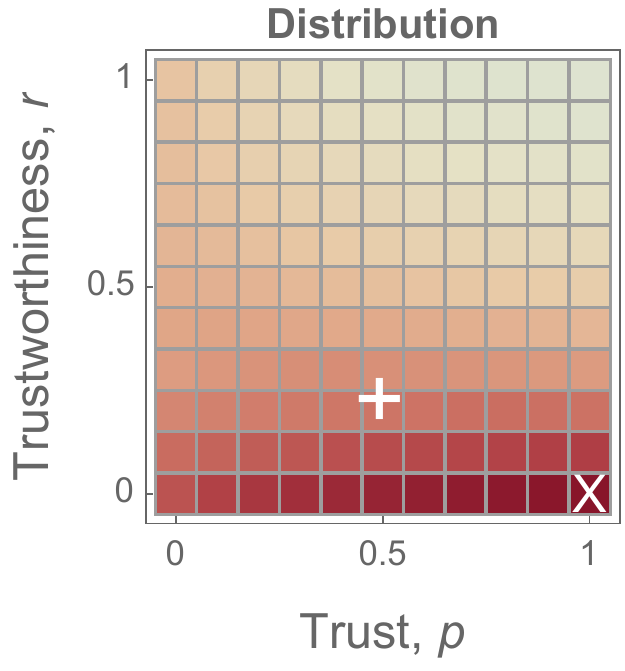}  
\hspace{-0.1cm}\adjustbox{raise=7ex}{\includegraphics[height=0.11\textwidth]{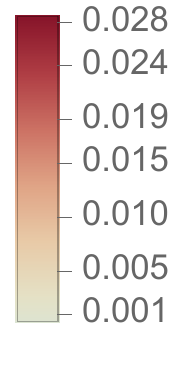}} & 
\includegraphics[width=0.19\textwidth]{AsymmetricTrustGames-429}  
\hspace{-0.1cm}\adjustbox{raise=7ex}{\includegraphics[height=0.11\textwidth]{AsymmetricTrustGames-327}} & 
\includegraphics[width=0.19\textwidth]{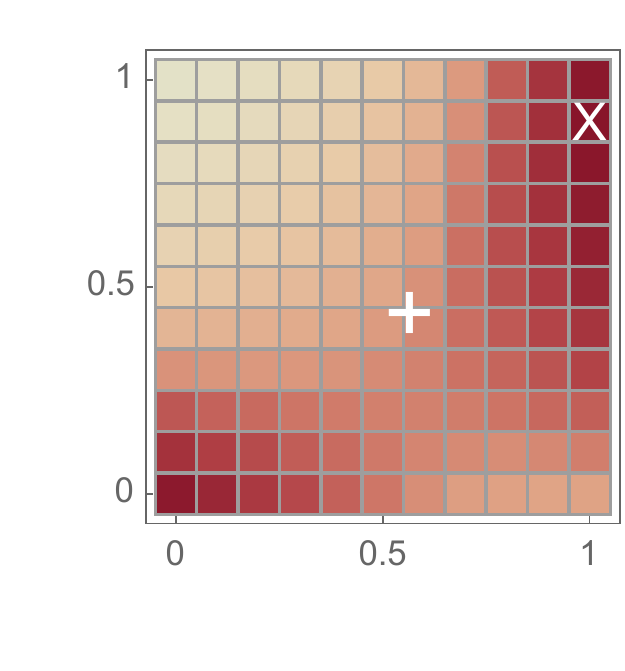} 
\hspace{-0.1cm}\adjustbox{raise=7ex}{\includegraphics[height=0.11\textwidth]{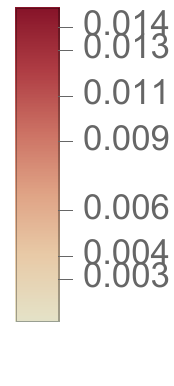}}  
\\
\includegraphics[width=0.19\textwidth]{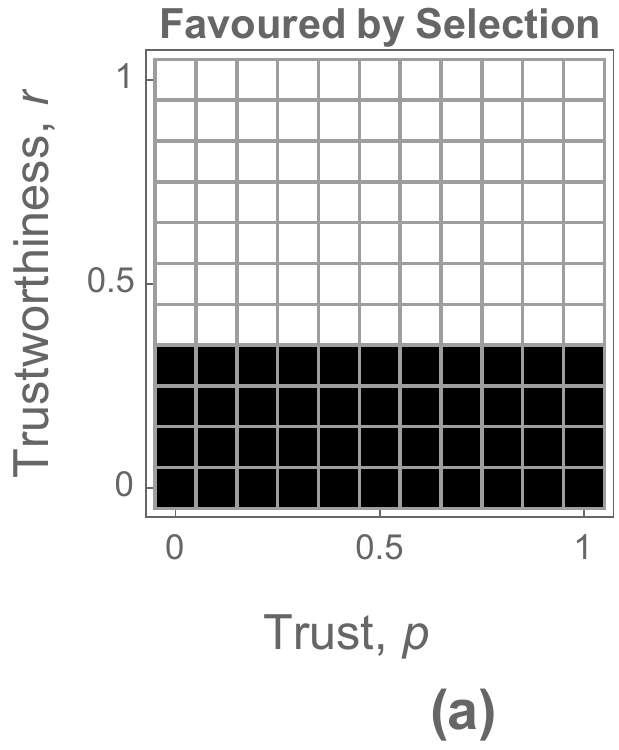} & 
\includegraphics[width=0.19\textwidth]{AsymmetricTrustGames-430} & 
\includegraphics[width=0.19\textwidth]{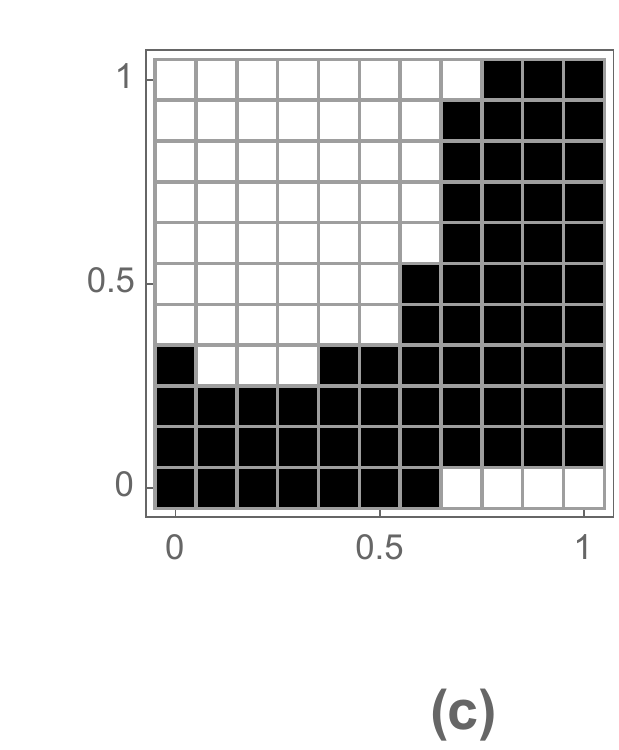}  
\end{tabular}   
\end{center}  
\caption{Asymmetry in selection strength.
(a) $\beta_{\scriptscriptstyle I}=0.005, \beta_{\scriptscriptstyle T}=0.05$.
When the selection strength in the investor population is weaker than that in the investor population, there is little difference compared to the symmetric case of (b).
(b) $\beta_{\scriptscriptstyle I}= \beta_{\scriptscriptstyle T}=0.05$.
(c)   $\beta_{\scriptscriptstyle I}=0.05, \beta_{\scriptscriptstyle T}=0.005$.
When the selection strength in the investor population is stronger than that  in the investor population
and the latter is weak,
the modal strategy is full trust and near full trustworthiness.
The mean strategy also rises to mid trust and mid trustworthiness
while  high trust and mid-to-high trustworthiness are favoured by selection.
The remaining parameters are the same as those in Fig.\,\ref{fig_base_line}\,(b).  
}   
\label{fig_selection_difference}
\end{figure*}

Asymmetric selection strengths can substantially promote the evolution of trust and trustworthiness. 
The stationary distribution becomes multi-modal
when the selection strength in the investor population is stronger than
that in the trustee population ($\beta_{\scriptscriptstyle I} >\beta_{\scriptscriptstyle T}$)
and the latter is weak ($\beta_{\scriptscriptstyle T} \ll 1$)
(Fig.\ref{fig_selection_difference}).
The modal strategy is full trust and near full trustworthiness
while the mean strategy rises to mid trust and mid trustworthiness.
Also, high trust and mid-to-high trustworthiness are favoured by selection.
These outcomes of the boosted trust and trustworthiness are substantially different from those of the baseline with a unimodal distribution.

\subsubsection{Asymmetric Population Sizes}

\begin{figure*}  
\begin{center}  
\begin{tabular}{lll}
\includegraphics[width=0.19\textwidth]{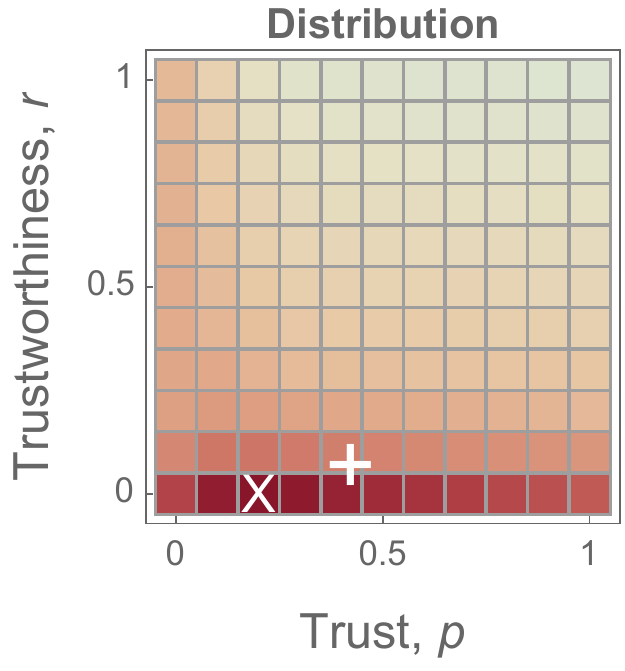}  
\hspace{-0.1cm}\adjustbox{raise=7ex}{\includegraphics[height=0.11\textwidth]{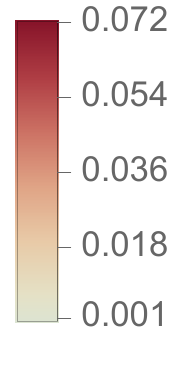}} & 
\includegraphics[width=0.19\textwidth]{AsymmetricTrustGames-429}  
\hspace{-0.1cm}\adjustbox{raise=7ex}{\includegraphics[height=0.11\textwidth]{AsymmetricTrustGames-327}} &
\includegraphics[width=0.19\textwidth]{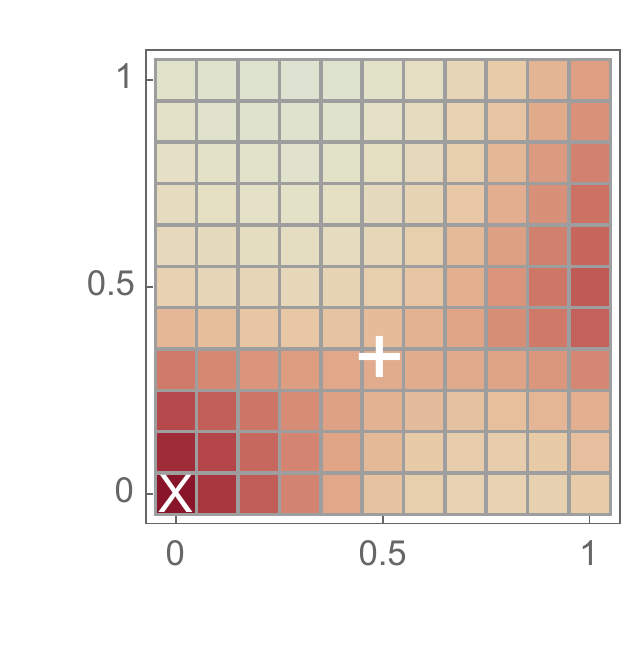} 
\hspace{-0.1cm}\adjustbox{raise=7ex}{\includegraphics[height=0.11\textwidth]{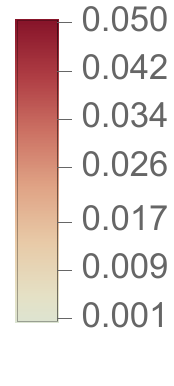}} 
\\
\includegraphics[width=0.19\textwidth]{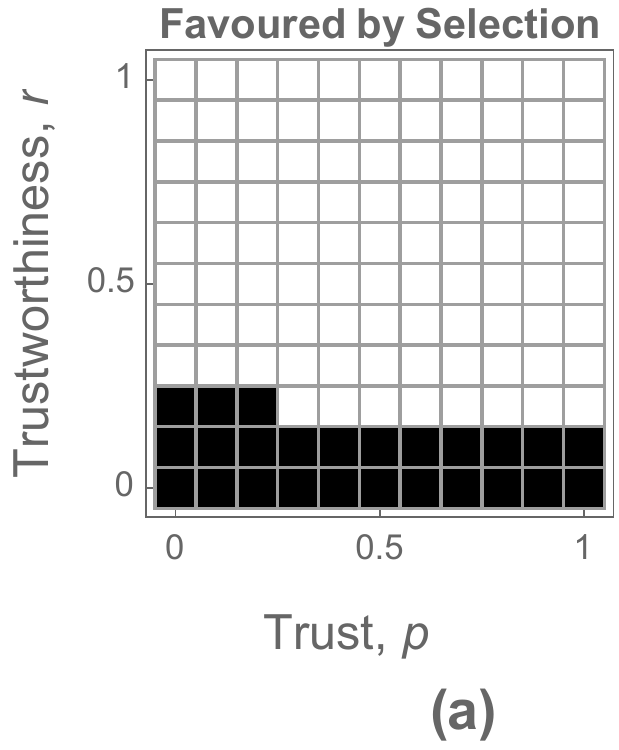} & 
\includegraphics[width=0.19\textwidth]{AsymmetricTrustGames-430} & 
\includegraphics[width=0.19\textwidth]{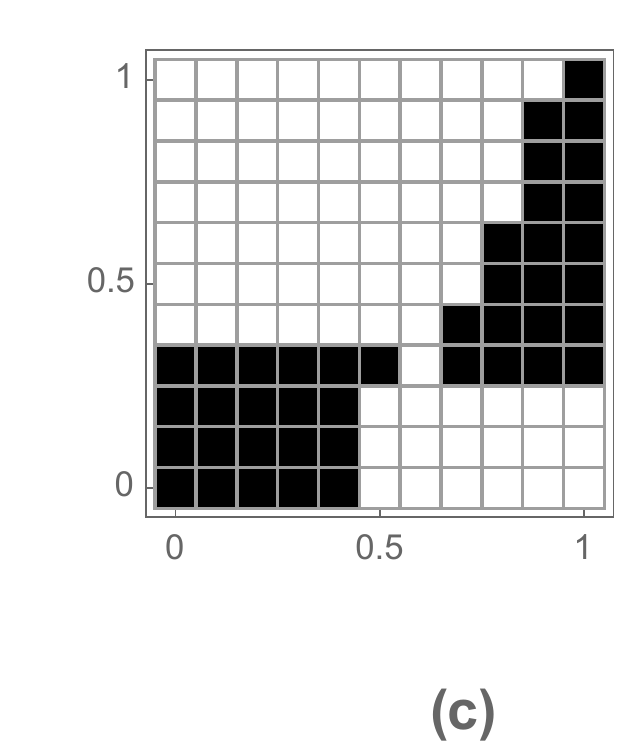}  
\end{tabular}   
\end{center}  
\caption{Asymmetry in population size.
(a) $N_{\scriptscriptstyle I}=25, N_{\scriptscriptstyle T}=250$.
When the size of the investor population is smaller than that of the trustee population,
trustworthiness of the mean and strategies favoured by selection is lowered than that of the baseline case of (b).
(b) $N_{\scriptscriptstyle I}=N_{\scriptscriptstyle T}=50$.
(c) $N_{\scriptscriptstyle I}=250, N_{\scriptscriptstyle T}=25$.
When the size of the trustee population is smaller,
the stationary distribution becomes multi-modal,
including the one encompassing full trust and mid-to-high trustworthiness.
High trust and mid-to-high trustworthiness are favoured by selection.
The remaining parameters are the same as those in Fig.\,\ref{fig_base_line}\,(b).  
}   
\label{fig_population_difference}
\end{figure*}

	Asymmetric population sizes can promote the evolution of trust and trustworthiness. 
When the population size of trustees is 
substantially smaller than that of investors ($N_{\scriptscriptstyle I} \gg N_{\scriptscriptstyle T}$), 
it leads to a multi-modal distribution
(Fig.\ref{fig_population_difference}).
	Analogous to the asymmetric selection strengths,
trust and trustworthiness are boosted.
The mean strategy rises to mid trust and mid trustworthiness,
while high trust and mid-to-high trustworthiness are favoured by selection.

\subsubsection{Product of Selection Strength and Population Size}

\begin{figure}[t]  
\begin{center}  
\includegraphics[height=0.159\textwidth]{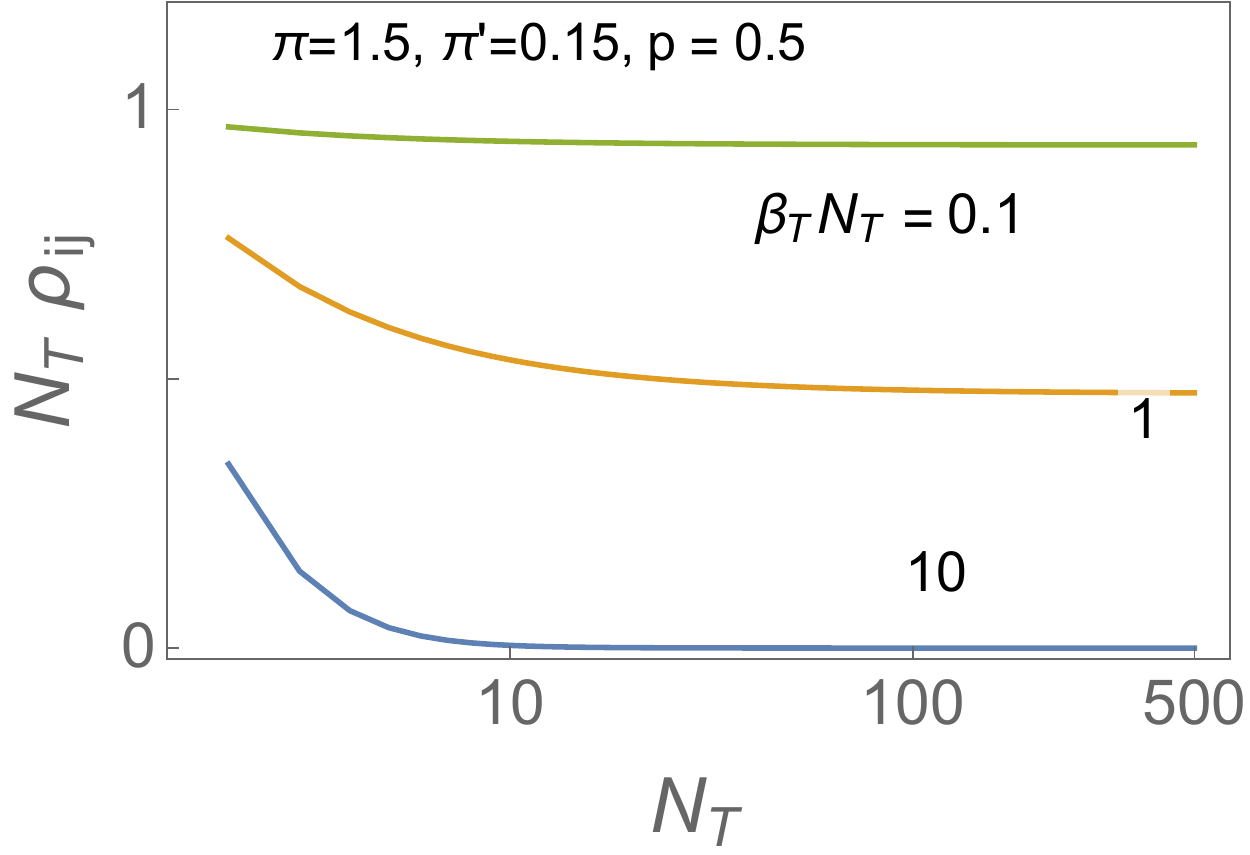} 
\includegraphics[height=0.159\textwidth]{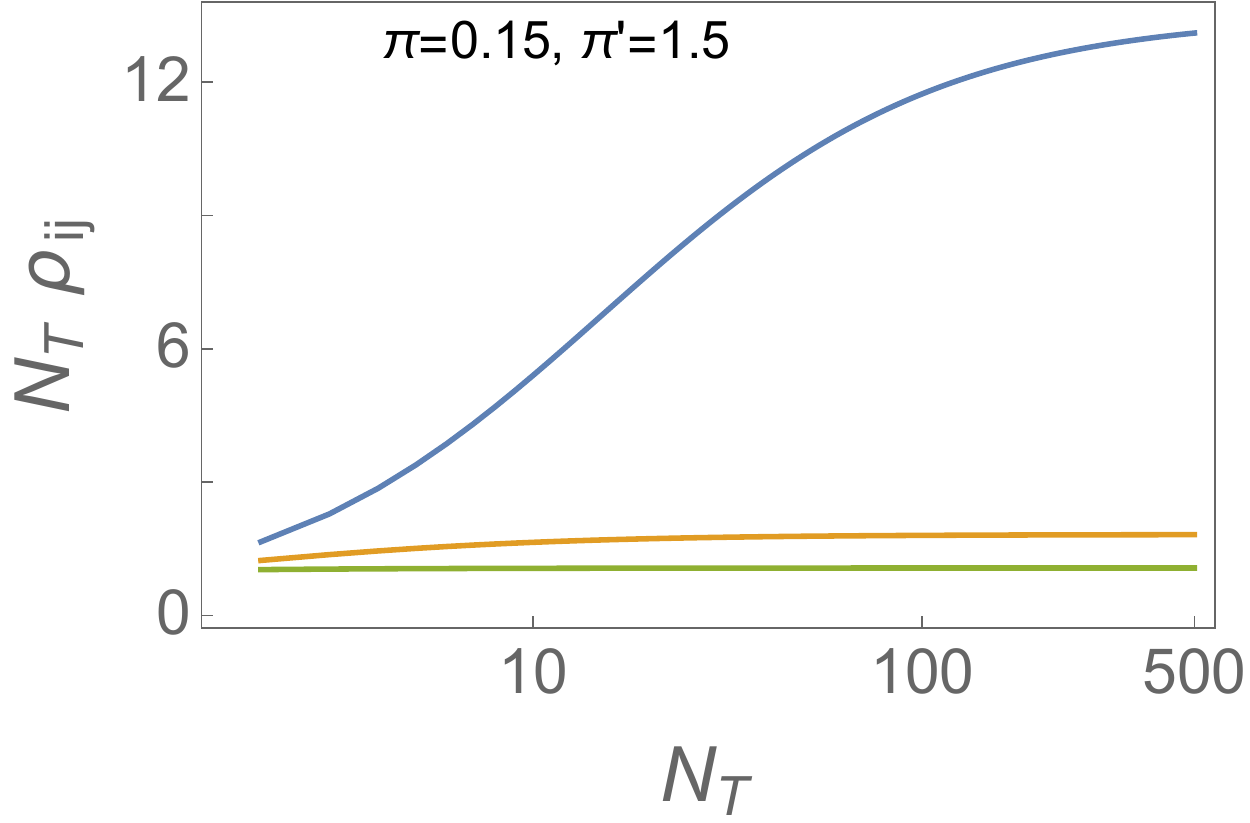}  
\end{center}   
\caption{
$N_{\scriptscriptstyle T} \rho_{ij}$ is invariant given $\beta_{\scriptscriptstyle T} N_{\scriptscriptstyle T}$, unless the population size $N_{\scriptscriptstyle T}$ is too small or the $\beta_{\scriptscriptstyle T} N_{\scriptscriptstyle T}$ is too high.
$\pi$ and $\pi^\prime$ denote the payoffs of the resident and the mutant, respectively.
$\beta_{\scriptscriptstyle I} =0.05, N_{\scriptscriptstyle I}=50$.
}   
\label{fig_population_times_fixation}  
\end{figure}

\begin{figure*}  
\begin{center}  
\begin{tabular}{llll}
\includegraphics[width=0.19\textwidth]{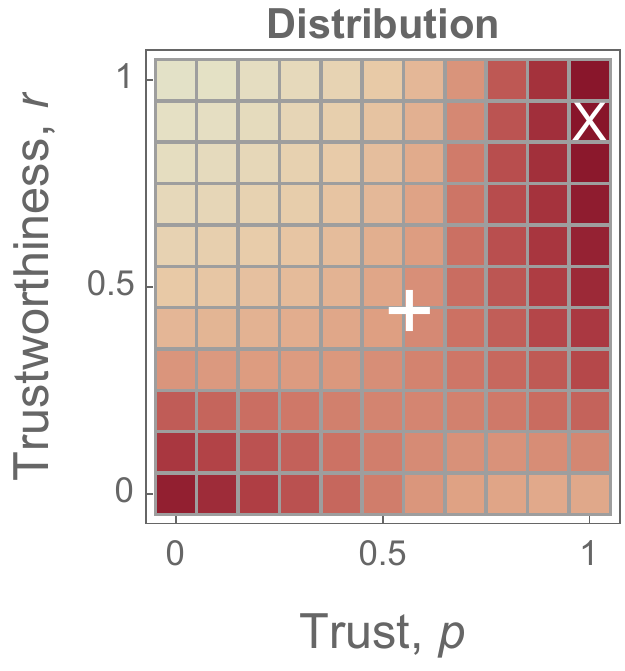}
\hspace{-0.1cm}\adjustbox{raise=7ex}{\includegraphics[height=0.11\textwidth]{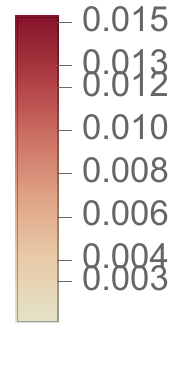}} &
\includegraphics[width=0.19\textwidth]{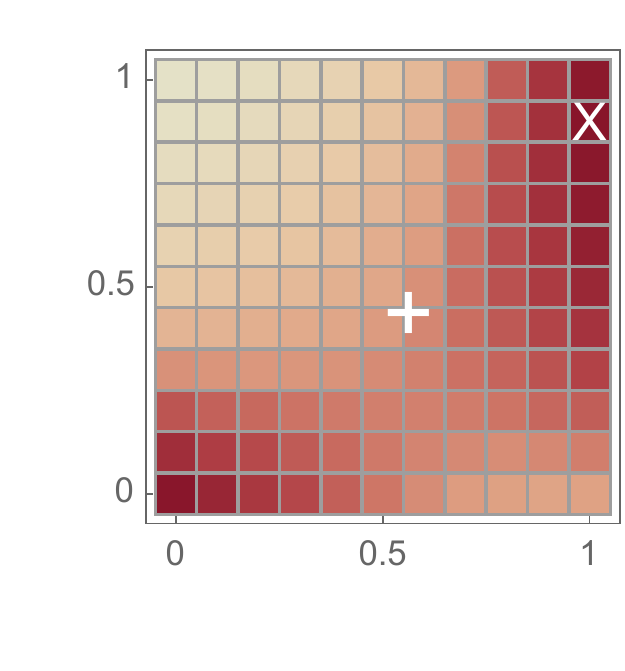}
\hspace{-0.1cm}\adjustbox{raise=7ex}{\includegraphics[height=0.11\textwidth]{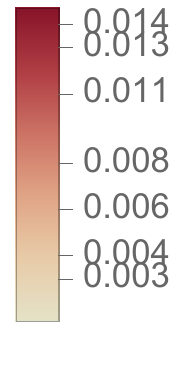}} &
\includegraphics[width=0.19\textwidth]{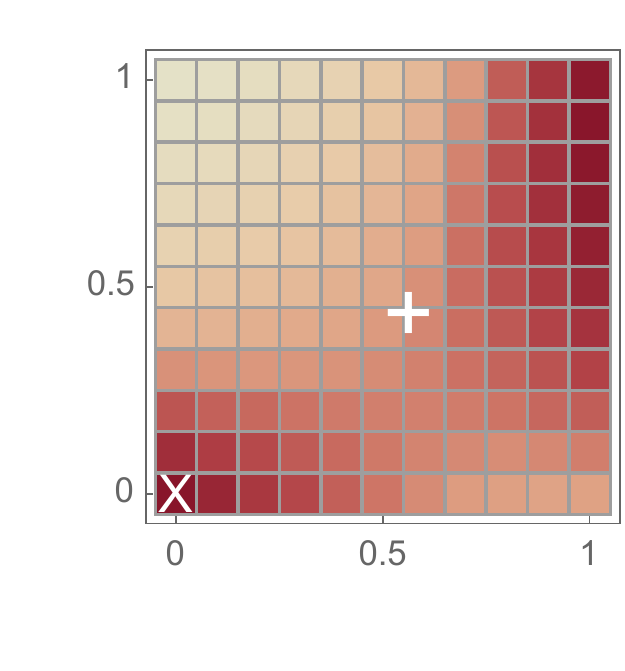}
\hspace{-0.1cm}\adjustbox{raise=7ex}{\includegraphics[height=0.11\textwidth]{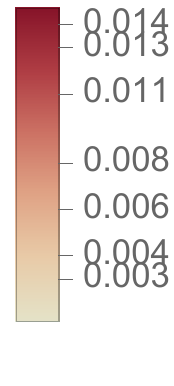}} &
\includegraphics[width=0.19\textwidth]{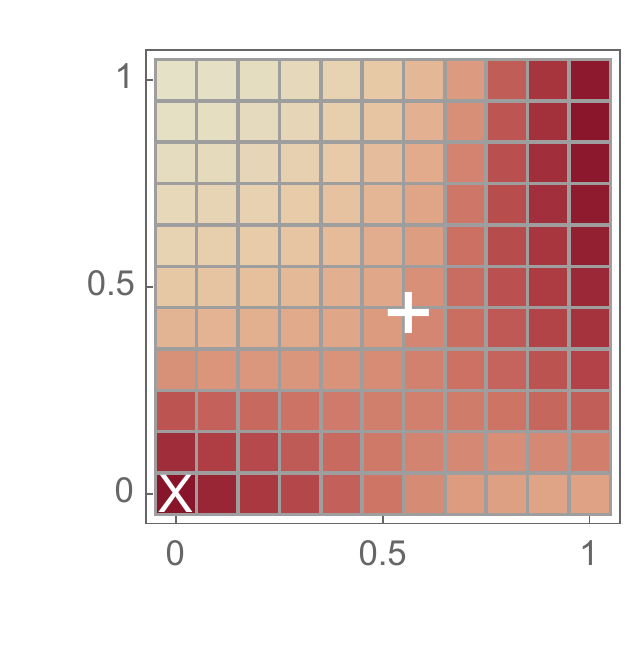}
\hspace{-0.1cm}\adjustbox{raise=7ex}{\includegraphics[height=0.11\textwidth]{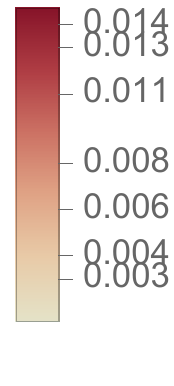}} 
\\
\includegraphics[width=0.19\textwidth]{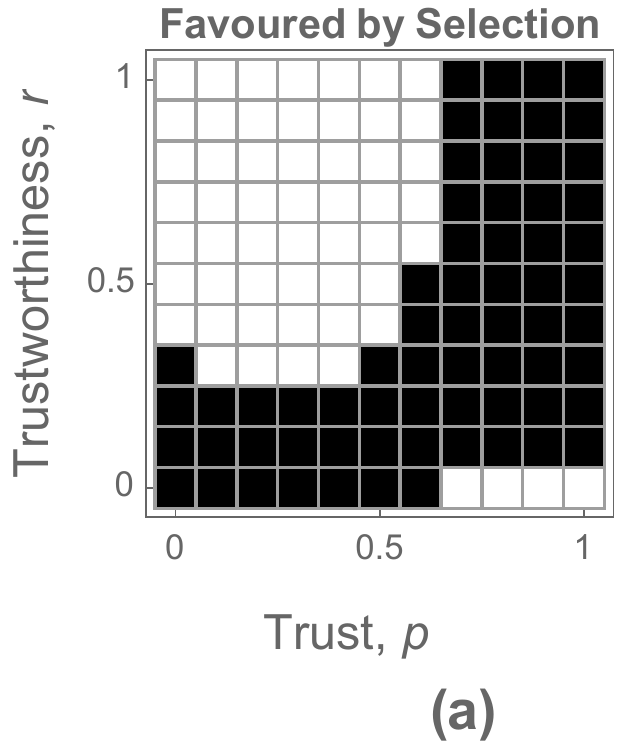}&
\includegraphics[width=0.19\textwidth]{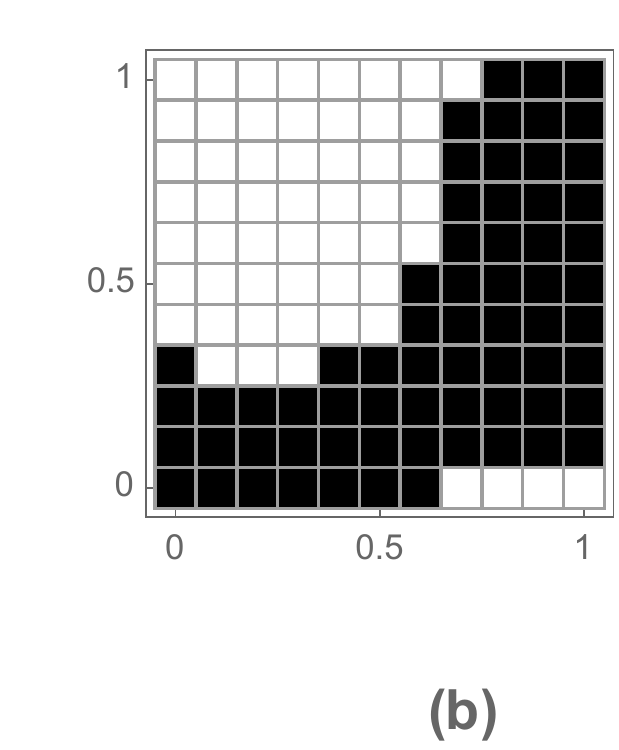}&  \includegraphics[width=0.19\textwidth]{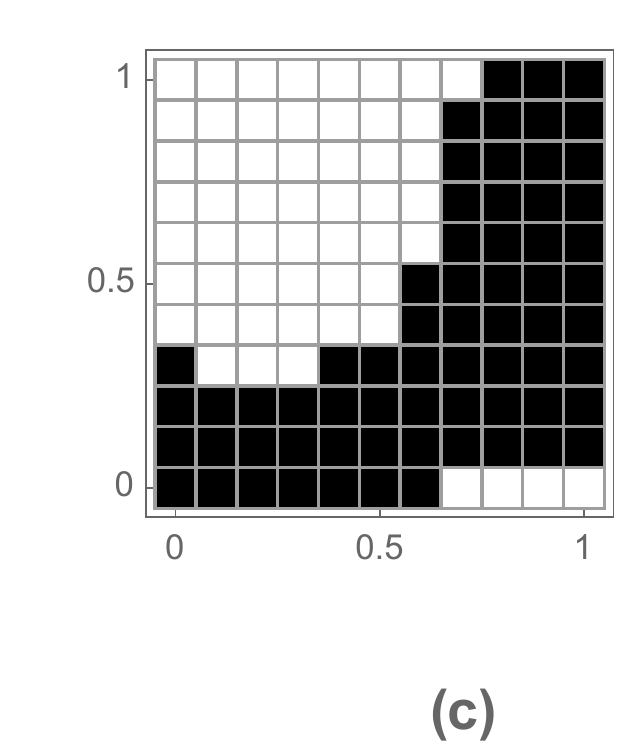}&
\includegraphics[width=0.19\textwidth]{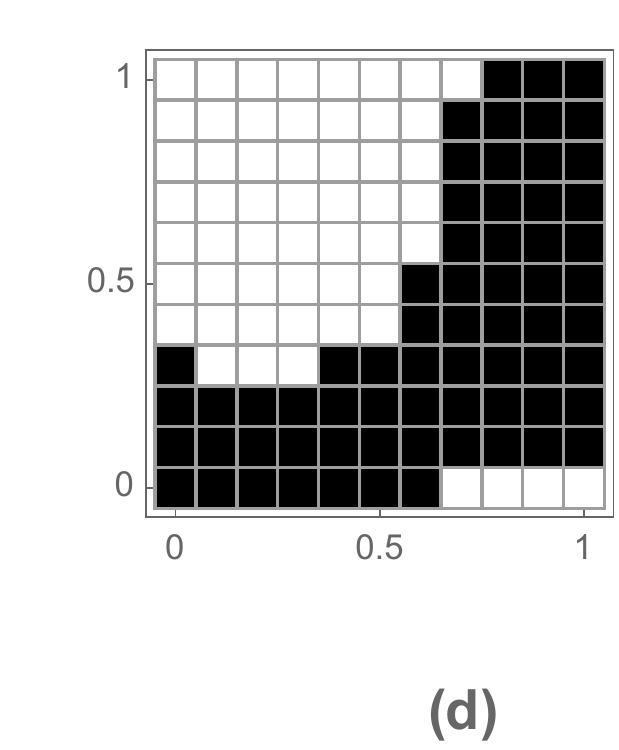}
\end{tabular}   
\end{center}  
\caption{Invariance of stationary distributions under $\beta_{\scriptscriptstyle T} N_{\scriptscriptstyle T} =0.25$  while varying $N_{\scriptscriptstyle T}$.
(a) $N_{\scriptscriptstyle T} =10$. 
(b) $N_{\scriptscriptstyle T} =200$. 
(c)  $N_{\scriptscriptstyle T} =500$. 
(d)  $N_{\scriptscriptstyle T} =1000$.
$\beta_{\scriptscriptstyle I} =0.05, N_{\scriptscriptstyle I}=50$.
}  
\label{fig_selection_times_population_invariance}.
\end{figure*}

\begin{figure*}  
\begin{center}  
\begin{tabular}{llll}
\includegraphics[width=0.19\textwidth]{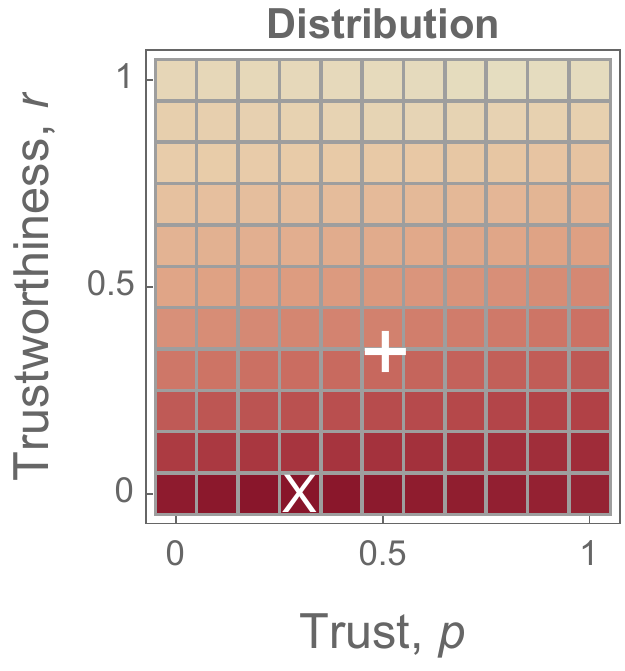} 
\hspace{-0.1cm}\adjustbox{raise=7ex}{\includegraphics[height=0.11\textwidth]{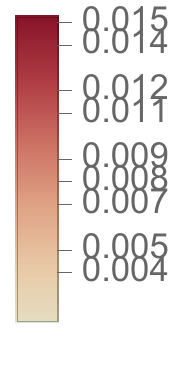}}  &\includegraphics[width=0.19\textwidth]{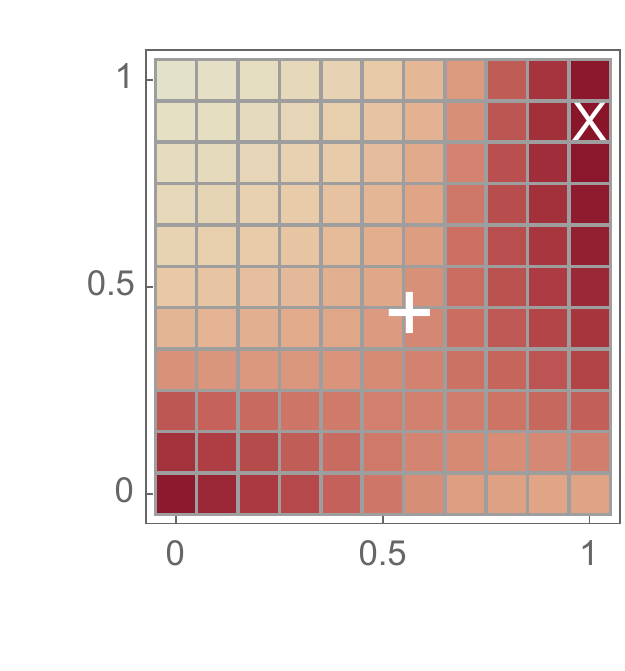} 
\hspace{-0.1cm}\adjustbox{raise=7ex}{\includegraphics[height=0.11\textwidth]{AsymmetricTrustGames-336}}  &
\includegraphics[width=0.19\textwidth]{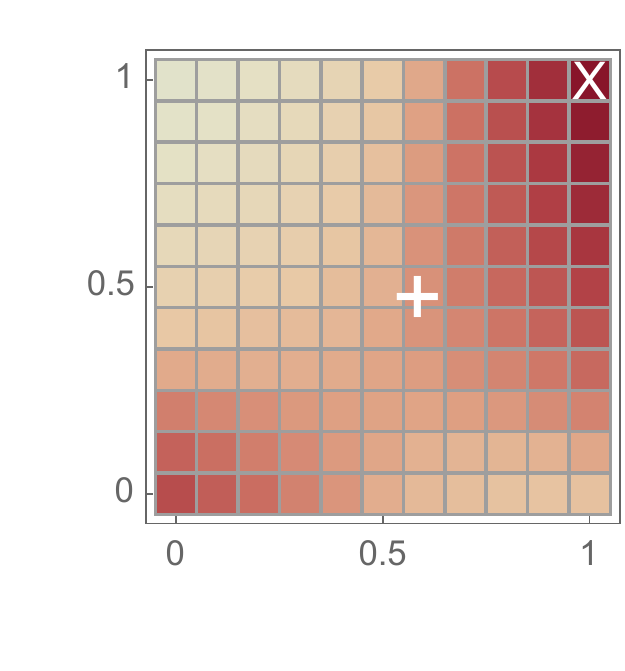} 
\hspace{-0.1cm}\adjustbox{raise=7ex}{\includegraphics[height=0.11\textwidth]{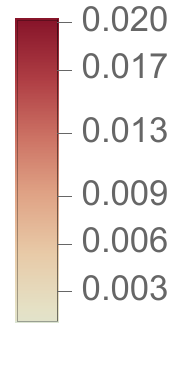}} &
\includegraphics[width=0.19\textwidth]{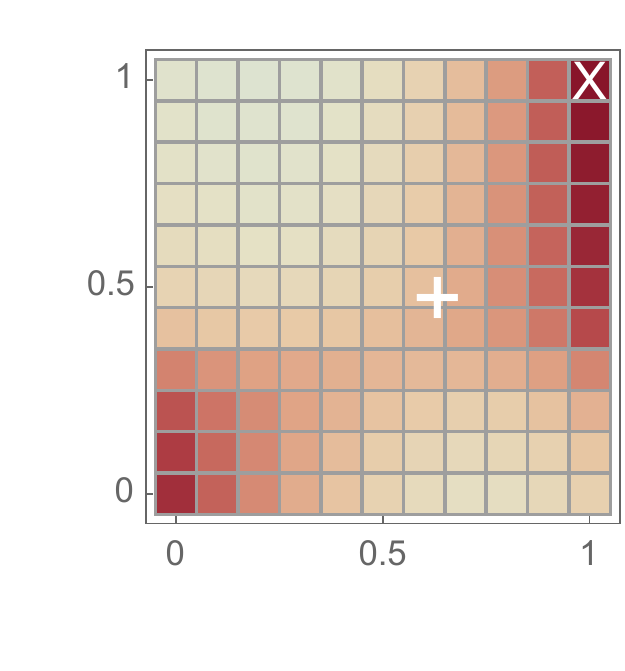} 
\hspace{-0.1cm}\adjustbox{raise=7ex}{\includegraphics[height=0.11\textwidth]{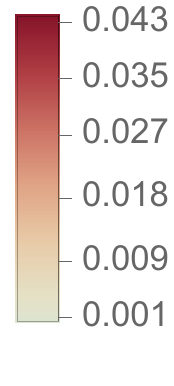}} 
\\
\includegraphics[width=0.19\textwidth]{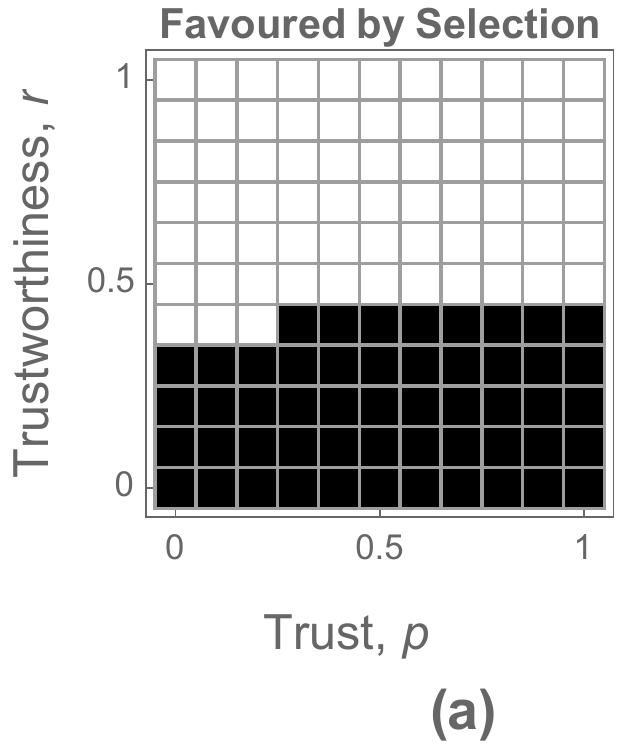} &
\includegraphics[width=0.19\textwidth]{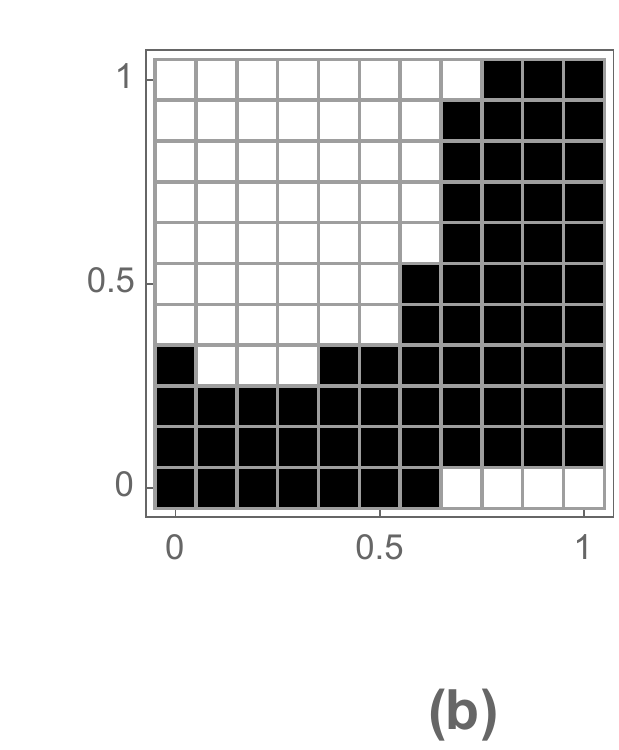} &
\includegraphics[width=0.19\textwidth]{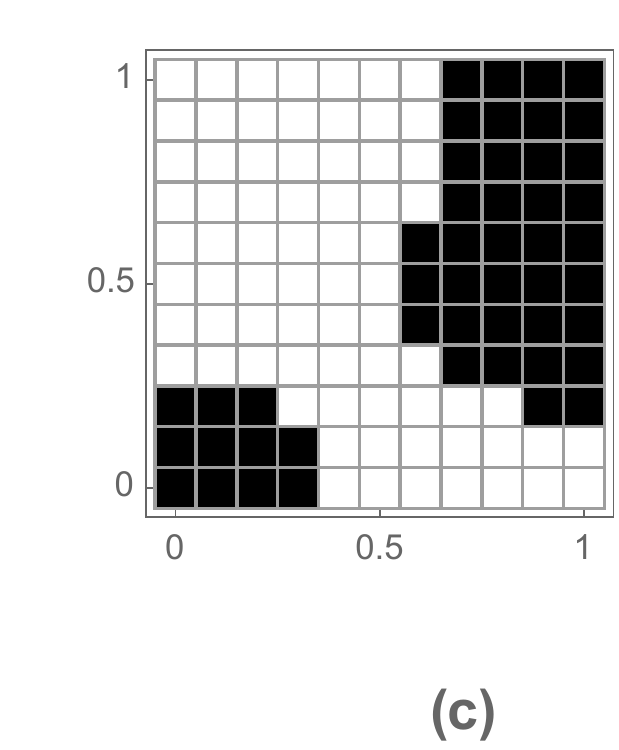} &
\includegraphics[width=0.19\textwidth]{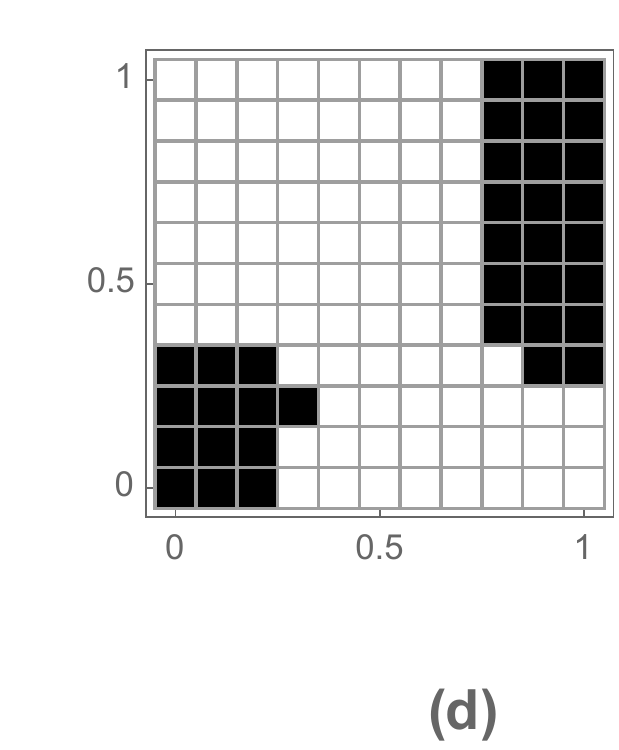}
\end{tabular}   
\end{center}  
\caption{The stronger of the product $\beta_{\scriptscriptstyle I} N_{\scriptscriptstyle I}$
and the weaker of the product $\beta_{\scriptscriptstyle T} N_{\scriptscriptstyle T}$,
the higher trust and trustworthiness evolve.
(a) $\beta_{\scriptscriptstyle I} N_{\scriptscriptstyle I} =1$, $\beta_{\scriptscriptstyle T} N_{\scriptscriptstyle T} =1$.
(b) $\beta_{\scriptscriptstyle I} N_{\scriptscriptstyle I} =2.5$, $\beta_{\scriptscriptstyle T} N_{\scriptscriptstyle T} =0.25$.
(c) $\beta_{\scriptscriptstyle I} N_{\scriptscriptstyle I} =2.5$, $\beta_{\scriptscriptstyle T} N_{\scriptscriptstyle T} =0.025$.
(d) $\beta_{\scriptscriptstyle I} N_{\scriptscriptstyle I} =25$, $\beta_{\scriptscriptstyle T} N_{\scriptscriptstyle T} =0.025$.
}  
\label{fig_selection_times_population}
\end{figure*}

The product $\beta_l N_l$ of selection strength and population size can act as a single parameter, in effect.
For weak selection $\beta_l \ll 1$ and $N_l \gg 1$,
$N_l \rho_{ij}$ is a function of $\beta_l N_l$ as seen in Eq.\eqref{eq_selection_times_population}.
Given the value of $\beta_l N_l$,
in other words,
$N_l \rho_{ij}$ is invariant
even if each of $\beta_l$ and $N_l$  varies.
We numerically demonstrate this invariance
(Fig.\,\ref{fig_population_times_fixation}).
It also leads to the invariance of transition probabilities and, consequently,
invariance of the stationary distribution.
Given the value of the product, indeed,
the stationary distribution hardly varies for a wide range of population size (and associated selection strength)
(Fig.\,\ref{fig_selection_times_population_invariance}).                                                        %
When selection acts stronger in the investor population than the trustee population
($\beta_{\scriptscriptstyle I} N_{\scriptscriptstyle I} \gg \beta_{\scriptscriptstyle T} N_{\scriptscriptstyle T}$)
and selection acts weakly in the trustee population ($\beta_{\scriptscriptstyle T} N_{\scriptscriptstyle T} <1$),
it can lead to the evolution of high trust and high trustworthiness.
The stronger selection among the investors ($\beta_{\scriptscriptstyle I} N_{\scriptscriptstyle I} \rightarrow \infty$) and the weaker selection among the trustees ($\beta_{\scriptscriptstyle T} N_{\scriptscriptstyle T} \rightarrow 0$), 
the 
higher trust and higher trustworthiness 
(Fig.\,\ref{fig_selection_times_population}).  

\begin{figure*} 
\begin{center}  
\begin{tabular}{lcl}
\includegraphics[width=0.19\textwidth]{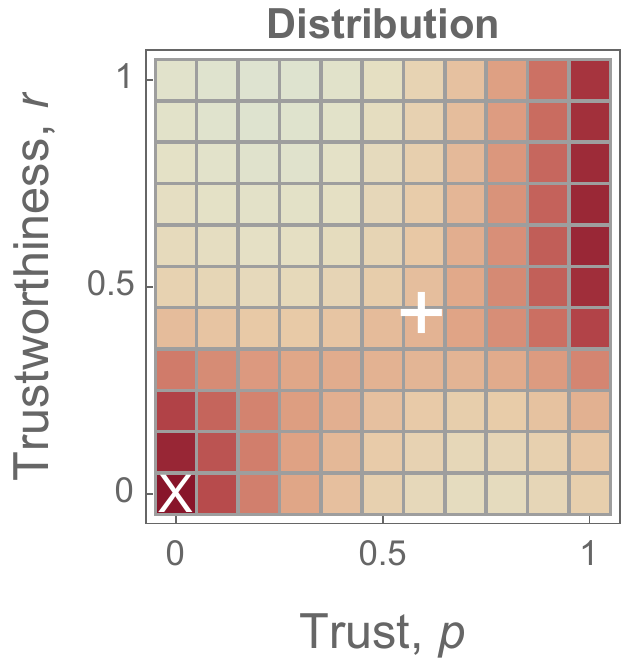}  
\hspace{-0.1cm}\adjustbox{raise=7ex}{\includegraphics[height=0.11\textwidth]{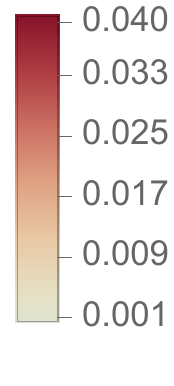}} &
\includegraphics[width=0.19\textwidth]{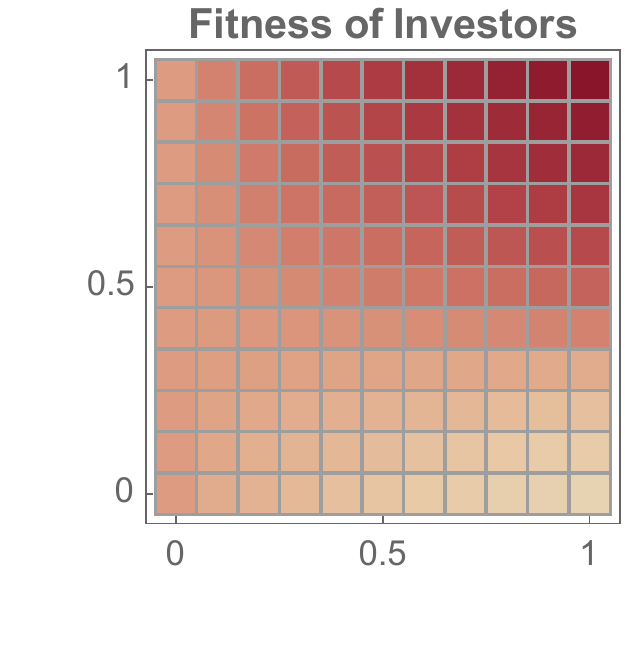}  
\hspace{-0.1cm}\adjustbox{raise=7ex}{\includegraphics[height=0.11\textwidth]{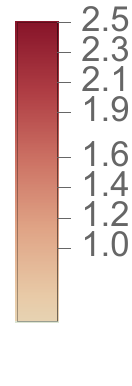}}
& 
\includegraphics[width=0.19\textwidth]{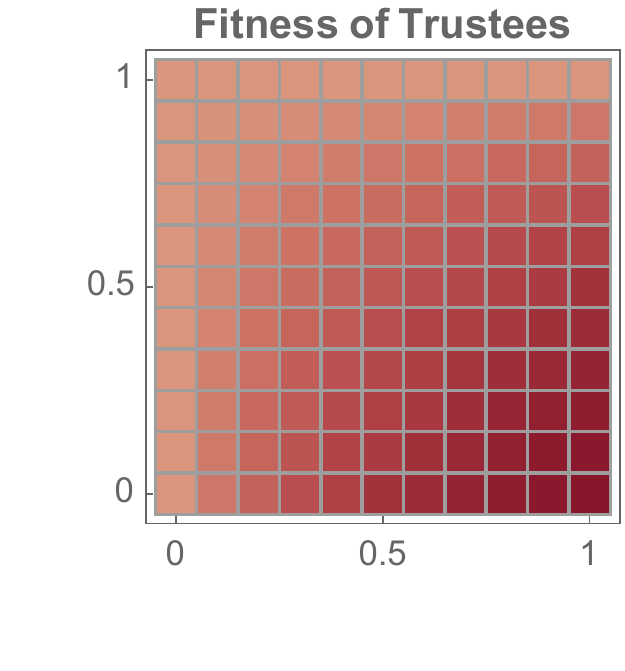}  
\hspace{-0.1cm}\adjustbox{raise=7ex}{\includegraphics[height=0.11\textwidth]{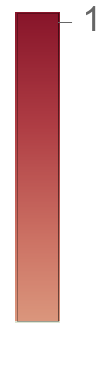}}
\\ 
\includegraphics[width=0.19\textwidth]{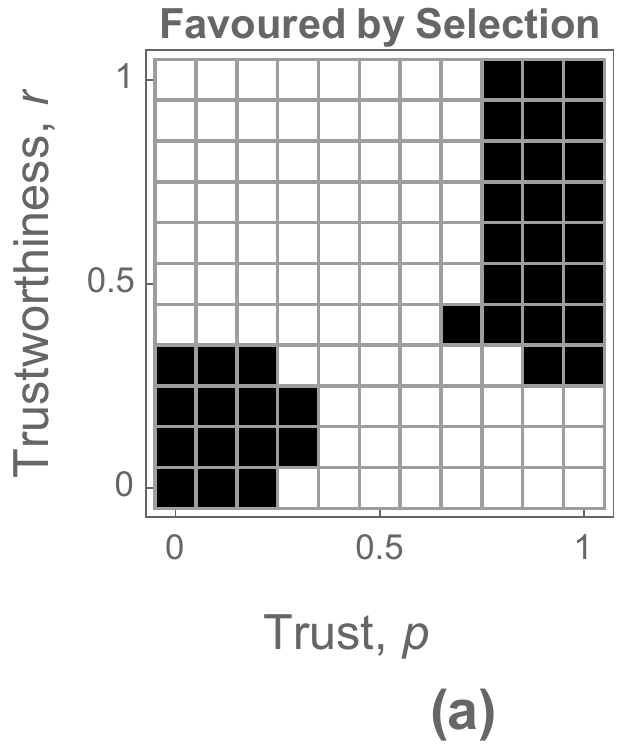} &
\includegraphics[height=0.215\textwidth]{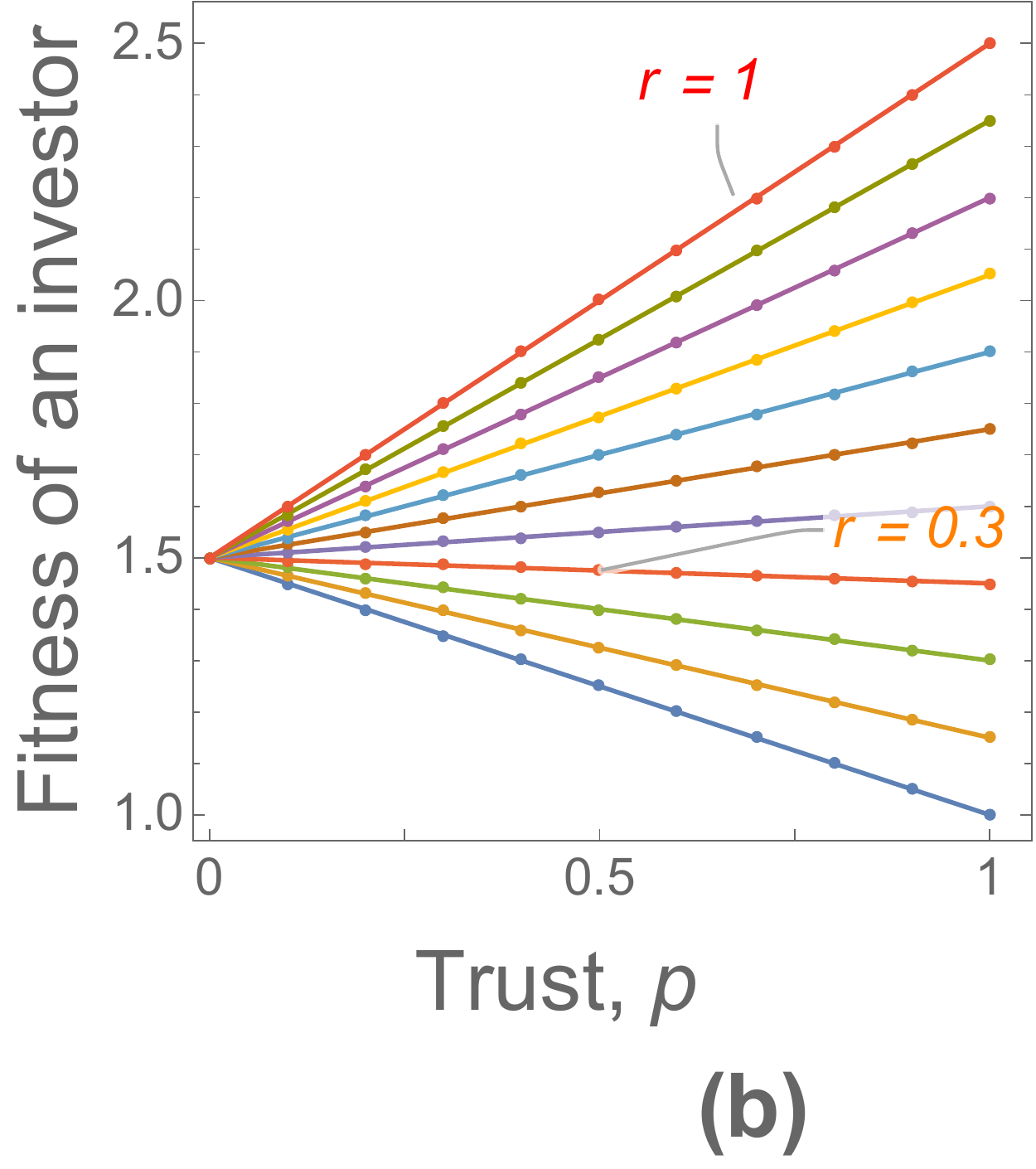} &      
\includegraphics[height=0.215
\textwidth]{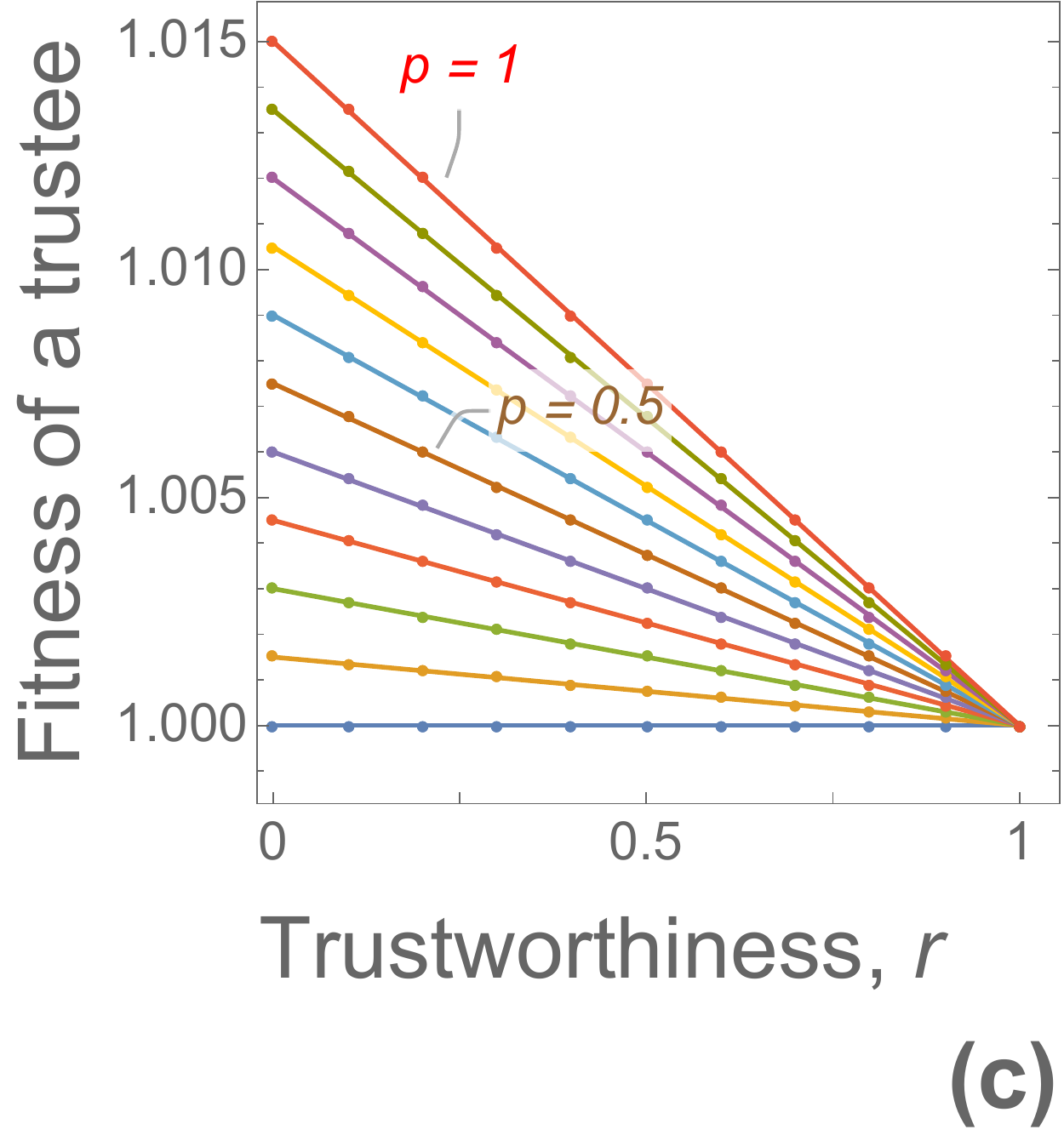}       
\end{tabular}   
\end{center}  
\caption{The reason for the evolution of high trust and high trustworthiness under stronger and weak selection in the investor and trustee populations, respectively.
(a) $\beta_{\scriptscriptstyle I} N_{\scriptscriptstyle I}=25,  \beta_{\scriptscriptstyle T} N_{\scriptscriptstyle T}=0.25$.
(b) The fitness of an investor increases with trust $p$ if trustworthiness $r > 1/b=1/3$ and decreases if $r < 1/b$.
Thus, higher $p$ is selected for if $r > 1/b$ and lower $p$ if $r < 1/b$.
(c) Although the fitness of trustees decreases with $r$, the selection acts weak in the trustee population
so that a wide range of $r$ (low to high) is selected for 
like neutral selection.
Due to the combination of these, high-trust $\otimes$ high-trustworthiness  and low-trust $\otimes$ lower-trustworthiness are selected for.
}   
\label{fig_reason_trust}
\end{figure*}

	An intuition for the evolution of high trust and high trustworthiness can be built from the switching monotonicity of an investor's fitness.
The fitness of an investor increases with trust $p$ if $r > 1/b$ but decreases if $r < 1/b$.
Under strong selection among investors,
high trust is thus favoured for trustworthiness higher than the threshold ($r > 1/b$)
and low trust for lower trustworthiness ($r < 1/b$).
Even though the fitness of a trustee decreases with trustworthiness $r$,
on the other hand,
a wide range of trustworthiness (from low to high) can be favoured under weak selection among trustees.
We thus expect the evolution of high-trust\,$\otimes$\,high-trustworthiness and low-trust\,$\otimes$\,low-trustworthiness
but not high-trust\,$\otimes$\,low-trustworthiness nor low-trust\,$\otimes$\,high-trustworthiness (Fig.\,\ref{fig_reason_trust}).

\subsubsection{Asymmetric Discretisation}

\begin{figure*}  
\begin{center}  
\begin{tabular}{lll}
\includegraphics[width=0.19\textwidth]{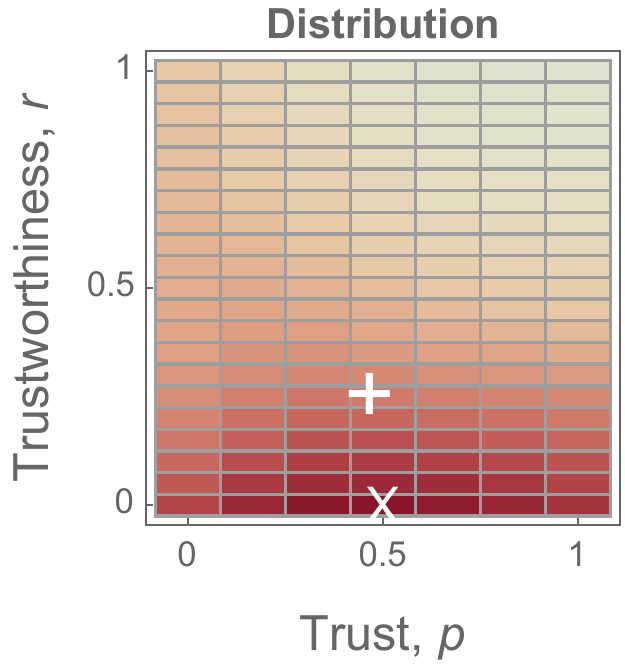}  
\hspace{-0.1cm}\adjustbox{raise=7ex}{\includegraphics[height=0.11\textwidth]{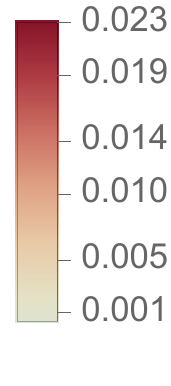}} & 
\includegraphics[width=0.19\textwidth]{AsymmetricTrustGames-429}  
\hspace{-0.1cm}\adjustbox{raise=7ex}{\includegraphics[height=0.11\textwidth]{AsymmetricTrustGames-327}} &
\includegraphics[width=0.19\textwidth]{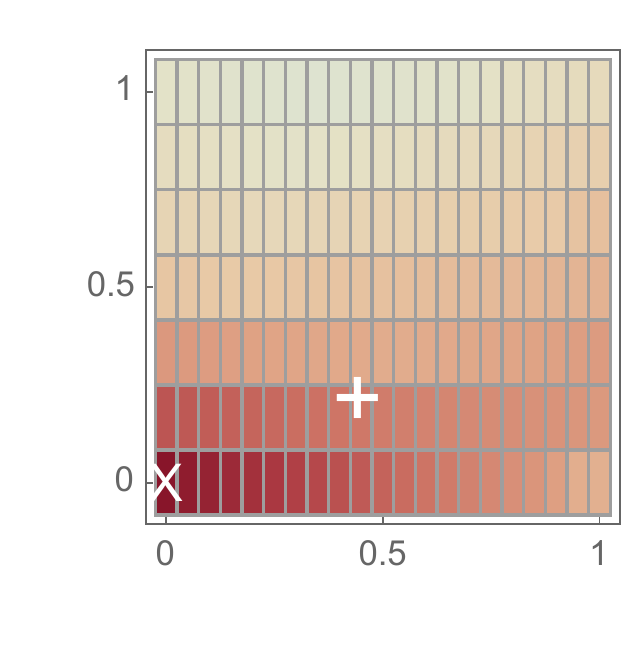} 
\hspace{-0.1cm}\adjustbox{raise=7ex}{\includegraphics[height=0.11\textwidth]{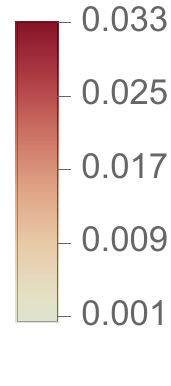}} 
\\
\includegraphics[width=0.19\textwidth]{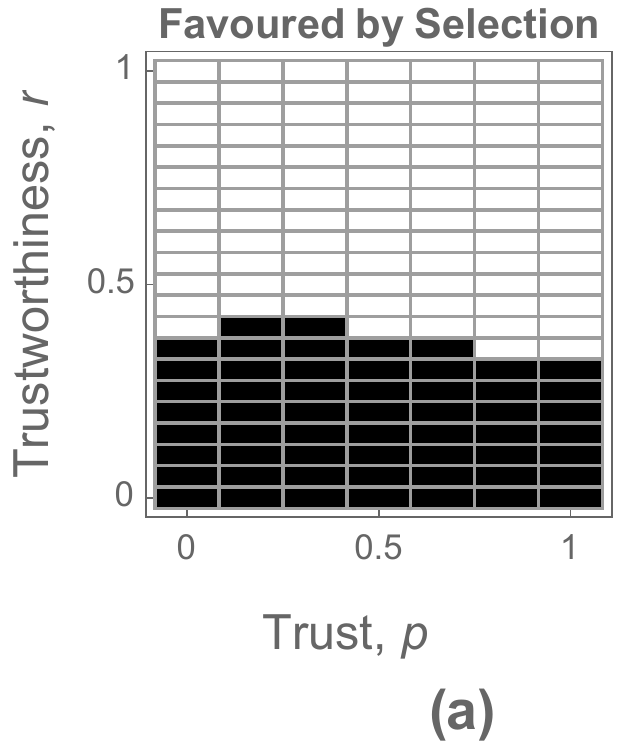} & 
\includegraphics[width=0.19\textwidth]{AsymmetricTrustGames-430} & 
\includegraphics[width=0.19\textwidth]{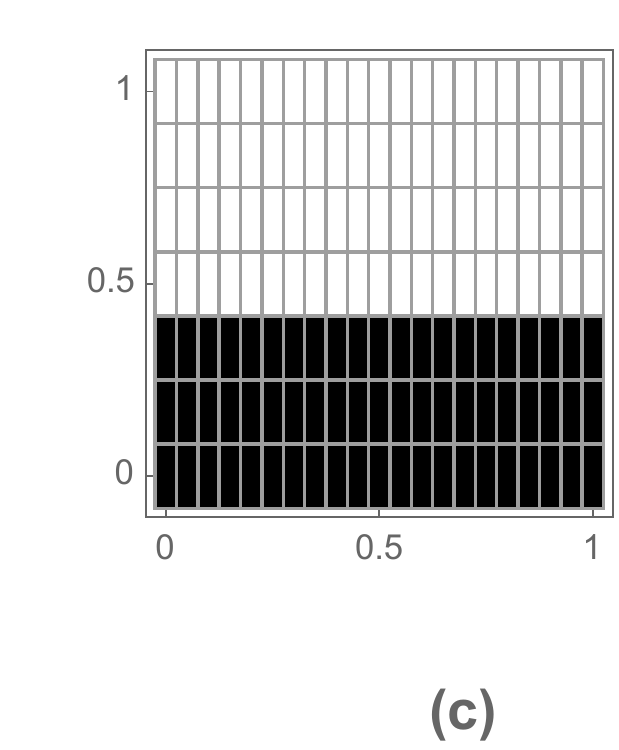} 
\end{tabular}   
\end{center}  
\caption{Asymmetry in the discretisation of strategy space.
(a) $|\mathcal{S}_{\scriptscriptstyle I}|=7, |\mathcal{S}_{\scriptscriptstyle T}|=21$. 
(b)   $|\mathcal{S}_{\scriptscriptstyle I}|=|\mathcal{S}_{\scriptscriptstyle T}|=11$.
(c)  $|\mathcal{S}_{\scriptscriptstyle I}|=21, |\mathcal{S}_{\scriptscriptstyle T}|=7$.
In terms of the mean strategy and the strategies favoured by selection,
there is little difference between asymmetric and symmetric cases.
Although the modal frequency varies,
the trustworthiness of it is null.
The remaining parameters are the same as those in Fig.\,\ref{fig_base_line}\,(b).  
}   
\label{fig_resolution_difference}
\end{figure*}

\begin{figure*}  
\begin{center}  
\begin{tabular}{llll}
\includegraphics[width=0.19\textwidth]{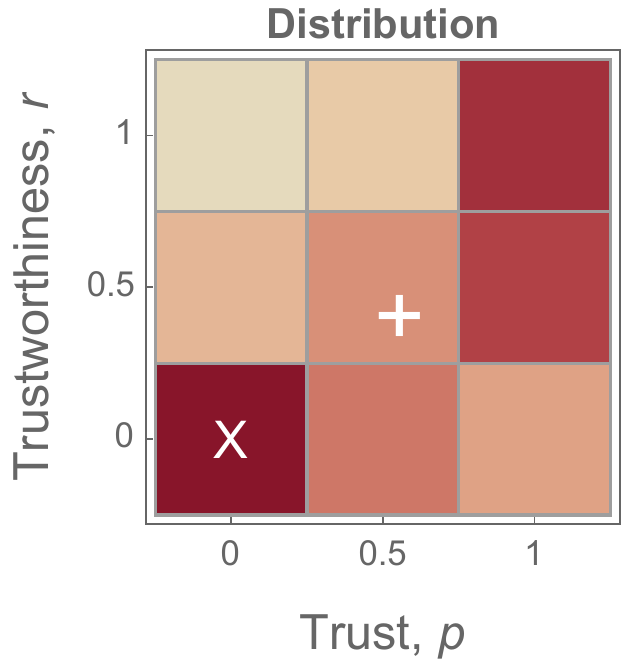}
\hspace{-0.1cm}\adjustbox{raise=7ex}{\includegraphics[height=0.11\textwidth]{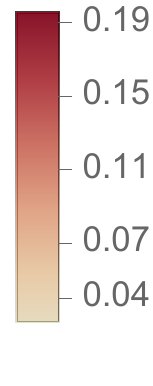}}  &
\includegraphics[width=0.19\textwidth]{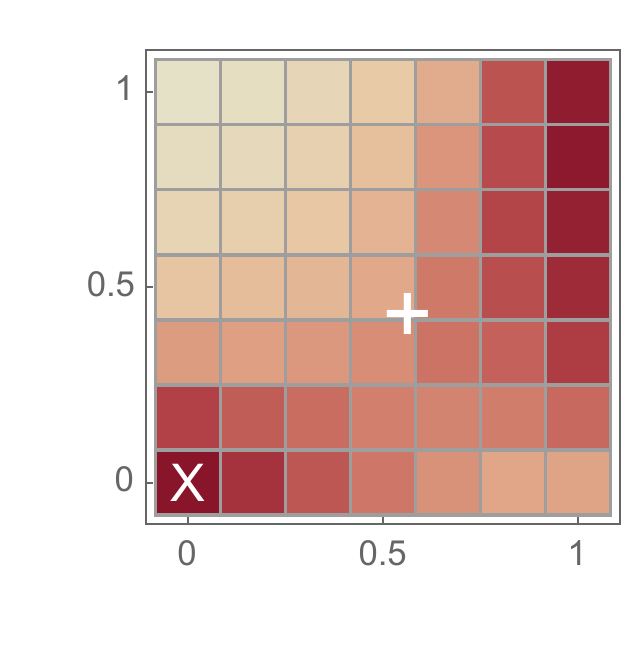}
\hspace{-0.1cm}\adjustbox{raise=7ex}{\includegraphics[height=0.11\textwidth]{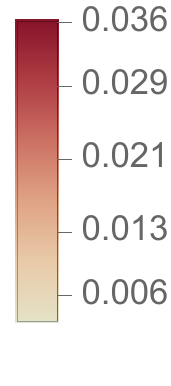}}  &
\includegraphics[width=0.19\textwidth]{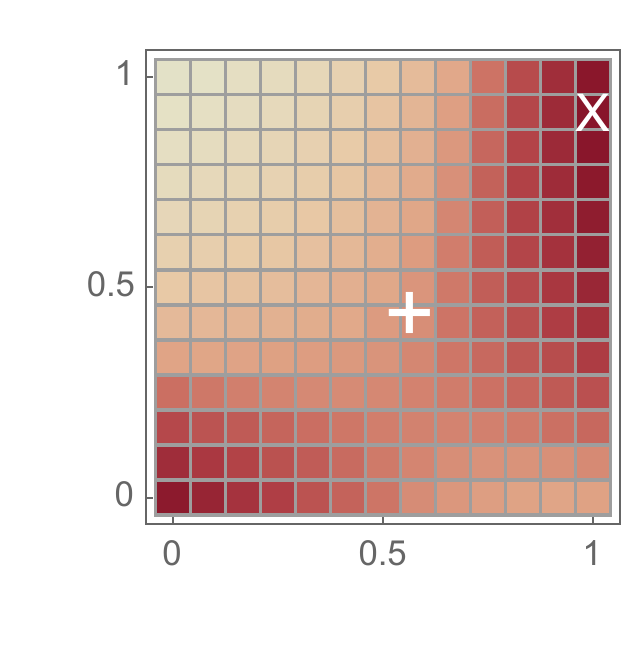} 
\hspace{-0.1cm}\adjustbox{raise=7ex}{\includegraphics[height=0.11\textwidth]{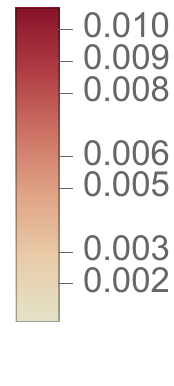}}  &
\includegraphics[width=0.19\textwidth]{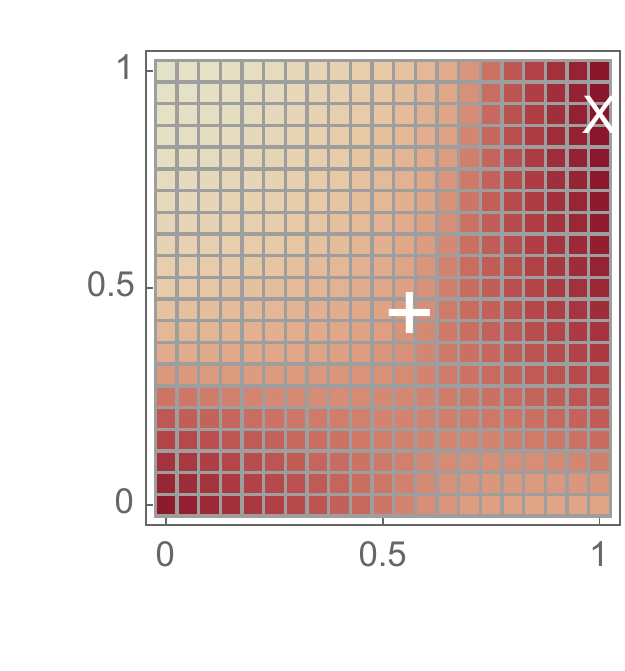}
\hspace{-0.1cm}\adjustbox{raise=7ex}{\includegraphics[height=0.11\textwidth]{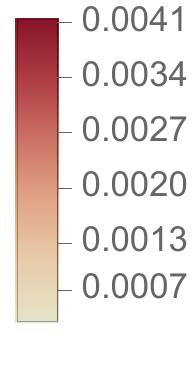}}  
\\
\includegraphics[width=0.19\textwidth]{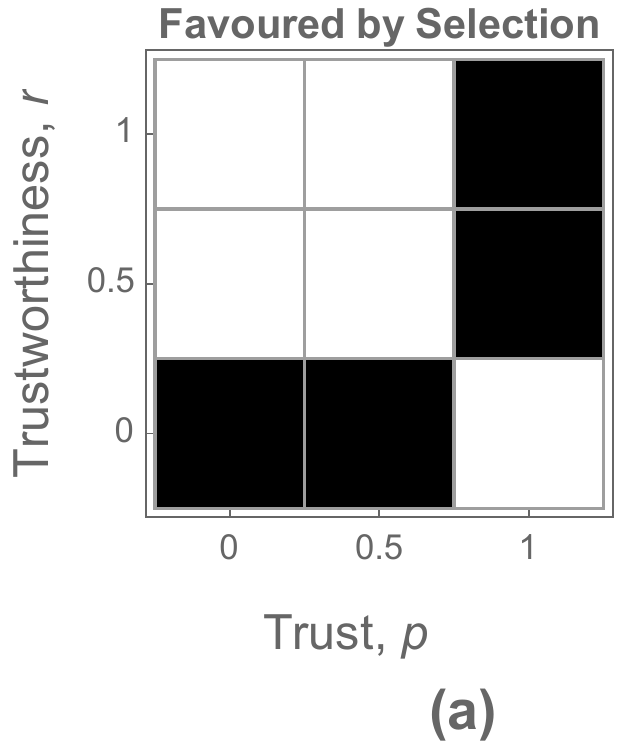}  &
\includegraphics[width=0.19\textwidth]{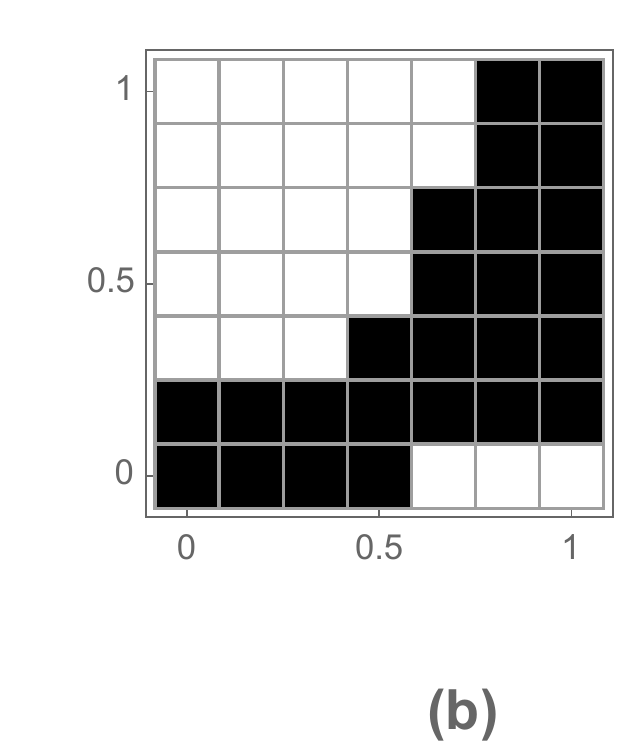}  & \includegraphics[width=0.19\textwidth]{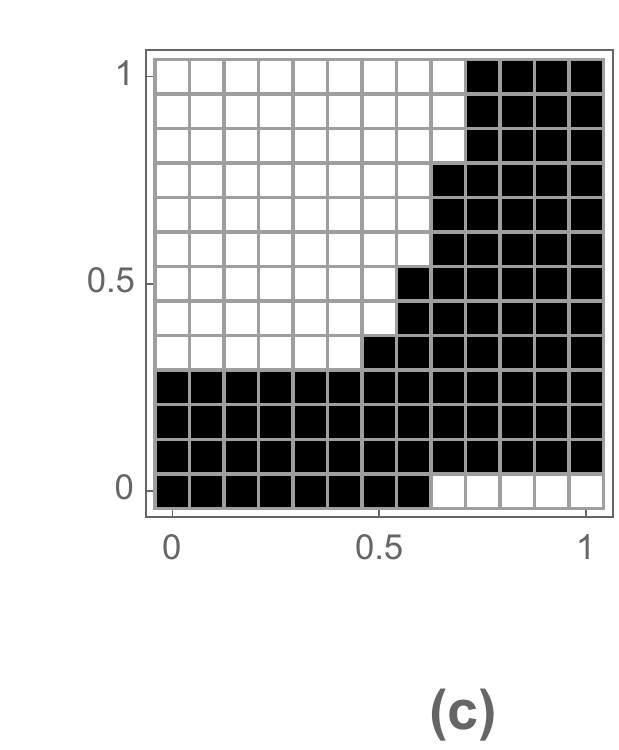}  &
\includegraphics[width=0.19\textwidth]{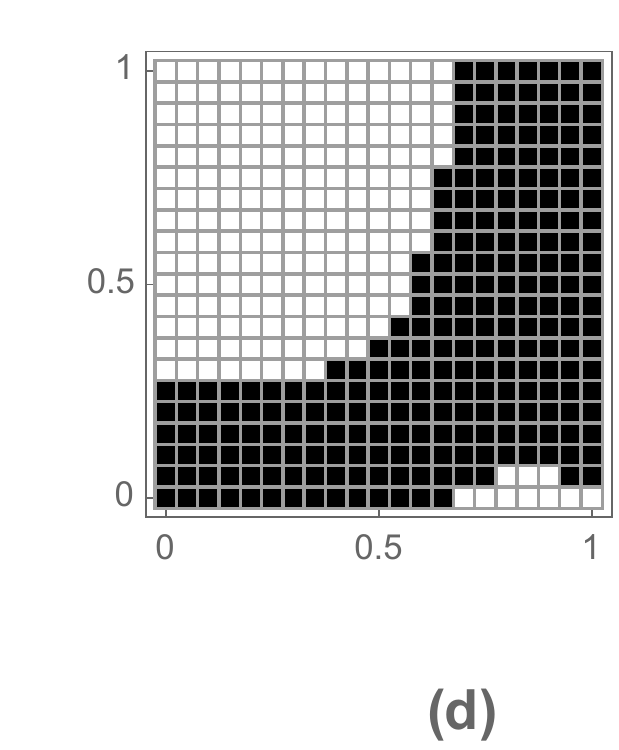}
\end{tabular}   
\end{center}  
\caption{Different resolutions in discretised strategy space. 
(a) $|\mathcal{S}_{\scriptscriptstyle I}|=|\mathcal{S}_{\scriptscriptstyle T}|=3$,
(b)  $|\mathcal{S}_{\scriptscriptstyle I}|=|\mathcal{S}_{\scriptscriptstyle T}|=7$, 
(c)  $|\mathcal{S}_{\scriptscriptstyle I}|=|\mathcal{S}_{\scriptscriptstyle T}|=13$, 
(d) $|\mathcal{S}_{\scriptscriptstyle I}|=|\mathcal{S}_{\scriptscriptstyle T}|=21$.
$\beta_{\scriptscriptstyle I} N_{\scriptscriptstyle I}=2.5, \beta_{\scriptscriptstyle T} N_{\scriptscriptstyle T}=0.25$.
Both the mean strategy and strategies favoured by selection generally remain unchanged with the varying resolutions. 
The modal strategy is not robust to the resolution.
}  
\label{fig_different_resolution}
\end{figure*}

Asymmetric discretisation or resolution in strategy space does not make a substantial difference, at least, in the mean strategy and the strategies favoured by selection (Fig.\,\ref{fig_resolution_difference}).
Note that the modal frequency varies or is not robust to both the asymmetric and symmetric resolutions of strategy discretisation,
whereas the mean strategy and the strategies favoured by selection generally remain unchanged  (Fig.\,\ref{fig_different_resolution}).

\subsubsection{Interference from the Ratio of Mutation Rate to Generation Time}

\begin{figure*}  
\begin{center}  
\begin{tabular}{lll}
\includegraphics[width=0.19\textwidth]{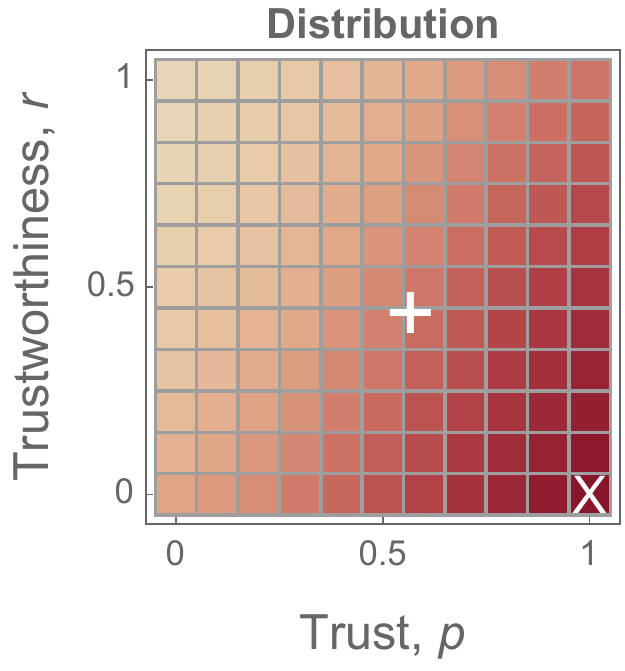}
\hspace{-0.1cm}\adjustbox{raise=7ex}{\includegraphics[height=0.11\textwidth]{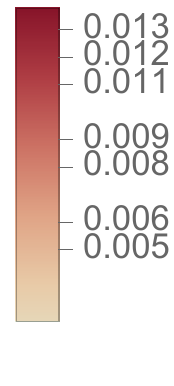}}  &
\includegraphics[width=0.19\textwidth]{AsymmetricTrustGames-459} 
\hspace{-0.1cm}\adjustbox{raise=7ex}{\includegraphics[height=0.11\textwidth]{AsymmetricTrustGames-336}}  &
\includegraphics[width=0.19\textwidth]{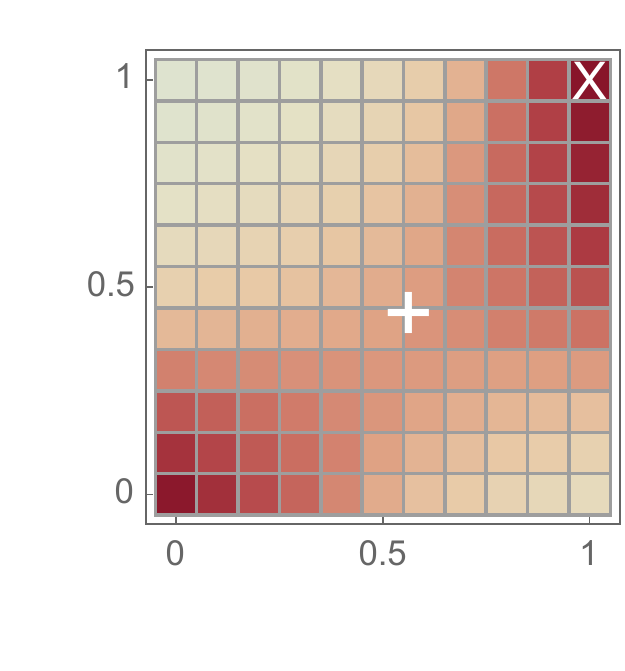}
\hspace{-0.1cm}\adjustbox{raise=7ex}{\includegraphics[height=0.11\textwidth]{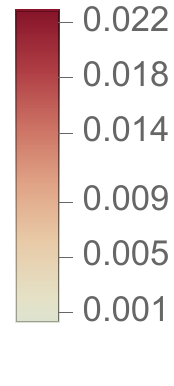}}  
\\
\includegraphics[width=0.19\textwidth]{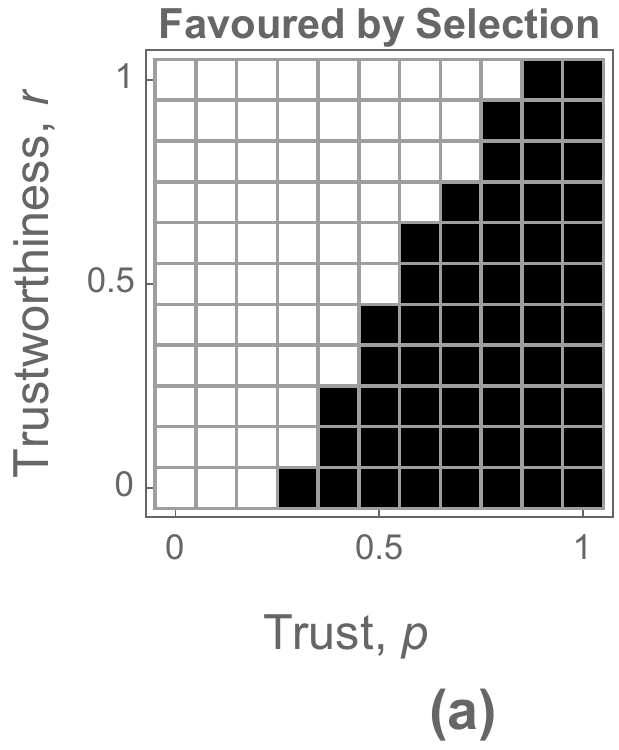}& 
\includegraphics[width=0.19\textwidth]{AsymmetricTrustGames-460}& 
\includegraphics[width=0.19\textwidth]{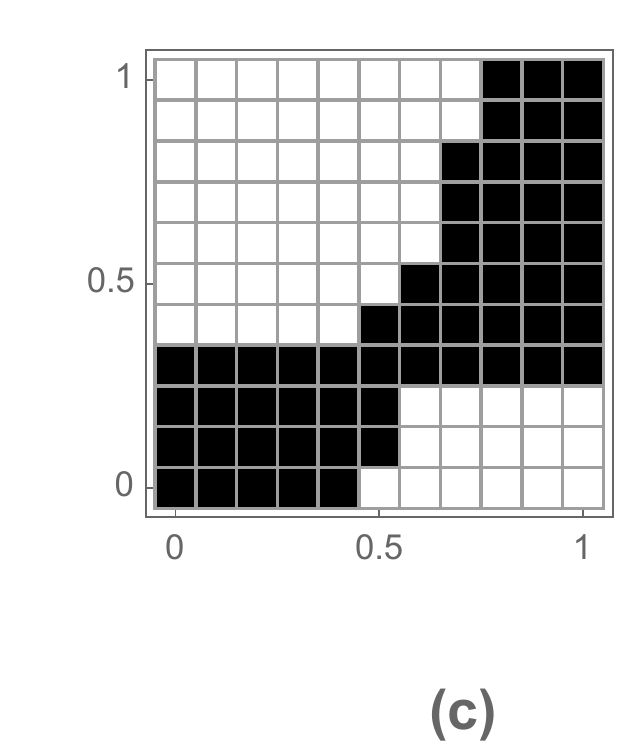}
\end{tabular}   
\end{center}  
\caption{Interactions between asymmetric ratio $u_l/g_l$ and asymmetric product $\beta_l N_l$.
(a)  $u_{\scriptscriptstyle I} / g_{\scriptscriptstyle I} =0.1, u_{\scriptscriptstyle T} / g_{\scriptscriptstyle T} =1$.
(b)  $u_{\scriptscriptstyle I} / g_{\scriptscriptstyle I} =u_{\scriptscriptstyle T} / g_{\scriptscriptstyle T} =1$. 
(c)   $u_{\scriptscriptstyle I} / g_{\scriptscriptstyle I} =1, u_{\scriptscriptstyle T} / g_{\scriptscriptstyle T} =0.1$. 
Given $\beta_{\scriptscriptstyle I} N_{\scriptscriptstyle I} =2.5$ and $\beta_{\scriptscriptstyle T} N_{\scriptscriptstyle T} =0.25$,
the asymmetric ratios do not significantly interfere with the evolution of high trust and high trustworthiness,
the latter of which is yielded by the asymmetric product.
The modal strategy can substantially change, though.
}  
\label{fig_no_interference}
\end{figure*}

We also test out a possible interaction between the asymmetric ratio $u_l/g_l$ and the asymmetric product $\beta_l N_l$.
The evolution of high trust and high trustworthiness due to the asymmetric $\beta_l N_l$ was demonstrated,
holding $u_l/g_l$ symmetric.
We now relax the all-else-equal constraint
and allows for asymmetry in both $u_l/g_l$ and $\beta_l N_l$ at the same time.
Despite the additional asymmetry,
the general outcome of  high trust and high trustworthiness remains unchanged
(Fig.\,\ref{fig_no_interference}).
In other words,
$u_l/g_l$ does not have  significant interaction with $\beta_l N_l$.
Thus, our main result that high trust and high trustworthiness is evolved by a combination of stronger and weak selections  in the investor and the trustee populations is robust.

\subsection{Individual-based Simulation of Moran Process}

\begin{figure*}  
\begin{center}  
\begin{tabular}{llll}
\includegraphics[width=0.19\textwidth]{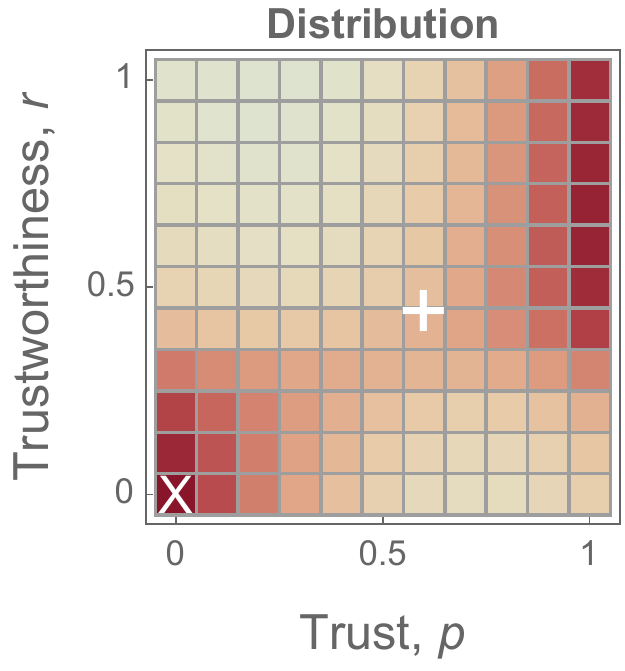}
\hspace{-0.1cm}\adjustbox{raise=7ex}{\includegraphics[height=0.11\textwidth]{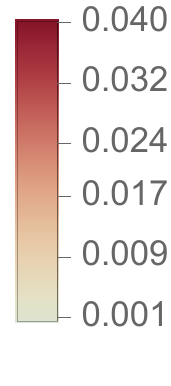}} &
\includegraphics[width=0.19\textwidth]{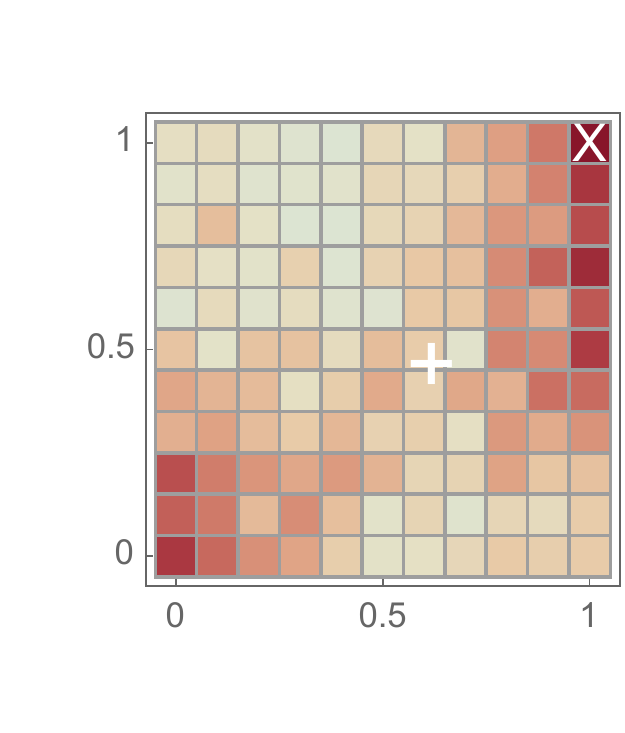}
\hspace{-0.1cm}\adjustbox{raise=7ex}{\includegraphics[height=0.11\textwidth]{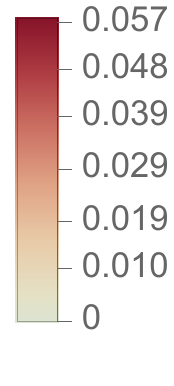}} &
\includegraphics[width=0.19\textwidth]{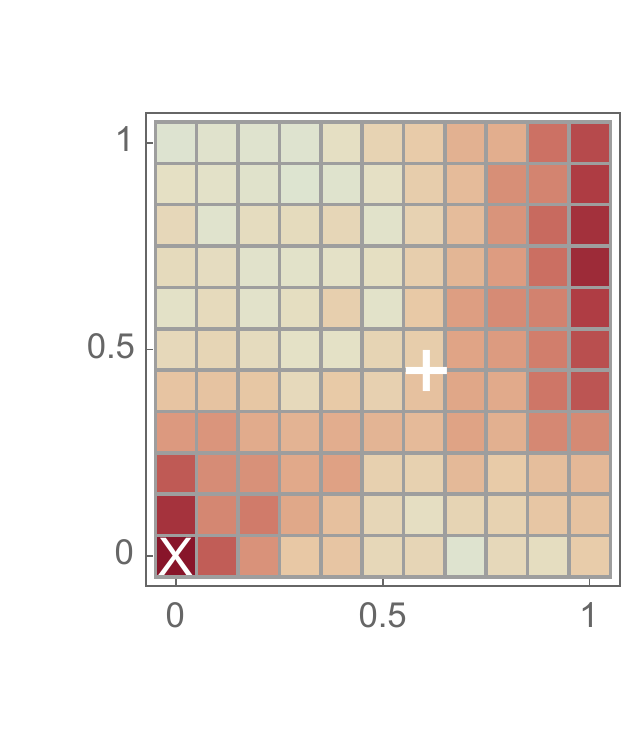}
\hspace{-0.1cm}\adjustbox{raise=7ex}{\includegraphics[height=0.11\textwidth]{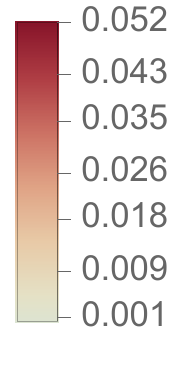}} &
\includegraphics[width=0.19\textwidth]{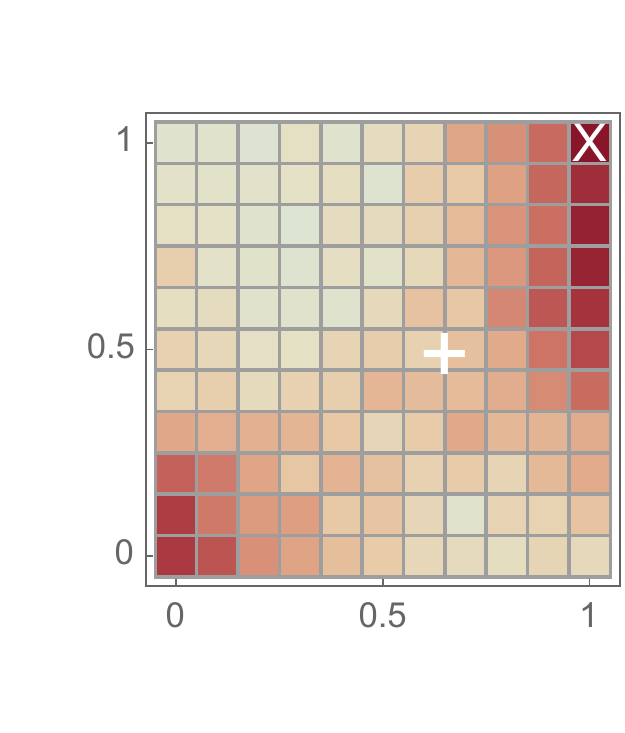}
\hspace{-0.1cm}\adjustbox{raise=7ex}{\includegraphics[height=0.11\textwidth]{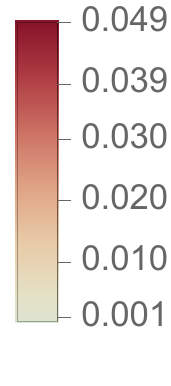}} 
\\
\includegraphics[width=0.19\textwidth]{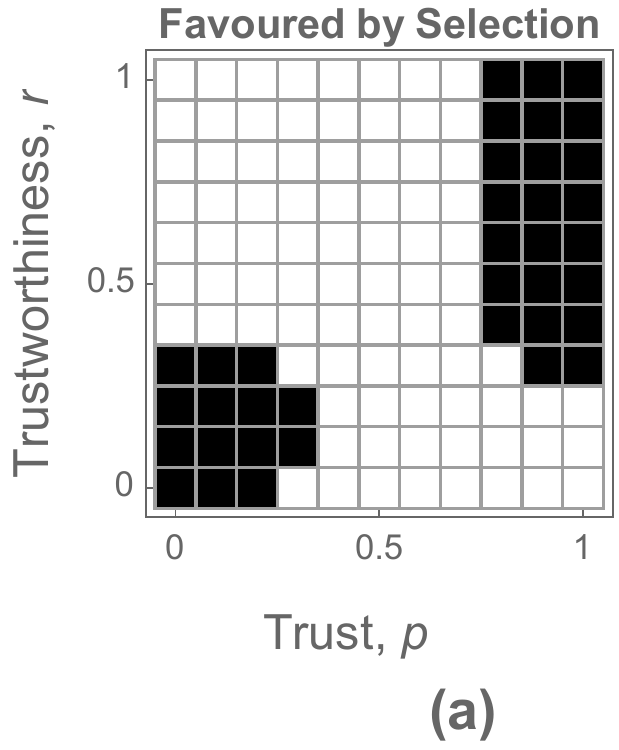} &
\includegraphics[width=0.19\textwidth]{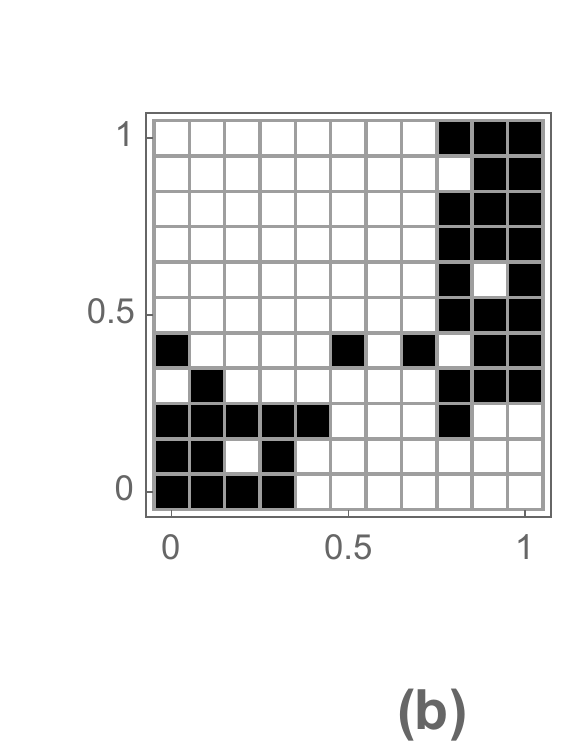} &
\includegraphics[width=0.19\textwidth]{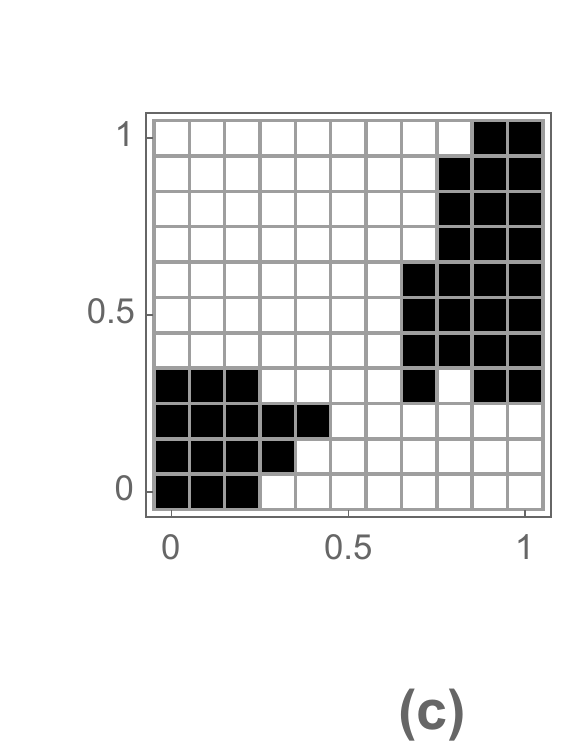} &
\includegraphics[width=0.19\textwidth]{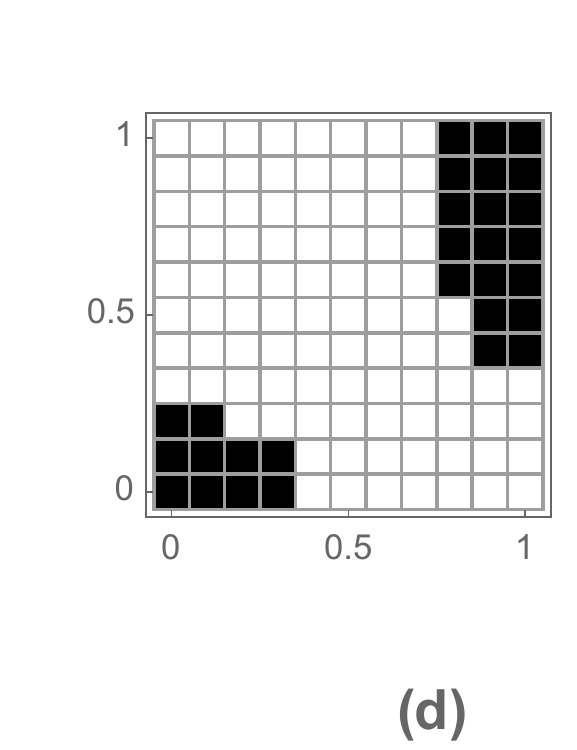}
\end{tabular}   
\end{center}  
\caption{
The distributions of the pure state by
the individual-based simulations of the Moran process.
(a)
The distribution obtained by solving the eigenequation.
The distributions are also obtained by the simulation of the Moran process at various mutation rates
(b)   $u =u_{\scriptscriptstyle I} =u_{\scriptscriptstyle T}  =10^{-4}$,  
(c)    $10^{-3}$,
and  
(d)    $10^{-2}$.
The proportion of the simulation time/generations when the populations are in the pure states
are
(b)    98\%, 
(c)    82\%,
and
(d)    15\%.
As the mutation rate increases,
the chance of the populations in pure states decreases.
The distribution predicted by the Markov chain matches the distributions obtained by the simulation well enough.
For $u =10^{-1}$ (not shown), virtually no pure state is realised, but only a mixture of different strategies.
$\beta_{\scriptscriptstyle I} =0.5$, $\beta_{\scriptscriptstyle T}=0.025$,
$N_{\scriptscriptstyle I} =50$ and $N_{\scriptscriptstyle T} =10$. 
}   
\label{fig_simulation}
\end{figure*}

So far, all the results of Fig.\,\ref{fig_base_line} to \ref{fig_no_interference} are obtained by numerically solving the eigenequation \eqref{eq_eigen}.
We also run the individual-based simulations of the Moran process at various mutations rates, while maintaining the asymmetry between   $\beta_{\scriptscriptstyle I} N_{\scriptscriptstyle I} $
and $\beta_{\scriptscriptstyle T} N_{\scriptscriptstyle T}$.
The stationary distribution obtained by solving the eigenequation associated with  the Markov chain Eq.\,\eqref{eq_transition_prob} is compared with those obtained by the simulations of the Moran process.
With the asymmetric parameters,
the individual-based simulations lead to the evolution of high trust and trustworthiness at the low mutation rates ($u =10^{-4}$ and $10^{-3}$),
as predicted by the Markov chain (Fig.\,\ref{fig_simulation}).

The individual-based simulation also enables us to examine the evolutionary dynamics of the Moran process 
even at (relatively) high mutation rates,
where the assumption for the weak-mutation limit may  not be met for most of the time.
At the mutation rate $u=10^{-2}$,
for instance,
the populations are in pure or homogeneous states
only for 15\% of the time,
unlike the lower mutation rates
where they are in pure states for most of the time.
For that 15\% of the time when the populations are in pure states,
however,
high trust and trustworthiness still evolve,
matching the prediction of the Markov chain.
In other words,
the weak-mutation limit approach well predicts the stationary distribution of pure state even at high mutation rates if the pure states exist.

\begin{figure*}  
\begin{center}  
\begin{tabular}{llll}
\includegraphics[width=0.19\textwidth]{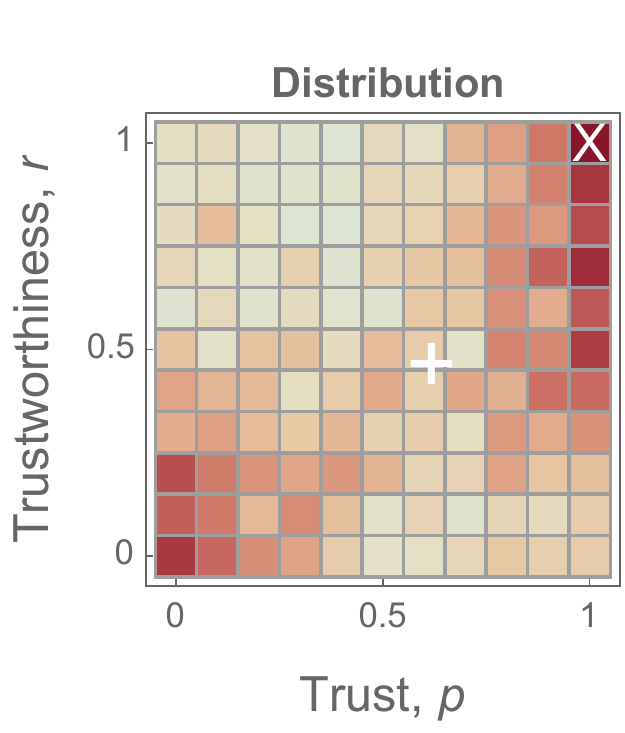}
\hspace{-0.1cm}\adjustbox{raise=7ex}{\includegraphics[height=0.11\textwidth]{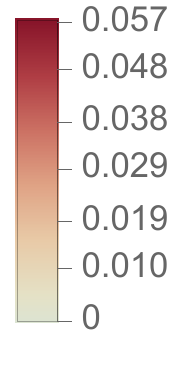}}  &
\includegraphics[width=0.19\textwidth]{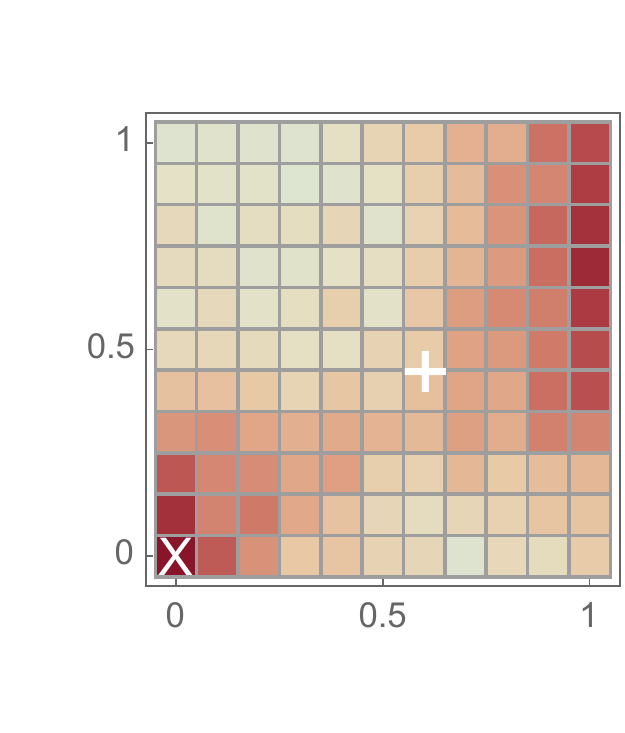}
\hspace{-0.1cm}\adjustbox{raise=7ex}{\includegraphics[height=0.11\textwidth]{AsymmetricTrustGames-411}} &
\includegraphics[width=0.19\textwidth]{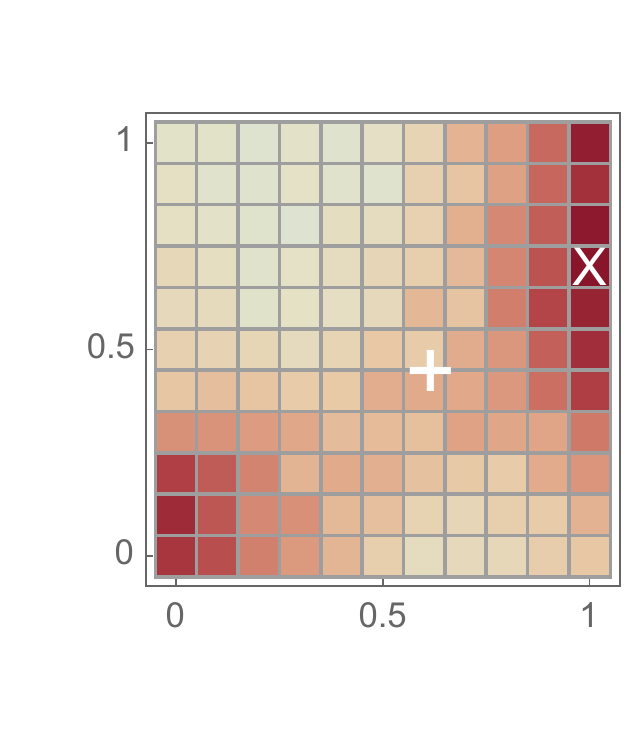}
\hspace{-0.1cm}\adjustbox{raise=7ex}{\includegraphics[height=0.11\textwidth]{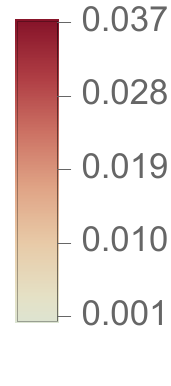}} &
\includegraphics[width=0.19\textwidth]{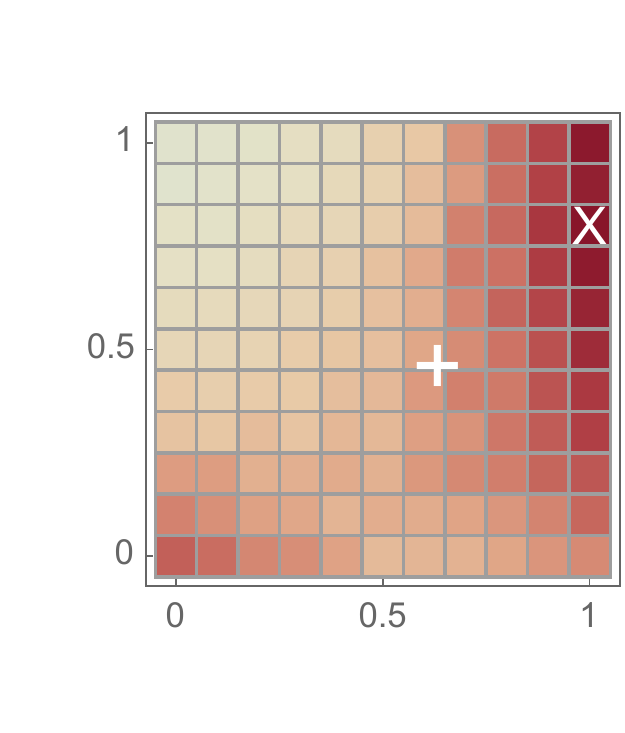}
\hspace{-0.1cm}\adjustbox{raise=7ex}{\includegraphics[height=0.11\textwidth]{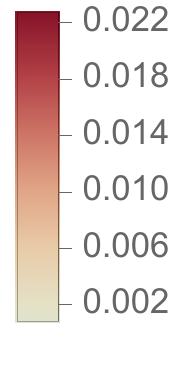}} 
\\
\includegraphics[width=0.19\textwidth]{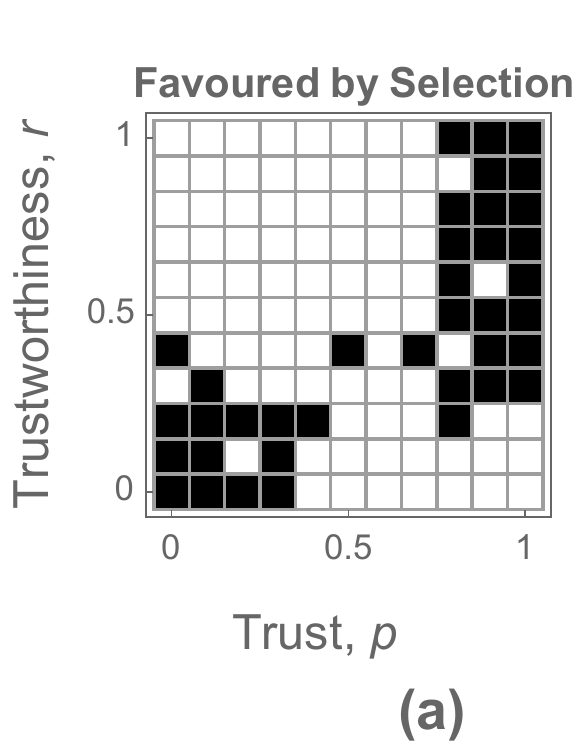} &
\includegraphics[width=0.19\textwidth]{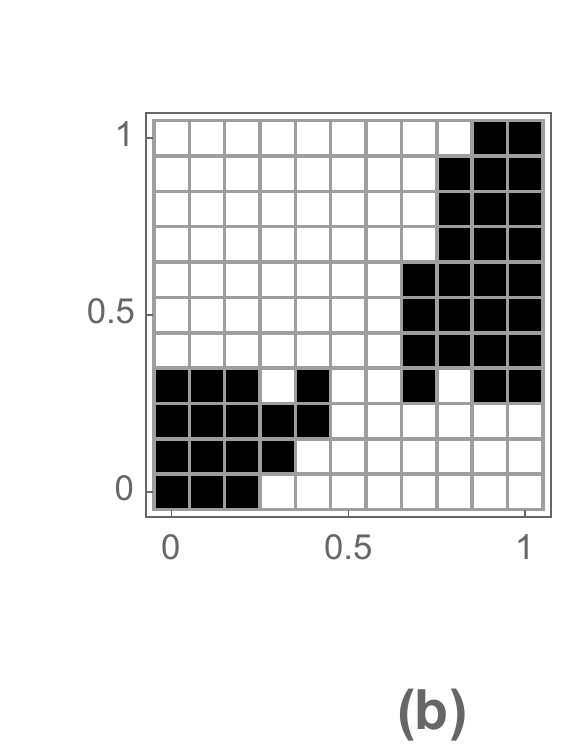} &
\includegraphics[width=0.19\textwidth]{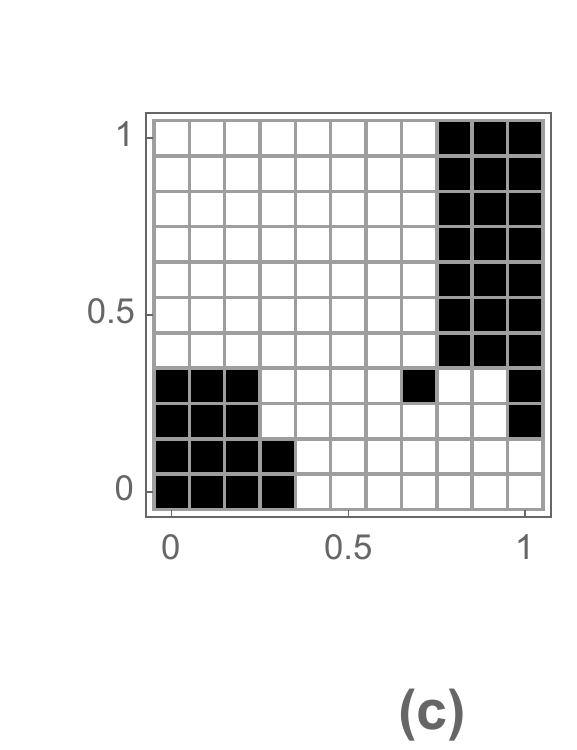} &
\includegraphics[width=0.19\textwidth]{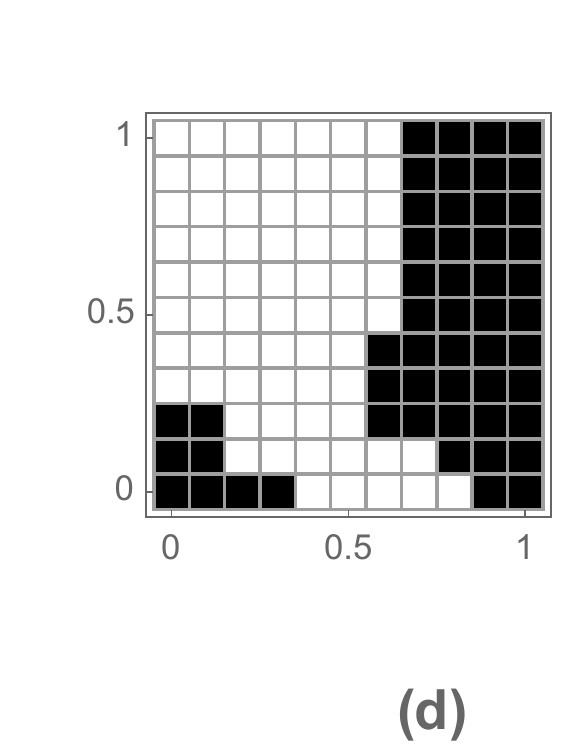} 
\end{tabular}   
\end{center}  
\caption{
The distributions of the most frequent strategy in the populations by the same simulations of
Fig.\,\ref{fig_simulation}.
(a)  $u =u_{\scriptscriptstyle I} =u_{\scriptscriptstyle T}  =10^{-4}$,
(b)   $10^{-3}$, 
(c)   $10^{-2}$, 
and
(d)   $10^{-1}$.
At low mutation rates (e.g.\,$u=10^{-4}$),
the distributions of the most frequent strategy and the pure state are virtually the same.
At high mutation rates (e.g.\,$u=10^{-1}$),
the distribution of the most frequent strategy is still well defined,
whereas the distribution of pure state is not since
the populations are hardly in any pure states.
The distributions of the most frequent strategy are (qualitatively) similar to each other regardless of mutation rates.
In other words, the asymmetric parameters  evolve high trust and trustworthiness regardless of mutation rates.
}  
\label{fig_simulation_commonest}
\end{figure*}

We can relax the examination of only the pure population states.
We instead monitor the most frequent strategy in a population at a time or per generation.
The pure state is a special case of the most frequent strategy, where there is only one (type of) strategy in a population at a time.
At low mutation rates, distributions of the most frequent strategy 
and pure state are very similar 
since each population is in pure states for most of the time.
At high mutation rates, each of the populations is hardly in any pure states (but mostly in heterogeneous states) 
but the most frequent strategy exists at any time.
Surprisingly,
the distributions of the most frequent strategy at high mutation rates are similar to those at low mutation rates (Fig.\,\ref{fig_simulation_commonest}).
In other words,
regardless of mutation rates,
the asymmetric parameters can evolve high trust and trustworthiness as the most frequent strategy in the populations at a time.

\section{Discussion}

	Stochastic evolutionary dynamics of the asymmetric TG 
can yield the evolution of substantial trust and trustworthiness  
when demographic parameters between the two populations are asymmetric. 
Specifically, a combination of 
stronger selection in the investor population (i.e.\,$\beta_{\scriptscriptstyle I} N_{\scriptscriptstyle I} \gg \beta_{\scriptscriptstyle T} N_{\scriptscriptstyle T}$)
and weak selection 
in the trustee population (i.e.\,$\beta_{\scriptscriptstyle T} N_{\scriptscriptstyle T}  <1$)
leads to the evolution of high trust and high trustworthiness.
Even full trust and near full trustworthiness can be the most common strategy.
	These predictions are qualitatively different from those of previous work with single- and two-population models, where the most common strategy always involved null trustworthiness \cite{Tarnita:2015aa};
enforcing the symmetric parameter values between the populations, 
the previous two-population model of the asymmetric TG missed the richer evolutionary dynamics
that would stem from the asymmetric parameters.
To evolve non-zero trustworthiness as the most common strategy in the symmetric TG,
previous works turned to additional mechanisms such as reputation-based interactions \cite{Manapat:2013aa}\cite{King-Casas:2005aa}\cite{Masuda:2012aa}.
It would apply to only humans but not lower animals 
that lack high cognitive complexity 
to process the reputation information.
However, we have demonstrated  that it does not necessarily require additional mechanisms or deterministic causes.
Life being a discrete phenomenon,
the consequent stochasticity may yield evolutionary outcomes that deterministic models cannot 
 \cite{Houchmandzadeh:2014aa}.
For the asymmetric TG,
we show that the stochasticity combined with the asymmetric demographic parameters suffices to yield high degrees of trustworthiness as the most common strategy,
which is applicable to lower animals
as well as humans.

	We effectively reduce the number of demographic parameters 
necessary for studying the impact of their asymmetry on evolutionary dynamics.
A previous work  independently varied mutation rate, generation time, selection strength and population size 
for an asymmetric game \cite{Veller:2017aa}.
Since the transition probability is a function of the ratio of  mutation rate to generation time,
one only needs to treat and vary the ratio as if a single parameter but not the two independent parameters.
We can also treat the product 
of selection strength and population size as if a single parameter
since 
transition probability can be well approximated as a function of  the product.
Hence,
we 
reduce the number of parameters 
from four 
to two in studying the effects of asymmetry in parameters.        
Note that this reduction applies to two-population models of asymmetric games in general.

	We also investigate possible interactions between parameters. 
The all-else-equal approach is effective
to pinpoint the parameter,
asymmetry in which
would significantly alter the evolutionary outcome compared to that of the symmetric case; 
 all parameters except for one are held equal and constant for the two populations 
\cite{Veller:2017aa}.
However,
the all-else-equal approach does not reveal possible interactions between the parameters. 
Indeed, the condition of the all-else-equal
may rarely be realised in the uncontrolled real world.
Relaxing the all-else-equal,
we consider all combinations of asymmetry in both the ratio $\mu_l/g_l$ and the product $\beta_l N_l$.
We find that there is no significant interaction between the ratio and the product.
Hence, we can conclude 
that the combination of stronger selection in the investor population and weak selection in the trustee population robustly promotes the evolution of trust and trustworthiness regardless of the other parameter.

	Weak-selection limit is often assumed to analytically study the stochastic dynamics in finite populations  \cite{Wu:2010aa}\cite{Ohtsuki:2010aa}.
The weak-selection limit was applied
to both the symmetric TG and the asymmetric TG \cite{Tarnita:2015aa}; only symmetric demographic parameters for the latter were considered, though.
An alternative to weak-selection limit is weak-mutation limit \cite{Fudenberg:2006aa},
which has been widely used for single-population models of symmetric games
\cite{Hauert:2007fk}\cite{Sigmund:2010aa}\cite{Van-Segbroeck:2012aa}\cite{Requejo:2012aa}\cite{Wang:2010aa}.
The weak-mutation limit for two-population models of asymmetric games has been recently proposed \cite{Veller:2016aa}.
To our best knowledge, our work is the first application of the weak-mutation approach to the asymmetric TG.
The weak-selection and weak-mutation approaches complement each other
in the sense that the former has no restriction on mutation rates, 
whereas the latter has no restriction on selection strengths.

	The mode of a stationary distribution would have a less predictive meaning 
if the distribution  is relatively flat and wide;
large deviations from it would be frequent
unlike a distribution sharply peaked at the mode.
The previous work on the TG used weak-selection limit to analytically derive the mode \cite{Tarnita:2015aa}.
Ironically, the weak-selection limit flattens and widens the stationary distribution significantly.
The modal strategy 
from such a flat distribution 
is of limited information
since a wide range of strategies significantly deviated from it can be selected for or observed.		
	The mean of a stationary distribution well complements the mode
especially when they are substantially different from each other. 
Our two-population model with asymmetric parameters yields  wide and multimodal distributions.
Although representing the highest one,
the mode of such a distribution represents only one of the multiple peaks in it,
offering less information especially 
when 
the strategies corresponding to
those picks substantially deviate from each other.
On the other hand,
the mean better reflects all the peaks
in that it takes all of them into consideration.
Also, the mean is robust to the (resolution of) discretisation of continuous strategies,
whereas the mode is not.	
Unlike the previous work not considering the mean \cite{Tarnita:2015aa},
we believe that both the mode and the mean had better be used together.

	The evolution of high trust and high trustworthiness by the asymmetric demographic parameters is robust to the resolution or granularity of discretised strategies,
regardless of symmetric or asymmetric resolution.
In a sense, the resolution reflects degrees of errors in perceiving continuous strategies; the lower resolution, the higher error in imitation of continuous strategies.
Hence, we can say that the evolution of high trust and trustworthiness by the asymmetric parameters is robust to the perception error.

The one-step transitions in the embedded dynamics over pure states in the two-population model are constrained;
given pure state, it can transit to another pure state that exclusively differs in  either trust or trustworthiness but not both.
In the single-population model,
there is no such constraint;
one-step transition to pure state differing in both trust and trustworthiness is possible.
It is one of the key differences between the single- and two-population models.
However, it turns out
that the constraint yields little difference in terms of evolutionary outcomes.
Despite  the constraint,
the two-population model yields outcomes similar to those of the single-population model if symmetric parameters are used.
Transitions between pure states different in both trust and trustworthiness are possible in two steps; for instance, one transition for different trust followed by another transition for different trustworthiness.
In other words, any pure state is possible to be reached in two one-step transitions
and being unable to do it in a one-step transition does not
make much difference in 
terms of evolutionary outcomes.

	The two-population model of the asymmetric TG   ironically yields both simplicity and richness in the weak-mutation limit.
It leads to simpler fixation probability 
 that is frequency-independent and easy to compute.
It yields rich evolutionary outcomes at the same time.
Naturally inducing asymmetric parameters between the populations,
it can yield high trust and high trustworthiness as well as low trust and low trustworthiness.
This contrasts to the single-population model 
that leads to complex fixation probability 
and yet simpler evolutionary outcomes,
unable to yield high trust and high trustworthiness.

The key assumption for the embedded Markov chain over the pure states to validly approximate the evolutionary dynamics is
that the mutation rate is so small
that the populations are in pure or homogeneous states most of the time.
How small should the mutation rate be?
This can be experimentally answered by running the individual-based simulation of the Moran process at various mutation rates
as we did with the rates of different orders of magnitude.
The key criterion for the mutation rate sufficiently small is whether the population is in pure homogeneous states most of the time.
According to our computer simulation results, the mutation $u=10^{-4}$ is sufficiently small in that the populations are in pure states for 98\% of the total simulation periods or generations as shown in  Fig.\,\ref{fig_simulation}.
Even $u=10^{-3}$ can be considered sufficiently small for practical purposes since the population are in pure states for 82\% of the simulation periods. 
There is also an analytical alternative to the experimental approach.
Although derived for a single population case,
there is a threshold mutation rate $(N \ln N)^{-1}$,
where a mutation rate lower than it is considered small enough for the embedded Markov chain to be a valid approximation
\cite{Wu:2012aa}.
For the threshold rate with our simulations of the two-population system,
we use the total population size $N =N_{\scriptscriptstyle I} + N_{\scriptscriptstyle T} =50 +10 =60$,
where $N_{\scriptscriptstyle I}$ and $ N_{\scriptscriptstyle T}$ are the sizes of the investor population and the trustee population, respectively.
The threshold mutation rate is $(N \ln N)^{-1} =(60 \ln 60)^{-1} \approx 4\times 10^{-3}$,
according to which $u=10^{-4}$ is small enough,
whereas $u=10^{-2}$ and $u=10^{-1}$ are not.
The analytical approach seems to match the experimental approach well.

By definition, the weak-mutation limit approach  applies to only low mutation rates,
where populations are mostly in pure states
and the simplified dynamics of the Markov chain over the pure states well approximates the evolutionary dynamics of the Moran process.
For higher mutation rates where the populations are not in pure states but heterogeneous states (i.e.\,a mixture of different strategies),
the assumption for the Markov chain over pure states is not met
and
the individual-based computer simulation of the Moran process is instead used to study the evolutionary dynamics of it \cite{Van-Segbroeck:2012aa}\cite{Requejo:2012aa}.
If we summarise a population state at a time with the most frequent strategy in it,
it applies to any mutation rates, 
unlike the pure population state
that applies to only low mutation rates.
At low mutation rates,
the most frequent strategy in a population is virtually the same as the pure state of a population.
Also,
the distributions of the most frequent strategy are similar to each other regardless of mutation rates.
This implies that
the distribution of pure state predicted by the weak-mutation limit  can also approximate the distribution of the most frequent strategy 
regardless of mutation rates.
In other words,
although it was originally devised to describe the distribution of pure state at low mutation rates,
the weak-mutation limit approach can have its predictive power 
to describe distributions of the most frequent strategy 
regardless of mutation rates.
If this conclusion also applies to other games (e.g.\,PD) as well,
it would mean that, regardless of mutation rates, one can study the stochastic evolutionary dynamics using the distributions obtained from the weak-mutation approach.
This would be an interesting future work
that has potentially wide implications.

The asymmetric parameters in our two-population model are of inter-population asymmetry but not intra-population asymmetry. Whereas the selection strength in one population is different from that in the other population, for instance, the selection strength among individuals in the same population is identical.
Thus, the (inter-population) asymmetry is not applicable to a single-population model such as PD as well as the symmetric TG.
However,
we could also break the symmetry or homogeneity of a parameter in a population, yielding intra-population asymmetry. 
For instance, a portion of individuals in a population can have a low selection strength,
whereas the remaining individuals in the same population have a high selection strength. 
This intra-population asymmetry of a parameter would then be applied to the single-population models as well.
This would be an interesting line of future work.


Rich evolutionary dynamics
induced by asymmetric parameters could be explored for other asymmetric games
such as the Ultimatum Game
that was studied with symmetric parameters only \cite{Rand:2013aa}.
With asymmetric parameters,
it would be also interesting
to analytically derive the modal strategy  in the weak-selection limit
and study the condition for it to be high trust and high trustworthiness in the asymmetric TG.
We hope that our work paves a way to explore rich 
game dynamics with asymmetric parameters.

\begin{acknowledgments}
We would like to thank  Naoki Matsuda for his helpful comments on the draft.
In memory of Peter Wittek
who prematurely passed away 
and would contribute more to physics, otherwise.
\end{acknowledgments}

%

\begin{thebibliography}{45}%
\makeatletter
\providecommand \@ifxundefined [1]{%
 \@ifx{#1\undefined}
}%
\providecommand \@ifnum [1]{%
 \ifnum #1\expandafter \@firstoftwo
 \else \expandafter \@secondoftwo
 \fi
}%
\providecommand \@ifx [1]{%
 \ifx #1\expandafter \@firstoftwo
 \else \expandafter \@secondoftwo
 \fi
}%
\providecommand \natexlab [1]{#1}%
\providecommand \enquote  [1]{``#1''}%
\providecommand \bibnamefont  [1]{#1}%
\providecommand \bibfnamefont [1]{#1}%
\providecommand \citenamefont [1]{#1}%
\providecommand \href@noop [0]{\@secondoftwo}%
\providecommand \href [0]{\begingroup \@sanitize@url \@href}%
\providecommand \@href[1]{\@@startlink{#1}\@@href}%
\providecommand \@@href[1]{\endgroup#1\@@endlink}%
\providecommand \@sanitize@url [0]{\catcode `\\12\catcode `\$12\catcode
  `\&12\catcode `\#12\catcode `\^12\catcode `\_12\catcode `\%12\relax}%
\providecommand \@@startlink[1]{}%
\providecommand \@@endlink[0]{}%
\providecommand \url  [0]{\begingroup\@sanitize@url \@url }%
\providecommand \@url [1]{\endgroup\@href {#1}{\urlprefix }}%
\providecommand \urlprefix  [0]{URL }%
\providecommand \Eprint [0]{\href }%
\providecommand \doibase [0]{https://doi.org/}%
\providecommand \selectlanguage [0]{\@gobble}%
\providecommand \bibinfo  [0]{\@secondoftwo}%
\providecommand \bibfield  [0]{\@secondoftwo}%
\providecommand \translation [1]{[#1]}%
\providecommand \BibitemOpen [0]{}%
\providecommand \bibitemStop [0]{}%
\providecommand \bibitemNoStop [0]{.\EOS\space}%
\providecommand \EOS [0]{\spacefactor3000\relax}%
\providecommand \BibitemShut  [1]{\csname bibitem#1\endcsname}%
\let\auto@bib@innerbib\@empty
\bibitem [{\citenamefont {Nowak}(2006{\natexlab{a}})}]{nowak2006five}%
  \BibitemOpen
  \bibfield  {author} {\bibinfo {author} {\bibfnamefont {M.~A.}\ \bibnamefont
  {Nowak}},\ }\href
  {http://science.sciencemag.org/content/314/5805/1560.abstract} {\bibfield
  {journal} {\bibinfo  {journal} {Science}\ }\textbf {\bibinfo {volume}
  {314}},\ \bibinfo {pages} {1560} (\bibinfo {year}
  {2006}{\natexlab{a}})}\BibitemShut {NoStop}%
\bibitem [{\citenamefont {Van~Segbroeck}\ \emph {et~al.}(2009)\citenamefont
  {Van~Segbroeck}, \citenamefont {Santos}, \citenamefont {Lenaerts},\ and\
  \citenamefont {Pacheco}}]{Van-Segbroeck:2009uq}%
  \BibitemOpen
  \bibfield  {author} {\bibinfo {author} {\bibfnamefont {S.}~\bibnamefont
  {Van~Segbroeck}}, \bibinfo {author} {\bibfnamefont {F.~C.}\ \bibnamefont
  {Santos}}, \bibinfo {author} {\bibfnamefont {T.}~\bibnamefont {Lenaerts}},\
  and\ \bibinfo {author} {\bibfnamefont {J.~M.}\ \bibnamefont {Pacheco}},\
  }\href {https://link.aps.org/doi/10.1103/PhysRevLett.102.058105} {\bibfield
  {journal} {\bibinfo  {journal} {Physical Review Letters}\ }\textbf {\bibinfo
  {volume} {102}},\ \bibinfo {pages} {058105} (\bibinfo {year}
  {2009})}\BibitemShut {NoStop}%
\bibitem [{\citenamefont {Helbing}\ \emph {et~al.}(2010)\citenamefont
  {Helbing}, \citenamefont {Szolnoki}, \citenamefont {Perc},\ and\
  \citenamefont {Szab{\'o}}}]{Helbing:2010bx}%
  \BibitemOpen
  \bibfield  {author} {\bibinfo {author} {\bibfnamefont {D.}~\bibnamefont
  {Helbing}}, \bibinfo {author} {\bibfnamefont {A.}~\bibnamefont {Szolnoki}},
  \bibinfo {author} {\bibfnamefont {M.}~\bibnamefont {Perc}},\ and\ \bibinfo
  {author} {\bibfnamefont {G.}~\bibnamefont {Szab{\'o}}},\ }\href
  {https://doi.org/10.1103/PhysRevE.81.057104} {\bibfield  {journal} {\bibinfo
  {journal} {Physical Review E}\ }\textbf {\bibinfo {volume} {81}},\ \bibinfo
  {pages} {057104} (\bibinfo {year} {2010})}\BibitemShut {NoStop}%
\bibitem [{\citenamefont {Perc}\ \emph {et~al.}(2017)\citenamefont {Perc},
  \citenamefont {Jordan}, \citenamefont {Rand}, \citenamefont {Wang},
  \citenamefont {Boccaletti},\ and\ \citenamefont {Szolnoki}}]{Perc:2017aa}%
  \BibitemOpen
  \bibfield  {author} {\bibinfo {author} {\bibfnamefont {M.}~\bibnamefont
  {Perc}}, \bibinfo {author} {\bibfnamefont {J.~J.}\ \bibnamefont {Jordan}},
  \bibinfo {author} {\bibfnamefont {D.~G.}\ \bibnamefont {Rand}}, \bibinfo
  {author} {\bibfnamefont {Z.}~\bibnamefont {Wang}}, \bibinfo {author}
  {\bibfnamefont {S.}~\bibnamefont {Boccaletti}},\ and\ \bibinfo {author}
  {\bibfnamefont {A.}~\bibnamefont {Szolnoki}},\ }\bibfield  {booktitle} {\emph
  {\bibinfo {booktitle} {Statistical physics of human cooperation}},\ }\href
  {https://doi.org/https://doi.org/10.1016/j.physrep.2017.05.004} {\bibfield
  {journal} {\bibinfo  {journal} {Physics Reports}\ }\textbf {\bibinfo {volume}
  {687}},\ \bibinfo {pages} {1} (\bibinfo {year} {2017})}\BibitemShut {NoStop}%
\bibitem [{\citenamefont {Lim}\ and\ \citenamefont
  {Wittek}(2018)}]{Lim:2018aa}%
  \BibitemOpen
  \bibfield  {author} {\bibinfo {author} {\bibfnamefont {I.~S.}\ \bibnamefont
  {Lim}}\ and\ \bibinfo {author} {\bibfnamefont {P.}~\bibnamefont {Wittek}},\
  }\href {https://doi.org/10.1103/PhysRevE.98.062113} {\bibfield  {journal}
  {\bibinfo  {journal} {Physical Review E}\ }\textbf {\bibinfo {volume} {98}},\
  \bibinfo {pages} {062113} (\bibinfo {year} {2018})}\BibitemShut {NoStop}%
\bibitem [{\citenamefont {Mittal}\ \emph {et~al.}(2020)\citenamefont {Mittal},
  \citenamefont {Mukhopadhyay},\ and\ \citenamefont
  {Chakraborty}}]{Mittal:2020aa}%
  \BibitemOpen
  \bibfield  {author} {\bibinfo {author} {\bibfnamefont {S.}~\bibnamefont
  {Mittal}}, \bibinfo {author} {\bibfnamefont {A.}~\bibnamefont
  {Mukhopadhyay}},\ and\ \bibinfo {author} {\bibfnamefont {S.}~\bibnamefont
  {Chakraborty}},\ }\href {https://doi.org/10.1103/PhysRevE.101.042410}
  {\bibfield  {journal} {\bibinfo  {journal} {Physical Review E}\ }\textbf
  {\bibinfo {volume} {101}},\ \bibinfo {pages} {042410} (\bibinfo {year}
  {2020})}\BibitemShut {NoStop}%
\bibitem [{\citenamefont {Smith}\ and\ \citenamefont
  {Price}(1973)}]{SMITH:1973aa}%
  \BibitemOpen
  \bibfield  {author} {\bibinfo {author} {\bibfnamefont {J.~M.}\ \bibnamefont
  {Smith}}\ and\ \bibinfo {author} {\bibfnamefont {G.~R.}\ \bibnamefont
  {Price}},\ }\href {https://doi.org/10.1038/246015a0} {\bibfield  {journal}
  {\bibinfo  {journal} {Nature}\ }\textbf {\bibinfo {volume} {246}},\ \bibinfo
  {pages} {15} (\bibinfo {year} {1973})}\BibitemShut {NoStop}%
\bibitem [{\citenamefont {Taylor}\ and\ \citenamefont
  {Jonker}(1978)}]{Taylor:1978aa}%
  \BibitemOpen
  \bibfield  {author} {\bibinfo {author} {\bibfnamefont {P.~D.}\ \bibnamefont
  {Taylor}}\ and\ \bibinfo {author} {\bibfnamefont {L.~B.}\ \bibnamefont
  {Jonker}},\ }\href
  {https://doi.org/https://doi.org/10.1016/0025-5564(78)90077-9} {\bibfield
  {journal} {\bibinfo  {journal} {Mathematical Biosciences}\ }\textbf {\bibinfo
  {volume} {40}},\ \bibinfo {pages} {145} (\bibinfo {year} {1978})}\BibitemShut
  {NoStop}%
\bibitem [{\citenamefont {Hofbauer}\ and\ \citenamefont
  {Sigmund}(1998)}]{hofbauer1998evolutionary}%
  \BibitemOpen
  \bibfield  {author} {\bibinfo {author} {\bibfnamefont {J.}~\bibnamefont
  {Hofbauer}}\ and\ \bibinfo {author} {\bibfnamefont {K.}~\bibnamefont
  {Sigmund}},\ }\href@noop {} {\emph {\bibinfo {title} {Evolutionary Games and
  Population Dynamics}}}\ (\bibinfo  {publisher} {Cambridge University Press},\
  \bibinfo {year} {1998})\BibitemShut {NoStop}%
\bibitem [{\citenamefont {Szab{\'o}}\ and\ \citenamefont
  {F{\'a}th}(2007)}]{Szabo:2007tg}%
  \BibitemOpen
  \bibfield  {author} {\bibinfo {author} {\bibfnamefont {G.}~\bibnamefont
  {Szab{\'o}}}\ and\ \bibinfo {author} {\bibfnamefont {G.}~\bibnamefont
  {F{\'a}th}},\ }\href
  {https://doi.org/https://doi.org/10.1016/j.physrep.2007.04.004} {\bibfield
  {journal} {\bibinfo  {journal} {Physics Reports}\ }\textbf {\bibinfo {volume}
  {446}},\ \bibinfo {pages} {97} (\bibinfo {year} {2007})}\BibitemShut
  {NoStop}%
\bibitem [{\citenamefont {Lieberman}\ \emph {et~al.}(2005)\citenamefont
  {Lieberman}, \citenamefont {Hauert},\ and\ \citenamefont
  {Nowak}}]{Lieberman:2005aa}%
  \BibitemOpen
  \bibfield  {author} {\bibinfo {author} {\bibfnamefont {E.}~\bibnamefont
  {Lieberman}}, \bibinfo {author} {\bibfnamefont {C.}~\bibnamefont {Hauert}},\
  and\ \bibinfo {author} {\bibfnamefont {M.~A.}\ \bibnamefont {Nowak}},\ }\href
  {https://doi.org/10.1038/nature03204} {\bibfield  {journal} {\bibinfo
  {journal} {Nature}\ }\textbf {\bibinfo {volume} {433}},\ \bibinfo {pages}
  {312} (\bibinfo {year} {2005})}\BibitemShut {NoStop}%
\bibitem [{\citenamefont {Fu}\ \emph {et~al.}(2008)\citenamefont {Fu},
  \citenamefont {Hauert}, \citenamefont {Nowak},\ and\ \citenamefont
  {Wang}}]{Fu:2008fk}%
  \BibitemOpen
  \bibfield  {author} {\bibinfo {author} {\bibfnamefont {F.}~\bibnamefont
  {Fu}}, \bibinfo {author} {\bibfnamefont {C.}~\bibnamefont {Hauert}}, \bibinfo
  {author} {\bibfnamefont {M.~A.}\ \bibnamefont {Nowak}},\ and\ \bibinfo
  {author} {\bibfnamefont {L.}~\bibnamefont {Wang}},\ }\href
  {https://link.aps.org/doi/10.1103/PhysRevE.78.026117} {\bibfield  {journal}
  {\bibinfo  {journal} {Physical Review E}\ }\textbf {\bibinfo {volume} {78}},\
  \bibinfo {pages} {026117} (\bibinfo {year} {2008})}\BibitemShut {NoStop}%
\bibitem [{\citenamefont {Nowak}\ and\ \citenamefont
  {Sigmund}(1998)}]{Nowak:1998aa}%
  \BibitemOpen
  \bibfield  {author} {\bibinfo {author} {\bibfnamefont {M.~A.}\ \bibnamefont
  {Nowak}}\ and\ \bibinfo {author} {\bibfnamefont {K.}~\bibnamefont
  {Sigmund}},\ }\href {https://doi.org/10.1038/31225} {\bibfield  {journal}
  {\bibinfo  {journal} {Nature}\ }\textbf {\bibinfo {volume} {393}},\ \bibinfo
  {pages} {573} (\bibinfo {year} {1998})}\BibitemShut {NoStop}%
\bibitem [{\citenamefont {Traulsen}\ \emph {et~al.}(2006)\citenamefont
  {Traulsen}, \citenamefont {Nowak},\ and\ \citenamefont
  {Pacheco}}]{traulsen2006stochastic}%
  \BibitemOpen
  \bibfield  {author} {\bibinfo {author} {\bibfnamefont {A.}~\bibnamefont
  {Traulsen}}, \bibinfo {author} {\bibfnamefont {M.~A.}\ \bibnamefont
  {Nowak}},\ and\ \bibinfo {author} {\bibfnamefont {J.~M.}\ \bibnamefont
  {Pacheco}},\ }\href {http://link.aps.org/doi/10.1103/PhysRevE.74.011909}
  {\bibfield  {journal} {\bibinfo  {journal} {Physical Review E}\ }\textbf
  {\bibinfo {volume} {74}},\ \bibinfo {pages} {011909} (\bibinfo {year}
  {2006})}\BibitemShut {NoStop}%
\bibitem [{\citenamefont {Taylor}\ \emph {et~al.}(2004)\citenamefont {Taylor},
  \citenamefont {Fudenberg}, \citenamefont {Sasaki},\ and\ \citenamefont
  {Nowak}}]{Taylor:2004aa}%
  \BibitemOpen
  \bibfield  {author} {\bibinfo {author} {\bibfnamefont {C.}~\bibnamefont
  {Taylor}}, \bibinfo {author} {\bibfnamefont {D.}~\bibnamefont {Fudenberg}},
  \bibinfo {author} {\bibfnamefont {A.}~\bibnamefont {Sasaki}},\ and\ \bibinfo
  {author} {\bibfnamefont {M.~A.}\ \bibnamefont {Nowak}},\ }\href
  {https://doi.org/https://doi.org/10.1016/j.bulm.2004.03.004} {\bibfield
  {journal} {\bibinfo  {journal} {Bulletin of Mathematical Biology}\ }\textbf
  {\bibinfo {volume} {66}},\ \bibinfo {pages} {1621} (\bibinfo {year}
  {2004})}\BibitemShut {NoStop}%
\bibitem [{\citenamefont {Johnson}\ and\ \citenamefont
  {Mislin}(2011)}]{Johnson:2011aa}%
  \BibitemOpen
  \bibfield  {author} {\bibinfo {author} {\bibfnamefont {N.~D.}\ \bibnamefont
  {Johnson}}\ and\ \bibinfo {author} {\bibfnamefont {A.~A.}\ \bibnamefont
  {Mislin}},\ }\href
  {https://doi.org/https://doi.org/10.1016/j.joep.2011.05.007} {\bibfield
  {journal} {\bibinfo  {journal} {Journal of Economic Psychology}\ }\textbf
  {\bibinfo {volume} {32}},\ \bibinfo {pages} {865} (\bibinfo {year}
  {2011})}\BibitemShut {NoStop}%
\bibitem [{\citenamefont {Manapat}\ \emph {et~al.}(2013)\citenamefont
  {Manapat}, \citenamefont {Nowak},\ and\ \citenamefont
  {Rand}}]{Manapat:2013aa}%
  \BibitemOpen
  \bibfield  {author} {\bibinfo {author} {\bibfnamefont {M.~L.}\ \bibnamefont
  {Manapat}}, \bibinfo {author} {\bibfnamefont {M.~A.}\ \bibnamefont {Nowak}},\
  and\ \bibinfo {author} {\bibfnamefont {D.~G.}\ \bibnamefont {Rand}},\
  }\bibfield  {booktitle} {\emph {\bibinfo {booktitle} {Evolution as a General
  Theoretical Framework for Economics and Public Policy}},\ }\href
  {https://doi.org/https://doi.org/10.1016/j.jebo.2012.10.018} {\bibfield
  {journal} {\bibinfo  {journal} {Journal of Economic Behavior \&
  Organization}\ }\textbf {\bibinfo {volume} {90}},\ \bibinfo {pages} {S57}
  (\bibinfo {year} {2013})}\BibitemShut {NoStop}%
\bibitem [{\citenamefont {Tarnita}(2015)}]{Tarnita:2015aa}%
  \BibitemOpen
  \bibfield  {author} {\bibinfo {author} {\bibfnamefont {C.}~\bibnamefont
  {Tarnita}},\ }\href@noop {} {\bibfield  {journal} {\bibinfo  {journal}
  {Games}\ }\textbf {\bibinfo {volume} {6}},\ \bibinfo {pages} {214} (\bibinfo
  {year} {2015})}\BibitemShut {NoStop}%
\bibitem [{\citenamefont {Berg}\ \emph {et~al.}(1995)\citenamefont {Berg},
  \citenamefont {Dickhaut},\ and\ \citenamefont {McCabe}}]{Berg:1995aa}%
  \BibitemOpen
  \bibfield  {author} {\bibinfo {author} {\bibfnamefont {J.}~\bibnamefont
  {Berg}}, \bibinfo {author} {\bibfnamefont {J.}~\bibnamefont {Dickhaut}},\
  and\ \bibinfo {author} {\bibfnamefont {K.}~\bibnamefont {McCabe}},\ }\href
  {https://doi.org/https://doi.org/10.1006/game.1995.1027} {\bibfield
  {journal} {\bibinfo  {journal} {Games and Economic Behavior}\ }\textbf
  {\bibinfo {volume} {10}},\ \bibinfo {pages} {122} (\bibinfo {year}
  {1995})}\BibitemShut {NoStop}%
\bibitem [{\citenamefont {McNamara}\ \emph {et~al.}(2009)\citenamefont
  {McNamara}, \citenamefont {Stephens}, \citenamefont {Dall},\ and\
  \citenamefont {Houston}}]{McNamara:2009aa}%
  \BibitemOpen
  \bibfield  {author} {\bibinfo {author} {\bibfnamefont {J.~M.}\ \bibnamefont
  {McNamara}}, \bibinfo {author} {\bibfnamefont {P.~A.}\ \bibnamefont
  {Stephens}}, \bibinfo {author} {\bibfnamefont {S.~R.~X.}\ \bibnamefont
  {Dall}},\ and\ \bibinfo {author} {\bibfnamefont {A.~I.}\ \bibnamefont
  {Houston}},\ }\href
  {http://rspb.royalsocietypublishing.org/content/276/1657/605.abstract}
  {\bibfield  {journal} {\bibinfo  {journal} {Proceedings of the Royal Society
  B: Biological Sciences}\ }\textbf {\bibinfo {volume} {276}},\ \bibinfo
  {pages} {605} (\bibinfo {year} {2009})}\BibitemShut {NoStop}%
\bibitem [{\citenamefont {Fehr}(2009)}]{Fehr:2009aa}%
  \BibitemOpen
  \bibfield  {author} {\bibinfo {author} {\bibfnamefont {E.}~\bibnamefont
  {Fehr}},\ }\bibfield  {booktitle} {\emph {\bibinfo {booktitle} {Journal of
  the European Economic Association}},\ }\href
  {https://doi.org/10.1162/JEEA.2009.7.2-3.235} {\bibfield  {journal} {\bibinfo
   {journal} {Journal of the European Economic Association}\ }\textbf {\bibinfo
  {volume} {7}},\ \bibinfo {pages} {235} (\bibinfo {year} {2009})}\BibitemShut
  {NoStop}%
\bibitem [{\citenamefont {Tzieropoulos}(2013)}]{Tzieropoulos:2013aa}%
  \BibitemOpen
  \bibfield  {author} {\bibinfo {author} {\bibfnamefont {H.}~\bibnamefont
  {Tzieropoulos}},\ }\bibfield  {booktitle} {\emph {\bibinfo {booktitle}
  {Social Neuroscience}},\ }\href
  {https://doi.org/10.1080/17470919.2013.832375} {\bibfield  {journal}
  {\bibinfo  {journal} {Social Neuroscience}\ }\textbf {\bibinfo {volume}
  {8}},\ \bibinfo {pages} {407} (\bibinfo {year} {2013})}\BibitemShut {NoStop}%
\bibitem [{\citenamefont {Abbass}\ \emph {et~al.}(2016)\citenamefont {Abbass},
  \citenamefont {Leu},\ and\ \citenamefont {Merrick}}]{Abbass:2016ac}%
  \BibitemOpen
  \bibfield  {author} {\bibinfo {author} {\bibfnamefont {H.~A.}\ \bibnamefont
  {Abbass}}, \bibinfo {author} {\bibfnamefont {G.}~\bibnamefont {Leu}},\ and\
  \bibinfo {author} {\bibfnamefont {K.}~\bibnamefont {Merrick}},\ }\bibfield
  {booktitle} {\emph {\bibinfo {booktitle} {IEEE Access}},\ }\href
  {https://doi.org/10.1109/ACCESS.2016.2571058} {\bibfield  {journal} {\bibinfo
   {journal} {IEEE Access}\ }\textbf {\bibinfo {volume} {4}},\ \bibinfo {pages}
  {2808} (\bibinfo {year} {2016})}\BibitemShut {NoStop}%
\bibitem [{\citenamefont {King-Casas}\ \emph {et~al.}(2005)\citenamefont
  {King-Casas}, \citenamefont {Tomlin}, \citenamefont {Anen}, \citenamefont
  {Camerer}, \citenamefont {Quartz},\ and\ \citenamefont
  {Montague}}]{King-Casas:2005aa}%
  \BibitemOpen
  \bibfield  {author} {\bibinfo {author} {\bibfnamefont {B.}~\bibnamefont
  {King-Casas}}, \bibinfo {author} {\bibfnamefont {D.}~\bibnamefont {Tomlin}},
  \bibinfo {author} {\bibfnamefont {C.}~\bibnamefont {Anen}}, \bibinfo {author}
  {\bibfnamefont {C.~F.}\ \bibnamefont {Camerer}}, \bibinfo {author}
  {\bibfnamefont {S.~R.}\ \bibnamefont {Quartz}},\ and\ \bibinfo {author}
  {\bibfnamefont {P.~R.}\ \bibnamefont {Montague}},\ }\href
  {https://doi.org/10.1126/science.1108062} {\bibfield  {journal} {\bibinfo
  {journal} {Science}\ }\textbf {\bibinfo {volume} {308}},\ \bibinfo {pages}
  {78} (\bibinfo {year} {2005})}\BibitemShut {NoStop}%
\bibitem [{\citenamefont {Masuda}\ and\ \citenamefont
  {Nakamura}(2012)}]{Masuda:2012aa}%
  \BibitemOpen
  \bibfield  {author} {\bibinfo {author} {\bibfnamefont {N.}~\bibnamefont
  {Masuda}}\ and\ \bibinfo {author} {\bibfnamefont {M.}~\bibnamefont
  {Nakamura}},\ }\href@noop {} {\bibfield  {journal} {\bibinfo  {journal} {PloS
  one}\ }\textbf {\bibinfo {volume} {7}},\ \bibinfo {pages} {e44169} (\bibinfo
  {year} {2012})}\BibitemShut {NoStop}%
\bibitem [{\citenamefont {Moran}(1962)}]{moran1962statistical}%
  \BibitemOpen
  \bibfield  {author} {\bibinfo {author} {\bibfnamefont {P.~A.~P.}\
  \bibnamefont {Moran}},\ }\href@noop {} {\emph {\bibinfo {title} {The
  statistical processes of evolutionary theory.}}}\ (\bibinfo  {publisher}
  {Clarendon Press},\ \bibinfo {address} {Oxford},\ \bibinfo {year}
  {1962})\BibitemShut {NoStop}%
\bibitem [{\citenamefont {Nowak}(2006{\natexlab{b}})}]{nowak2006evolutionary}%
  \BibitemOpen
  \bibfield  {author} {\bibinfo {author} {\bibfnamefont {M.~A.}\ \bibnamefont
  {Nowak}},\ }\href@noop {} {\emph {\bibinfo {title} {Evolutionary Dynamics:
  Exploring the Equations of Life}}}\ (\bibinfo  {publisher} {Harvard
  University Press},\ \bibinfo {year} {2006})\BibitemShut {NoStop}%
\bibitem [{\citenamefont {Fudenberg}\ and\ \citenamefont
  {Imhof}(2006)}]{Fudenberg:2006aa}%
  \BibitemOpen
  \bibfield  {author} {\bibinfo {author} {\bibfnamefont {D.}~\bibnamefont
  {Fudenberg}}\ and\ \bibinfo {author} {\bibfnamefont {L.~A.}\ \bibnamefont
  {Imhof}},\ }\href {https://doi.org/https://doi.org/10.1016/j.jet.2005.04.006}
  {\bibfield  {journal} {\bibinfo  {journal} {Journal of Economic Theory}\
  }\textbf {\bibinfo {volume} {131}},\ \bibinfo {pages} {251} (\bibinfo {year}
  {2006})}\BibitemShut {NoStop}%
\bibitem [{\citenamefont {Veller}\ and\ \citenamefont
  {Hayward}(2016)}]{Veller:2016aa}%
  \BibitemOpen
  \bibfield  {author} {\bibinfo {author} {\bibfnamefont {C.}~\bibnamefont
  {Veller}}\ and\ \bibinfo {author} {\bibfnamefont {L.~K.}\ \bibnamefont
  {Hayward}},\ }\href
  {https://doi.org/https://doi.org/10.1016/j.jet.2015.12.005} {\bibfield
  {journal} {\bibinfo  {journal} {Journal of Economic Theory}\ }\textbf
  {\bibinfo {volume} {162}},\ \bibinfo {pages} {93} (\bibinfo {year}
  {2016})}\BibitemShut {NoStop}%
\bibitem [{\citenamefont {Hauert}\ \emph {et~al.}(2007)\citenamefont {Hauert},
  \citenamefont {Traulsen}, \citenamefont {Brandt}, \citenamefont {Nowak},\
  and\ \citenamefont {Sigmund}}]{Hauert:2007fk}%
  \BibitemOpen
  \bibfield  {author} {\bibinfo {author} {\bibfnamefont {C.}~\bibnamefont
  {Hauert}}, \bibinfo {author} {\bibfnamefont {A.}~\bibnamefont {Traulsen}},
  \bibinfo {author} {\bibfnamefont {H.}~\bibnamefont {Brandt}}, \bibinfo
  {author} {\bibfnamefont {M.~A.}\ \bibnamefont {Nowak}},\ and\ \bibinfo
  {author} {\bibfnamefont {K.}~\bibnamefont {Sigmund}},\ }\href
  {http://science.sciencemag.org/content/316/5833/1905.abstract} {\bibfield
  {journal} {\bibinfo  {journal} {Science}\ }\textbf {\bibinfo {volume}
  {316}},\ \bibinfo {pages} {1905} (\bibinfo {year} {2007})}\BibitemShut
  {NoStop}%
\bibitem [{\citenamefont {Veller}\ \emph {et~al.}(2017)\citenamefont {Veller},
  \citenamefont {Hayward}, \citenamefont {Hilbe},\ and\ \citenamefont
  {Nowak}}]{Veller:2017aa}%
  \BibitemOpen
  \bibfield  {author} {\bibinfo {author} {\bibfnamefont {C.}~\bibnamefont
  {Veller}}, \bibinfo {author} {\bibfnamefont {L.~K.}\ \bibnamefont {Hayward}},
  \bibinfo {author} {\bibfnamefont {C.}~\bibnamefont {Hilbe}},\ and\ \bibinfo
  {author} {\bibfnamefont {M.~A.}\ \bibnamefont {Nowak}},\ }\href
  {https://doi.org/10.1073/pnas.1702020114} {\bibfield  {journal} {\bibinfo
  {journal} {Proceedings of the National Academy of Sciences}\ }\textbf
  {\bibinfo {volume} {114}},\ \bibinfo {pages} {E5396} (\bibinfo {year}
  {2017})}\BibitemShut {NoStop}%
\bibitem [{\citenamefont {Rand}\ \emph {et~al.}(2013)\citenamefont {Rand},
  \citenamefont {Tarnita}, \citenamefont {Ohtsuki},\ and\ \citenamefont
  {Nowak}}]{Rand:2013aa}%
  \BibitemOpen
  \bibfield  {author} {\bibinfo {author} {\bibfnamefont {D.~G.}\ \bibnamefont
  {Rand}}, \bibinfo {author} {\bibfnamefont {C.~E.}\ \bibnamefont {Tarnita}},
  \bibinfo {author} {\bibfnamefont {H.}~\bibnamefont {Ohtsuki}},\ and\ \bibinfo
  {author} {\bibfnamefont {M.~A.}\ \bibnamefont {Nowak}},\ }\href
  {https://doi.org/10.1073/pnas.1214167110} {\bibfield  {journal} {\bibinfo
  {journal} {Proceedings of the National Academy of Sciences}\ }\textbf
  {\bibinfo {volume} {110}},\ \bibinfo {pages} {2581} (\bibinfo {year}
  {2013})}\BibitemShut {NoStop}%
\bibitem [{\citenamefont {Tarnita}\ \emph {et~al.}(2011)\citenamefont
  {Tarnita}, \citenamefont {Wage},\ and\ \citenamefont
  {Nowak}}]{Tarnita:2011aa}%
  \BibitemOpen
  \bibfield  {author} {\bibinfo {author} {\bibfnamefont {C.~E.}\ \bibnamefont
  {Tarnita}}, \bibinfo {author} {\bibfnamefont {N.}~\bibnamefont {Wage}},\ and\
  \bibinfo {author} {\bibfnamefont {M.~A.}\ \bibnamefont {Nowak}},\ }\href@noop
  {} {\bibfield  {journal} {\bibinfo  {journal} {Proceedings of the National
  Academy of Sciences}\ }\textbf {\bibinfo {volume} {108}},\ \bibinfo {pages}
  {2334} (\bibinfo {year} {2011})}\BibitemShut {NoStop}%
\bibitem [{\citenamefont {Manapat}\ \emph {et~al.}(2012)\citenamefont
  {Manapat}, \citenamefont {Rand}, \citenamefont {Pawlowitsch},\ and\
  \citenamefont {Nowak}}]{Manapat:2012ab}%
  \BibitemOpen
  \bibfield  {author} {\bibinfo {author} {\bibfnamefont {M.~L.}\ \bibnamefont
  {Manapat}}, \bibinfo {author} {\bibfnamefont {D.~G.}\ \bibnamefont {Rand}},
  \bibinfo {author} {\bibfnamefont {C.}~\bibnamefont {Pawlowitsch}},\ and\
  \bibinfo {author} {\bibfnamefont {M.~A.}\ \bibnamefont {Nowak}},\ }\href
  {https://doi.org/https://doi.org/10.1016/j.jtbi.2012.03.014} {\bibfield
  {journal} {\bibinfo  {journal} {Journal of Theoretical Biology}\ }\textbf
  {\bibinfo {volume} {303}},\ \bibinfo {pages} {119} (\bibinfo {year}
  {2012})}\BibitemShut {NoStop}%
\bibitem [{\citenamefont {Hilbe}\ \emph {et~al.}(2013)\citenamefont {Hilbe},
  \citenamefont {Nowak},\ and\ \citenamefont {Sigmund}}]{Hilbe:2013aa}%
  \BibitemOpen
  \bibfield  {author} {\bibinfo {author} {\bibfnamefont {C.}~\bibnamefont
  {Hilbe}}, \bibinfo {author} {\bibfnamefont {M.~A.}\ \bibnamefont {Nowak}},\
  and\ \bibinfo {author} {\bibfnamefont {K.}~\bibnamefont {Sigmund}},\ }\href
  {https://doi.org/10.1073/pnas.1214834110} {\bibfield  {journal} {\bibinfo
  {journal} {Proceedings of the National Academy of Sciences}\ }\textbf
  {\bibinfo {volume} {110}},\ \bibinfo {pages} {6913} (\bibinfo {year}
  {2013})}\BibitemShut {NoStop}%
\bibitem [{\citenamefont {Du}\ \emph {et~al.}(2015)\citenamefont {Du},
  \citenamefont {Wu},\ and\ \citenamefont {Wang}}]{du2015aspiration}%
  \BibitemOpen
  \bibfield  {author} {\bibinfo {author} {\bibfnamefont {J.}~\bibnamefont
  {Du}}, \bibinfo {author} {\bibfnamefont {B.}~\bibnamefont {Wu}},\ and\
  \bibinfo {author} {\bibfnamefont {L.}~\bibnamefont {Wang}},\ }\href
  {http://dx.doi.org/10.1038/srep08014} {\bibfield  {journal} {\bibinfo
  {journal} {Scientific Reports}\ }\textbf {\bibinfo {volume} {5}},\ \bibinfo
  {pages} {8014 EP } (\bibinfo {year} {2015})}\BibitemShut {NoStop}%
\bibitem [{\citenamefont {Hilbe}\ \emph {et~al.}(2018)\citenamefont {Hilbe},
  \citenamefont {{\v S}imsa}, \citenamefont {Chatterjee},\ and\ \citenamefont
  {Nowak}}]{Hilbe:2018ab}%
  \BibitemOpen
  \bibfield  {author} {\bibinfo {author} {\bibfnamefont {C.}~\bibnamefont
  {Hilbe}}, \bibinfo {author} {\bibfnamefont {{\v S}.}~\bibnamefont {{\v
  S}imsa}}, \bibinfo {author} {\bibfnamefont {K.}~\bibnamefont {Chatterjee}},\
  and\ \bibinfo {author} {\bibfnamefont {M.~A.}\ \bibnamefont {Nowak}},\ }\href
  {https://doi.org/10.1038/s41586-018-0277-x} {\bibfield  {journal} {\bibinfo
  {journal} {Nature}\ }\textbf {\bibinfo {volume} {559}},\ \bibinfo {pages}
  {246} (\bibinfo {year} {2018})}\BibitemShut {NoStop}%
\bibitem [{\citenamefont {Houchmandzadeh}(2014)}]{Houchmandzadeh:2014aa}%
  \BibitemOpen
  \bibfield  {author} {\bibinfo {author} {\bibfnamefont {B.}~\bibnamefont
  {Houchmandzadeh}},\ }\href {https://doi.org/10.1007/s12038-013-9350-7}
  {\bibfield  {journal} {\bibinfo  {journal} {Journal of Biosciences}\ }\textbf
  {\bibinfo {volume} {39}},\ \bibinfo {pages} {249} (\bibinfo {year}
  {2014})}\BibitemShut {NoStop}%
\bibitem [{\citenamefont {Wu}\ \emph {et~al.}(2010)\citenamefont {Wu},
  \citenamefont {Altrock}, \citenamefont {Wang},\ and\ \citenamefont
  {Traulsen}}]{Wu:2010aa}%
  \BibitemOpen
  \bibfield  {author} {\bibinfo {author} {\bibfnamefont {B.}~\bibnamefont
  {Wu}}, \bibinfo {author} {\bibfnamefont {P.~M.}\ \bibnamefont {Altrock}},
  \bibinfo {author} {\bibfnamefont {L.}~\bibnamefont {Wang}},\ and\ \bibinfo
  {author} {\bibfnamefont {A.}~\bibnamefont {Traulsen}},\ }\href
  {https://doi.org/10.1103/PhysRevE.82.046106} {\bibfield  {journal} {\bibinfo
  {journal} {Physical Review E}\ }\textbf {\bibinfo {volume} {82}},\ \bibinfo
  {pages} {046106} (\bibinfo {year} {2010})}\BibitemShut {NoStop}%
\bibitem [{\citenamefont {Ohtsuki}(2010)}]{Ohtsuki:2010aa}%
  \BibitemOpen
  \bibfield  {author} {\bibinfo {author} {\bibfnamefont {H.}~\bibnamefont
  {Ohtsuki}},\ }\href
  {https://doi.org/https://doi.org/10.1016/j.jtbi.2010.01.016} {\bibfield
  {journal} {\bibinfo  {journal} {Journal of Theoretical Biology}\ }\textbf
  {\bibinfo {volume} {264}},\ \bibinfo {pages} {136} (\bibinfo {year}
  {2010})}\BibitemShut {NoStop}%
\bibitem [{\citenamefont {Sigmund}\ \emph {et~al.}(2010)\citenamefont
  {Sigmund}, \citenamefont {De~Silva}, \citenamefont {Traulsen},\ and\
  \citenamefont {Hauert}}]{Sigmund:2010aa}%
  \BibitemOpen
  \bibfield  {author} {\bibinfo {author} {\bibfnamefont {K.}~\bibnamefont
  {Sigmund}}, \bibinfo {author} {\bibfnamefont {H.}~\bibnamefont {De~Silva}},
  \bibinfo {author} {\bibfnamefont {A.}~\bibnamefont {Traulsen}},\ and\
  \bibinfo {author} {\bibfnamefont {C.}~\bibnamefont {Hauert}},\ }\href
  {https://doi.org/10.1038/nature09203} {\bibfield  {journal} {\bibinfo
  {journal} {Nature}\ }\textbf {\bibinfo {volume} {466}},\ \bibinfo {pages}
  {861} (\bibinfo {year} {2010})}\BibitemShut {NoStop}%
\bibitem [{\citenamefont {Van~Segbroeck}\ \emph {et~al.}(2012)\citenamefont
  {Van~Segbroeck}, \citenamefont {Pacheco}, \citenamefont {Lenaerts},\ and\
  \citenamefont {Santos}}]{Van-Segbroeck:2012aa}%
  \BibitemOpen
  \bibfield  {author} {\bibinfo {author} {\bibfnamefont {S.}~\bibnamefont
  {Van~Segbroeck}}, \bibinfo {author} {\bibfnamefont {J.~M.}\ \bibnamefont
  {Pacheco}}, \bibinfo {author} {\bibfnamefont {T.}~\bibnamefont {Lenaerts}},\
  and\ \bibinfo {author} {\bibfnamefont {F.~C.}\ \bibnamefont {Santos}},\
  }\href {https://doi.org/10.1103/PhysRevLett.108.158104} {\bibfield  {journal}
  {\bibinfo  {journal} {Physical Review Letters}\ }\textbf {\bibinfo {volume}
  {108}},\ \bibinfo {pages} {158104} (\bibinfo {year} {2012})}\BibitemShut
  {NoStop}%
\bibitem [{\citenamefont {Requejo}\ \emph {et~al.}(2012)\citenamefont
  {Requejo}, \citenamefont {Camacho}, \citenamefont {Cuesta},\ and\
  \citenamefont {Arenas}}]{Requejo:2012aa}%
  \BibitemOpen
  \bibfield  {author} {\bibinfo {author} {\bibfnamefont {R.~J.}\ \bibnamefont
  {Requejo}}, \bibinfo {author} {\bibfnamefont {J.}~\bibnamefont {Camacho}},
  \bibinfo {author} {\bibfnamefont {J.~A.}~\bibnamefont {Cuesta}},\ and\ \bibinfo
  {author} {\bibfnamefont {A.}~\bibnamefont {Arenas}},\ }\href
  {https://doi.org/10.1103/PhysRevE.86.026105} {\bibfield  {journal} {\bibinfo
  {journal} {Physical Review E}\ }\textbf {\bibinfo {volume} {86}},\ \bibinfo
  {pages} {026105} (\bibinfo {year} {2012})}\BibitemShut {NoStop}%
\bibitem [{\citenamefont {Wang}\ \emph {et~al.}(2010)\citenamefont {Wang},
  \citenamefont {Wu}, \citenamefont {Chen},\ and\ \citenamefont
  {Wang}}]{Wang:2010aa}%
  \BibitemOpen
  \bibfield  {author} {\bibinfo {author} {\bibfnamefont {J.}~\bibnamefont
  {Wang}}, \bibinfo {author} {\bibfnamefont {B.}~\bibnamefont {Wu}}, \bibinfo
  {author} {\bibfnamefont {X.}~\bibnamefont {Chen}},\ and\ \bibinfo {author}
  {\bibfnamefont {L.}~\bibnamefont {Wang}},\ }\href
  {https://doi.org/10.1103/PhysRevE.81.056103} {\bibfield  {journal} {\bibinfo
  {journal} {Physical Review E}\ }\textbf {\bibinfo {volume} {81}},\ \bibinfo
  {pages} {056103} (\bibinfo {year} {2010})}\BibitemShut {NoStop}%
\bibitem [{\citenamefont {Wu}\ \emph {et~al.}(2012)\citenamefont {Wu},
  \citenamefont {Gokhale}, \citenamefont {Wang},\ and\ \citenamefont
  {Traulsen}}]{Wu:2012aa}%
  \BibitemOpen
  \bibfield  {author} {\bibinfo {author} {\bibfnamefont {B.}~\bibnamefont
  {Wu}}, \bibinfo {author} {\bibfnamefont {C.~S.}\ \bibnamefont {Gokhale}},
  \bibinfo {author} {\bibfnamefont {L.}~\bibnamefont {Wang}},\ and\ \bibinfo
  {author} {\bibfnamefont {A.}~\bibnamefont {Traulsen}},\ }\href
  {https://doi.org/10.1007/s00285-011-0430-8} {\bibfield  {journal} {\bibinfo
  {journal} {Journal of Mathematical Biology}\ }\textbf {\bibinfo {volume}
  {64}},\ \bibinfo {pages} {803} (\bibinfo {year} {2012})}\BibitemShut
  {NoStop}%
\end{thebibliography}


%

\end{document}